\documentclass[acmsmall,final, screen]{acmart}

\usepackage{amsmath}
\usepackage{amsthm}
\usepackage{stmaryrd}
\usepackage{relsize}
\usepackage{hyperref}
\usepackage[htt]{hyphenat}
\usepackage{relsize}
\usepackage{mathtools}
\usepackage{graphicx}
\usepackage{algorithm}
\usepackage{algpseudocode}
\usepackage{algorithmicx}
\usepackage{tikz}
\usetikzlibrary{decorations.pathmorphing}
\usetikzlibrary{shapes}
\usetikzlibrary{automata,positioning}
\usetikzlibrary {arrows.meta}
\usepackage{enumerate}

\makeatletter
\DeclareOldFontCommand{\rm}{\normalfont\rmfamily}{\mathrm}
\DeclareOldFontCommand{\sf}{\normalfont\sffamily}{\mathsf}
\DeclareOldFontCommand{\tt}{\normalfont\ttfamily}{\mathtt}
\DeclareOldFontCommand{\bf}{\normalfont\bfseries}{\mathbf}
\DeclareOldFontCommand{\it}{\normalfont\itshape}{\mathit}
\DeclareOldFontCommand{\sl}{\normalfont\slshape}{\@nomath\sl}
\DeclareOldFontCommand{\sc}{\normalfont\scshape}{\@nomath\sc}
\makeatother

\newcommand{\intchoice}{\sqcap}

\newcommand{\refinedby}[1]{\sqsubseteq_{#1}}
\newcommand{\trace}[1]{\langle{}#1\rangle}
\newcommand{\cspm}{CSP$_{\textrm M}$}

\newcommand{\then}{\rightarrow}
\newcommand{\extchoice}{{\mathord{\Box}}} 

\begin{document}

\title[]{A Fair Kernel-Lock-Free Claim/Release Protocol for Shared Object Access in Cooperatively Scheduled Runtimes}

\author{Kevin Chalmers}
\orcid{0000-0002-3409-432X}
\affiliation{%
   \institution{Ravensbourne University}
   \department{School of Computing, Architecture, and Emerging Technologies, Ravensbourne University}
   \streetaddress{6 Penrose Way}
   \city{London}
   \postcode{SE10 0EW}
   \country{United Kingdom}}
\email{k.chalmers@rave.ac.uk}
\author{Jan B\ae{}kgaard Pedersen}
\orcid{0000-0002-2800-5095}
\affiliation{%
   \institution{University of Nevada Las Vegas}
   \department{Department of Computer Science}
   \streetaddress{4505 South Maryland Parkway}
   \city{Las Vegas}
   \state{NV}
   \postcode{89154}
   \country{United States of America}}
\email{matt.pedersen@unlv.edu}



\begin{abstract}
We present the first spin-free, kernel-lock-free mutex that cooperates with user-mode schedulers and is formally proven FIFO-fair and linearizable using CSP/FDR. Our fairness oracle and stability-based proof method are reusable across coroutine runtime designs. We designed the claim/release protocol for a process-oriented language --- ProcessJ --- to manage the race for claiming shared inter-process communication channels. Internally, we use a lock-free queue to park waiting processes for gaining access to a shared object, such as exclusive access to a shared channel to read from or write to. The queue ensures control and fairness for processes wishing to access a shared resource, as the protocol handles claim requests in the order they are inserted into the queue. We produce CSP models of our protocol and a mutex specification, demonstrating with FDR that our protocol behaves as a locking mutex.
\end{abstract}

\maketitle

\renewcommand{\shortauthors}{K.Chalmers and J.B.Pedersen}

\section{Introduction}

Concurrency enables faster task completion by exploiting available hardware resources. However, enabling concurrency introduces additional challenges often resolved via exclusive access controls to shared resources. Unfortunately, exclusive access leads to reduced concurrency, with concurrent behaviour potentially minimising to sequential behaviour. As such, developers are interested in lock-free mechanisms that increase concurrency by transforming exclusive access into safe transactional interactions between threads.

In this paper, we describe a claim/release protocol for cooperatively scheduled runtimes that is {\it system lock-free}, {\it spinlock-free}, {\it fair}, while ensuring {\it exclusive access to shared resources}. That is:
\begin{itemize}
    \item The runtime's {\it cooperatively scheduled processes} do not wait on {\it system/kernel} objects, nor do the system thread process runners block during claiming.
    \item The claim/release does not use busy claim loops.
    \item A FIFO queue ensures no process claims before existing waiters.
    \item Only one cooperatively scheduled process owns the protected resource.
\end{itemize}

We designed our protocol for a specific problem --- shared channel access control for a lock-free runtime under-development for ProcessJ~\cite{PedersenChalmers}. However, we present the protocol generally as it should be applicable to other cooperatively scheduled runtimes, given:
\begin{itemize}
    \item The runtime has a process state variable to determine process liveness that includes states equivalent to {\it active, engaging/claiming, scheduled,} and {\it waiting}.
    \item The runtime provides a {\it schedule} mechanism to wake waiting processes.
    \item The runtime has access to lock-free queues to park waiting processes.
    \item The runtime can use the result from {\it claim} to determine process {\it yielding} --- although we verify with an integrated yield.
\end{itemize}

Furthermore, our algorithm relies on visible happens-before memory model ordering for shared objects. Our modelling assumes {\it sequential consistency} --- the standard CAS memory semantics for Java~\cite{JavaMemModel}. We require no additional memory fences for correctness.

We have verified our claim/release protocol using the FDR~\cite{FDR4} model checker, using techniques from our previous work on specifying lock-free data structures in CSP~\cite{software4030015}. The protocol only uses {\it compare-and-swap} (CAS) operations with no spinlocks (busy waits). We present a simple algorithm where cooperatively scheduled processes wait to access a shared resource via queue parking. On parking, the runtime should free system thread resources to allow other process execution.

In this paper, we use the term {\it thread} to refer to a system-level thread (e.g., provided by the operating system), and {\it process} to refer to cooperatively scheduled threads. We will also use the term {\it mutex} to refer to any mechanism ensuring mutual exclusion, including our claim/release protocol. When referring to {\it lock-free}, we mean system-level lock-freedom rather than lock-freedom within processes. For example, our claim/release protocol is obviously a locking mechanism, but that mechanism requires no system-level locks that would block runner threads.

\subsection{Context and Motivation}

ProcessJ is a process-oriented programming language developed in collaboration between the University of Nevada, Las Vegas, and Ravensbourne University, London. ProcessJ has a JVM-based runtime and the ProcessJ compiler translates ProcessJ programs into Java. A runtime system, also written in Java, executes alongside the resulting ProcessJ Java program. The runtime has a {\it cooperative scheduler} with lightweight processes and fast context-switching. ProcessJ has a multithreaded runtime, with multiple runners (Java threads) executing concurrent process code. 

ProcessJ inter-process communication occurs via synchronous, sometimes shared, channels. ProcessJ currently uses {\tt synchronized} Java methods for shared object access control, creating bottlenecks by blocking runner threads during operations.

To mitigate this issue, we are developing a lock-free runtime for ProcessJ. This paper addresses a specific issue for our lock-free runtime: {\it accessing shared channels}. Any shared channel must control access to its ends (reading or writing). With preemptively scheduled Java threads, we would use {\tt synchronized} methods. However, for ProcessJ we must build a mutual exclusion mechanism based on the scheduler's behaviour. Currently, there is a queue of waiting processes and {\tt synchronized} methods for claim and release. These methods lock the ProcessJ channel object, meaning only one process (i.e., exclusively either one reader or one writer) can attempt to claim ownership at a time. Shared channel access is therefore a sequential bottleneck which we aim to remove. Within this paper is our solution for ensuring safe, system-lock-free access to shared channels in the ProcessJ runtime. We build a lock-free mutex for our cooperatively scheduled runtime, and prove that it is correct. As we do not reference the ProcessJ runtime in the protocol, our solution is applicable to other cooperatively scheduled runtimes given the conditions defined above.

\subsection{Contribution}

Our work has the following contributions:

\begin{itemize}
    \item Completely spin-free, CAS-only and scheduler-cooperative mutex for coroutine runtimes. The lock never busy waits --- a waiter always yields back to the runtime, releasing system resources.
    \item Interleaving the mutex with a cooperative runtime's process state interaction and a queue-based lock, proving this scheduler interaction preserves mutual exclusion and FIFO fairness.
    \item Compositional CSP specification of a linearizable mutex, supporting overlapping claim and release operations.
    \item Verification strategy that is rooted in linearization but avoids explicit linearization point observation, proving refinement to a specification where commit points are hidden then arguing on system stability. 
    \item Use of a fairness oracle ($FAIR$) that checks FIFO order without embedding fairness into the mutex specification.
    \item Cost-path analysis for the coroutine mutex lock that lets runtime implementers predict CAS operations under lock contention.
    \item Survey table contrasting lock-freedom, spin-freedom, fairness and verification across mainstream coroutine mutexes and two classic queue locks.
\end{itemize}

\subsection{Breakdown of the Paper}

In Section~\ref{sec:background} we provide background information to our work, including the ProcessJ runtime operational context and an overview of Communicating Sequential Processes. Section~\ref{sec:protocol} presents our claim/release protocol, including a discussion of its runtime environment requirements. This section includes an overview of the behaviour of the algorithm and why we undertake the checks we do to ensure safe ownership transfer. In Section~\ref{sec:verify} we describe how we derived the CSP models for our protocol. Section~\ref{sec:spec} discusses how we go about defining a suitable specification for our claim/release protocol, as well as the additional properties tested. We discuss the results of our verification in Section~\ref{sec:discussion}, including limitation considerations. Related work is presented in Section~\ref{sec:related}, and we provide a summary of conclusions, limitations, and future work in Section~\ref{sec:conclusion}. 
\section{Background}
\label{sec:background}

In this section we provide background information to our work. First, we will examine the ProcessJ runtime to understand the operational context of our work, including the shared channel problem that motivates our work. Finally, we end the background section by presenting CSP, the underpinning method validating our work.

\subsection{ProcessJ Runtime}
\label{sec:ProcessJRuntime}

The ProcessJ runtime architecture is important to appreciate the issues that arise as we transition to a lock-free runtime. We consider what a ProcessJ process is, how the ProcessJ runtime executes processes, and how the runtime manages shared channel access.

\subsubsection{Processes in the ProcessJ Runtime}

ProcessJ represents a process as a Java object ({\tt PJProcess}) with a run method for process code and three metadata properties:
\begin{itemize}
    \item A {\tt runLabel} (instruction pointer) integer tracking the last yield point.
    \item A {\tt ready} flag signifying if the process is ready to run.
    \item A {\tt running} flag signifying if the process is actively executing.
\end{itemize}

A process can be not-ready {\it and} running while it prepares to potentially wait during a synchronisation. {\tt yield}() will release the runner thread executing the process. If the process is ready, yielding will lead the process to be returned to the run queue, as {\tt ready} is set independently of {\tt yield}().

{\tt schedule}({\tt pid}) will attempt to add a process {\tt pid} to the run queue. It will first attempt to set {\tt ready} to true. If the process was not-ready and not-running, {\tt schedule}({\tt pid}) will enqueue the process to the run queue.

ProcessJ processes yield to release runner thread resources. During code generation, the ProcessJ compiler inserts {\it yield points} at synchronisation points --- programmers never explicitly call {\tt yield}().

\subsubsection{The Runner}

A {\it runner} loops, dequeuing a process from the run queue and executing its {\tt run}() method. When {\tt run}() returns (by yielding or terminating), the runner dequeues a new process to execute. Currently, the ProcessJ run queue is concurrent and blocking. Only ready-to-run processes are kept in the run queue. \textit{schedule} will enqueue not-ready-to-run processes, as will \textit{yield}. The main task of \textit{yield} is to set a process not-running.

\begin{figure}[h!]
\begin{tabbing}
==\===\===\===\=\kill
\>{\it Queue}$<${\it Process}$>$ {\it processQueue};\\
\>...\\
\>// enqueue 1 or more processes to run.\\
\>...\\
\>{\bf while} (!{\it processQueue.isEmpty}()) \{\\
\>\>{\it Process} {\it p} = {\it processQueue.dequeue}();\\
\>\>{\it p.run}();\\
\>\}
\end{tabbing}
\caption{The ProcessJ Scheduler (Simplified).}
\label{fig:PJScheduler}
\end{figure}

\paragraph{Modelling Processes and Runners}

There are two different approaches to modelling scheduling within FDR~\cite{PedersenChalmers}:
\begin{itemize}
	\item Model the scheduler system with a fixed number of runners. This approach provides an accurate representation of cooperative scheduling.
	\item Allow any {\it ready} process to execute, thus modelling a scheduler with unlimited resources.
\end{itemize}

We take the second approach in this paper as it is more general. Our mutex must ensure that only one process can own it. Therefore, we model the mutex in the worst-case scenario. For our protocol, we combine {\it ready} and {\it running} into a single {\it state} variable with four values. We will discuss this further when presenting the protocol in Section~\ref{sec:protocol} and examine the modelling approach in Section~\ref{sec:verify}.

\subsubsection{Shared Channels}

Only a single writer and reader can use a one-to-one channel. Numerous processes can access a shared channel, but only two (one reader and one writer) can use it simultaneously. A shared channel keeps track of its current reader and writer while maintaining lists of ({\it not ready}) processes awaiting (read or write) access. Currently, ProcessJ protects access ownership via {\tt synchronized} methods. When removing this locking, we must ensure consistency between the owner state and the waitlist state. Making the owner field atomic and the queue lock-free is insufficient as it introduces a race condition. Between a process releasing ownership and looking at the wait queue another process can attempt to claim ownership by adding itself to the wait queue. Either process in this interaction might not observe the other, meaning lock ownership is not transferred but rather one process waits in the queue with no process to notify it. We explore this particular interaction within our protocol in Section~\ref{sec:diagrammatic}.

\subsection{Communicating Sequential Processes}
\label{sec:CSP}

Communicating Sequential Processes (CSP)~\cite{Ho78,Ho85,theoryAndPractice,awr135} is a process algebra for specifying concurrent systems through {\em processes} and {\em events}. Processes are abstract components defined by the events they perform. Events in CSP are {\em atomic}, {\em synchronous}, and {\em instantaneous}. A CSP model is a collection of processes composed together to form a system.

The simplest CSP processes are {\sf STOP}, performing no events without terminating, and SKIP, performing no events and terminating. We introduce process events using the {\em prefix} operator ($\then$). For example, $P = x \then STOP$ performs $x$ and then stops. The general form is $Process = event \then Process'$, and processes can be recursive (e.g., $P = x \then P$).

\subsubsection{Choice}
\label{sec:choice}

CSP provides several {\em choice} operators for branching behaviour. The standard operators are {\em external (or deterministic) choice}, {\em internal (or non-deterministic) choice}, and {\em prefix choice}.

Given the processes $P$ and $Q$, $P \ \extchoice \ Q$ (external choice) will behave as $P$ or $Q$ depending on the events the environment offers. For example, the process:
\begin{align*}
P = (a \then Q) \ \extchoice \ (b \then R)
\end{align*}

\noindent can accept $a$ and behave as $Q$, or accept $b$ and behave as $R$. If both events are available, the system will non-deterministically choose one option.

$P \ \sqcap \ Q$ (internal choice) will behave as $P$ or $Q$ without considering the environment --- the choice is non-deterministic. We can also apply external and internal choice to a set of events, allowing simpler model expression.

Prefix choice is an event with a parameter. If we define an event set $\{ c.v \ | \ v \in Values \}$, $c$ is a channel communicating a value $v$. We can use input and output operations ($?$ and $!$) to define communication and bind variables. CSP channels can communicate multiple parameters at once (e.g., $c!x!y$) and can mix input and output operations in a single communication (e.g., $c!x!y?z$).

\paragraph{Pre-Guards}

We can control choice branch availability using Boolean expressions:
\begin{align*}
e_1\ \&\ c?x \then \ldots\ \extchoice\ e_2\ \&\ d?x \then \ldots
\end{align*}

If $e_1$ is true and $e_2$ is false, only $c?x$ is considered.

\subsubsection{Process Composition}
\label{sec:processComposition}

We combine processes through parallel and sequential composition. Parallel composition, $P \ \big|\big| \ Q$, comes in three forms:
\begin{description}
	\item[Generalised parallel] $P \ \underset{A}{\big|\big|} \ Q$ means $P$ and $Q$ synchronise on the event set $A$, but each can perform any other event independently.
	\item[Alphabetised parallel] $P \ \underset{A \hspace{8pt} B}{\big|\big|} \ Q$ means $P$ can only perform events in $A$, $Q$ in $B$, and $P$ and $Q$ must synchronise on the intersection of $A$ and $B$.
	\item[Interleaving parallel] $P \ \big|\big|\big| \ Q$ means $P$ and $Q$ do not synchronise on any event. 
\end{description}

Sequential composition, $P \ ; \ Q$, means $P$ must terminate before $Q$ can begin. As with choice, process composition can be replicated across a set.

\subsubsection{Traces and Hiding}
\label{sec:tracesHiding}

The {\it trace set} of a process is {\it all} the observed event sequences of a process. Given the process $P = a \then P$, its shortest observable trace is empty, $\trace{}$. The trace set of $P$ also contains different sequence lengths of $a$, thus $traces(P) = \{\trace{}, \trace{a}, \trace{a, a}, \trace{a, a, a}, \ldots\}$.

When comparing traces, we can conceal events via the hiding operator $\backslash$, replacing hidden events with $\tau$ which are ignored in traces. For instance, $(a \then b \then a \then SKIP) \ \backslash \ \{ a \}$ has traces $\{ \trace{}, \trace{b} \}$.

\subsubsection{Models}
\label{sec:models}

CSP defines three semantic behavioural models to analyse systems: the {\it traces}, {\it failures}, and {\it failures/divergence} models. Each model builds on the previous one, providing more stringent refinement checks.

\paragraph{The Traces Model}
\label{sec:tracesModel}

For an implementation model $Q$ and a specification model $P$, if $traces(Q) \subseteq traces(P)$, then $P \refinedby{T} Q$ ($Q$ {\it trace refines} $P$, the implementation trace refines the specification). The specification defines the set of allowable traces.

\paragraph{The Stable Failures Model}
\label{sec:failuresModel}

{\em Failures} are events that a process refuses to engage with after a given trace. A {\em stable state} is one where a process cannot make internal progress (i.e., via hidden events) and must engage externally. A {\em refusal} is an event a process will not participate in when in a stable state. Stable failures comparison overcomes limitations in trace comparison. $(P = a \then P) \refinedby{T} ((P = a \then P) \ \intchoice \ {\sf STOP})$ will pass, but the right-hand side may non-deterministically choose to {\it STOP} and perform no events. We are testing if the implementation performs good behaviour.

A {\em failure} is a pair $(s, X)$, where $s$ is a trace, and $X$ is the set of events that the process refuses after performing trace $s$. We define a process $P$'s complete set of failures as $failures(P)$. $P \refinedby{F} Q$ means that, for all stable states, $P$ refuses an event when $Q$ refuses an event.

\paragraph{The Failures-Divergences Model}
\label{sec:failuresDivergenceModel}

{\em Divergences} address any livelock scenarios where a process continuously performs internal actions without making external progress. For example:
\begin{align*}
P &= a \then {\sf STOP} \\
Q &= (a \then {\sf STOP}) \ \big|\big|\big| \ {\sf DIV}
\end{align*}

\noindent where ${\sf DIV}$ is a process that immediately diverges. $P \refinedby{T} Q$ and $P \refinedby{F} Q$, but $Q$ can continuously not accept $a$ by performing $\tau$.

Formally, we define the failures and divergences of a process as: $({\it failures}_\bot(P), {\it divergences}(P))$, where refinement requires:
\begin{align*}
failures(Q) \subseteq failues(P) \land divergences(Q) \subseteq divergences(P)
\end{align*}

${\it failures}_\bot(P)$ is ${\it failures}(P) \cup \{ (x, X) \ | \ x \in {\it divergences}(P) \}$ --- the set of traces leading to divergences are added to the stable failures set to form an extended failure set.

\paragraph{Simplified Refinement Checking}
\label{sec:simplifiedRefinementChecking}

If both the specification and the implementation are free from divergence, the analysis only needs to establish equivalence in the stable failures model. Our models are all divergence-free, and thus we only verify in the stable failures model.

\subsubsection{FDR}
\label{sec:fdr}

FDR~\cite{FDR4} is a refinement checker for verifying CSP models written in \cspm\space ({\it machine-readable} CSP). FDR can check properties (deadlock and divergence freedom, determinism) and compare implementation models against specifications in the three semantic models.

FDR also supports a module system to simplify the encapsulation of definitions. We can parameterise modules on instantiation, allowing the creation of different experiment configurations.
\section{System-Lock-Free Claim/Release for ProcessJ}
\label{sec:protocol}

In this section we present our lock-free claim/release protocol. We begin by presenting the current ProcessJ claim/release mechanism, providing context to how our the protocol must interact with the ProcessJ runtime. We then present the lock-free claim/release protocol devised for ProcessJ shared channel access control. This section concludes with an analysis of the protocol's behaviour, including a diagrammatic analysis of the different claim-release interactions.

\subsection{Current Locking Mechanism}

In the lock-based ProcessJ runtime we use Java {\tt synchronized} methods and code blocks:

\vspace{\lineskip}
\begin{tabbing}
=\==\==\==\==\==\==\======================\=\kill
\>{\bf synchronized} ({\tt thisChannel}) \{\\
\>\>// code \\
\>\}
\end{tabbing}
\vspace{\lineskip}

The Java Runtime Environment (JRE) associates a monitor with objects that have {\tt synchronized} code. Effectively, {\tt synchronized} code causes a claim then release of the monitor lock:

\vspace{\lineskip}
\begin{tabbing}
=\==\==\==\==\==\==\======================\=\kill
\>{\bf claim}({\tt lock});\\
\>// code\\
\>{\bf release}({\tt lock});
\end{tabbing}
\vspace{\lineskip}

Within the ProcessJ runtime, we control access to a shared channel in two stages:
\begin{enumerate}
  \item Communication operations ({\tt read}() or {\tt write}()) are {\tt synchronized}, permitting one thread access to the channel. Thus, there is only ever one process claiming a shared channel.
  \item A communication operation ({\tt read}() or {\tt write}()) calls a {\it claim operation} (e.g., {\tt claimWrite}()) to check channel operation ownership. If owned, the runtime parks the process in a wait queue for the operation (e.g., {\tt writeQueue}). If there is no owner, the runtime permits full operation invocation.
\end{enumerate}

On completion, the process frees the channel via a {\it release operation} (e.g., {\tt unclaimWrite}()), waking any next waiting process.

The ProcessJ compiler generates the Java code for these steps. For example, when we call {\tt write}():

\begin{tabbing}
==\===\kill
\>{\tt out.write(42);}
\end{tabbing}

\noindent
the Java code generated by the ProcessJ compiler is similar to:\\

\begin{tabbing}
==\===\============================\===\====\===\kill
\>{\tt ...}\\
\>{\tt if (!out.claimWrite(this) \{} \>\>\> // try to claim the channel end\\
\>\>{\tt this.runLabel = 1;}           \>\> // \>if not successful\\
\>\>{\tt this.yield();}                \>\> // \>yield and resume at label(1)\\
\>{\tt \}}\\
\>{\tt label(1);}                    \>\>\> // resume point 1\\
$ $\\
\>{\tt out.write(this, 42);}         \>\>\> // at this point we have the lock\\
\>                                   \>\>\> // and we can write on the channel\\
\>{\tt this.runLabel = 2;}           \>\>\> // yield and resume at label(2)\\
\>{\tt this.yield();}\\
\>{\tt label(2);}                    \>\>\> // resume point 2\\
$ $\\
\>{\tt out.unclaimWrite();}          \>\>\> // unclaim the channel end\\
\end{tabbing}

Our goal is lock-freedom, and therefore our claim/release protocol must be able to:
\begin{enumerate}
  \item Ensure access management to the shared channel by multiple concurrent readers and writers --- removing the sequential bottleneck.
  \item Provide a {\it claim} method that returns operation success, allowing continuation or yielding as appropriate.
\end{enumerate}

(1) is a trivial outcome of a lock-free algorithm --- both {\tt read} and {\tt write} operations have wait queues. The operations ({\tt write}() and {\tt read}()) will provide their own means of ensuring correct behaviour under lock-free conditions which is outside the scope of this paper. For (2), we must ensure that {\it claim} correctly returns true or false to support correct yielding behaviour.

\subsection{Lock-Free Claim/Release Protocol}

We present the pseudocode for our new claim/release protocol below. In Algorithm~\ref{alg:claim} we present the claim method, and in Algorithm~\ref{alg:release} we present the release method. Both are lock-free, using atomic variables and compare-and-swap (CAS) atomic operations. In addition, the queue is lock-free. CAS operations control access to the {\tt owner} field, queue, and process state.

For ProcessJ, CAS operations use the JVM {\tt VarHandle} {\tt compareAndSet} primitive operation (via {\tt AtomicReference} and {\tt compareAndSet}), which is {\it sequentially consistent}, requiring no additional memory fences for correctness~\cite{JavaMemModel}.

\alglanguage{pseudocode}
\begin{algorithm}
\caption{Claim Operation Pseudo-code}
\label{alg:claim}
\small
\begin{flushleft}
\textbf{Input}: {\tt pid}, a process object\\
\textbf{Output}: {\tt true} if claim succeeded, {\tt false} otherwise\\
\textbf{Data}: {\tt waitQueue}, a lock-free queue, {\tt owner}, process currently owning the resource.
\end{flushleft}
\begin{algorithmic}[1]
\State $pid.state \gets ENGAGING$
\State \Call{Enqueue}{waitQueue, pid}
\State $head \gets$ \Call{Peek}{waitQueue} 
\If{$head = pid$}
  \If {\Call{CAS}{$owner$, \textbf{null}, $pid$}}
    \State $pid.state \gets ACTIVE$
    \State \textbf{return true}
  \Else
    \State $local\_owner \gets owner$
    \If {$local\_owner =$ \textbf{null}}
      \State $owner \gets pid$
      \State $pid.state \gets ACTIVE$
      \State \textbf{return true}
    \ElsIf {$local\_owner = pid$ \textbf{AND} \Call{CAS}{$pid.state, ENGAGING, WAITING$}}
        \State \textbf{return false}
    \ElsIf {$local\_owner = pid$}
        \State $pid.state \gets ACTIVE$
        \State \textbf{return true}
    \ElsIf {\Call{CAS}{$owner, local\_owner, pid$}}
      \State $pid.state \gets ACTIVE$
      \State \textbf{return true}
    \ElsIf {\Call{CAS}{$owner$, \textbf{null}, $pid$}}
      \State $pid.state \gets ACTIVE$
      \State \textbf{return true}
    \ElsIf{\Call{CAS}{$pid.state, ENGAGING, WAITING$}}
      \State \textbf{return false}
    \Else
      \State $pid.state \gets ACTIVE$
      \State \textbf{return true}
    \EndIf
  \EndIf
\ElsIf {\Call{CAS}{$pid.state, ENGAGING, WAITING$}}
  \State \textbf{return false}
\Else
  \State $pid.state \gets ACTIVE$
  \State \textbf{return true}
\EndIf
\end{algorithmic}
\end{algorithm}

There are three mutex states \textit{claim} attempts to coordinate with --- \textit{unclaimed}, \textit{claimed}, and \textit{releasing}. Algorithm~\ref{alg:claim} works as follows:
\begin{enumerate}
  \item First (line 1), we set process \texttt{pid} \textit{ENGAGING} to support later scheduling behaviour. The algorithm may set the process \textit{ACTIVE} again if the claim is successful.
  \item Next (line 2), we enqueue process \texttt{pid} to the lock-free \texttt{waitQueue}.
  \item We test if the mutex is \textit{claimed}. First, we retrieve the head of the queue (line 3) and check if it is process \texttt{pid} (line 4). If not, we know the mutex is \textit{claimed}. To resolve:
  \begin{enumerate}[1. ]
    \item At line 32 we attempt to \texttt{CAS} process \texttt{pid} state to \textit{WAITING} assuming it is still \textit{ENGAGING}. On success process \texttt{pid} must yield and the claim fails (line 33).
    \item If the \texttt{CAS} fails, the releasing process has set process \texttt{pid} \textit{SCHEDULED} as the new owner. We set process \texttt{pid} \textit{ACTIVE} (line 35) and return true on a successful claim (line 36).
  \end{enumerate}
  \item When process \texttt{pid} is the queue head (tested on line 4), the mutex is either in state \textit{unclaimed} or \textit{releasing}. \textit{unclaimed} is easiest to resolve, so first we attempt a \texttt{CAS owner} to process \texttt{pid} assuming \texttt{owner} is \texttt{null} (line 5). On success, we set process \texttt{pid} \textit{ACTIVE} (line 6) and return true on a successful claim (line 7).
  \item On a failed \texttt{CAS} attempt at line 5, the mutex must be in a \textit{releasing} state. The claiming and releasing processes try to resolve ownership transition.
  \begin{enumerate}[1. ]
    \item The claiming process copies the \texttt{owner} field (line 9) to test as it may have changed since the failed \texttt{CAS} attempt.
    \item The simplest test is \texttt{owner} being \texttt{null} (line 10), which means we can safely set process \texttt{pid} as the owner (line 11), and complete the successful claim (lines 12 and 13).
    \item Next we check if process \texttt{pid} is now \texttt{owner} (lines 14 to 18). If it is, we need to determine if the releasing process has scheduled process \texttt{pid}.
    \begin{itemize}
      \item On line 14, if \texttt{owner} is process \texttt{pid}, we \texttt{CAS} the process to \textit{WAITING} assuming it is still \textit{ENGAGING}. On success, we know the schedule signal has not arrived so return false as the claim is incomplete (line 15).
      \item On line 16, if \texttt{owner} is process \texttt{pid}, we failed the \texttt{CAS} on line 14. Thus the schedule signal has arrived and we complete the succesful claim (lines 17 and 18).
      \item[] \textit{Note we have provided these two cases in this manner for algorithm presentation purposes. Normally, we would test if $owner = pid$ and then perform the \texttt{CAS}, thus only testing \texttt{owner} once.}
    \end{itemize}
    \item If process \texttt{pid} is not the copied owner value, it must have been the releasing process. \textit{claim} attempts to steal ownership (line 19) by a \texttt{CAS} on \texttt{owner} assuming it is the same. On success, we can complete the successful claim (lines 20 and 21).
    \item At this point, we know that the releasing process completed changing \texttt{owner} after {\it claim} copied it. \texttt{owner} must be \texttt{null} or process \texttt{pid}. {\it claim} attempts to resolve this final case (lines 22 to 29):
    \begin{itemize}
      \item First (line 22) we attempt a \texttt{CAS} to change \texttt{owner} from \texttt{null} to process \texttt{pid}. On success, we can complete the claim successfully (lines 23 and 24).
      \item If \texttt{owner} was not \texttt{null}, the releasing process set it to process \texttt{pid} and we must determine scheduling. First, we \texttt{CAS} to change state from \textit{ENGAGING} to \textit{WAITING} (line 25). On success, we have received no scheduling signal return false on a partially successful claim (line 26). Otherwise, the schedule signal was received, and we can complete a successful claim (lines 28 and 29).
    \end{itemize}
  \end{enumerate}
\end{enumerate}

The {\it release} operation defined in Algorithm~\ref{alg:release} is simpler than claim as it is only required to schedule the next process in the queue if there is one. The release operation works as follows:
\begin{enumerate}
  \item First (line 1), we dequeue from {\tt waitQueue} --- head is guaranteed to be {\tt pid} as a process only dequeues itself when releasing.
  \item Next (line 2), we check if another process is waiting. If there is no process, we {\tt CAS} {\tt owner} to {\tt null} (line 3). It may be that between peeking at the queue and the {\tt CAS}, another process has enqueued. However, {\it claim} will handle this case correctly as the enqueuing process will observe it is the head. {\it claim} manages the case where the owner is either another process or null.
  \item If a process is waiting, we {\it CAS} it to the owner (line 6). On success, the releasing process must schedule the new owner (line 7) as {\it claim} will not do this. If the claiming process has changed the owner to itself it will have set itself \textit{ACTIVE} and continued.
\end{enumerate}

\begin{algorithm}
\caption{Release Operation Pseudo-code}
\label{alg:release}
\small
\begin{flushleft}
\textbf{Input}: {\tt pid}, a process object\\
\textbf{Output}: None\\
\textbf{Data}: {\tt waitQueue}, a lock-free queue, {\tt owner}, process currently owning the resource.
\end{flushleft}
\begin{algorithmic}[1]
\State \Call{Dequeue}{waitQueue}
\State $head \gets$ \Call{Peek}{waitQueue}
\If {$head =$ \textbf{null}}
  \State \Call{CAS}{$owner, pid$, \textbf{null}}
\ElsIf {\Call{CAS}{$owner, pid, head$}}
  \State \Call{Schedule}{$head$}
\EndIf
\end{algorithmic}
\end{algorithm}

The protocol requires {\it happens-before}~\cite{Lamport:1978} memory operation relationships when processes are concurrently acting on the \texttt{owner}, \texttt{waitQueue}, and \texttt{state} shared objects. Section~\ref{sec:diagrammatic} describes how different operation permutations lead to different behaviours. We take a sequential consistency view to memory ordering, which is simplest to model and implement, but stricter than required. Certain operations permit an acquire-release ordering --- for example, when peeking at the queue and then performing a CAS on the {\it owner}.

\subsubsection{Runtime Environment Conditions}
\label{sec:runtime}

Our protocol requires two assumed runtime conditions:
\begin{enumerate}
  \item The scheduling approach is cooperative. The protocol works by parking waiting processes within a queue, not via other locking mechanisms.
  \item The memory model ensures strict operation ordering. We rely on operations globally occurring in the specified order. We have assumed sequential consistency.
\end{enumerate}

In addition, our claims about the protocol relies on three key assumptions:
\begin{enumerate}
  \item That processes have a \textit{state} that indicates whether a process is \textit{running}, \textit{engaging}, \textit{waiting}, or \textit{scheduled}. 
  \item The \texttt{waitQueue} is a lock-free queue allowing head peeking.
  \item A \textit{schedule} operation can notify a potentially \textit{waiting or engaging} process.
\end{enumerate}

If the runtime environment tightly couples the scheduler with state (that is, there is no equivalent to an \textit{engaging} state), our algorithm will fail. A process continues while it is \textit{engaging} (equivalent to \textit{not-ready and running} in current ProcessJ), and the protocol uses state to determine {\it schedule} signal success (we define a state \textit{scheduled} --- see Section~\ref{sec:behaviour}). There must also be some state that if a process yields it becomes \textit{waiting} if it is not-ready.

For {\it schedule}, we assume a process will eventually resume execution and not terminate or cancel. Our algorithm makes no guarantees unless processes complete interactions --- a {\it claim} completes on {\it release}. This is a minimal requirement for liveness.

\subsection{Behavioural Overview}
\label{sec:behaviour}

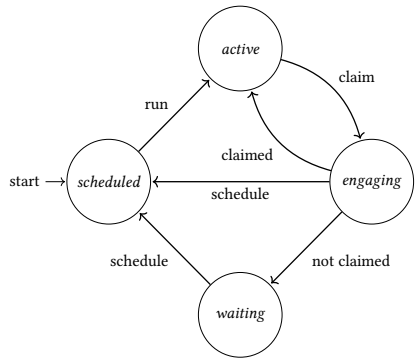
\begin{figure}
\begin{center}
\resizebox{0.4\textwidth}{!}{
\begin{tikzpicture}[shorten >=1pt,node distance=4cm, on grid,auto, state/.style={circle, draw, minimum size=1.8cm}]
\node[state, initial] (scheduled) {\textit{scheduled}};
\node[state] (active) [above right=of scheduled] {\textit{active}};
\node[state] (engaging) [below right=of active] {\textit{engaging}};
\node[state] (waiting) [below right=of scheduled] {\textit{waiting}};

\path[->, thick] (scheduled) edge node {run} (active)
          (active) edge [bend left] node {claim} (engaging)
          (engaging) edge [bend left] node {claimed} (active)
          (engaging) edge node {schedule} (scheduled)
          (engaging) edge node {not claimed} (waiting)
          (waiting) edge node {schedule} (scheduled);

\end{tikzpicture}
}
\end{center}
\caption{Process State Automata.}
\label{fig:state-machine}
\end{figure}

Processes are in one of four states: \textit{active} (running), \textit{engaging} (claiming), \textit{waiting}, or \textit{scheduled} (ready-to-run). Figure~\ref{fig:state-machine} provides the process state automata.

{\it claim} controls most of the behaviour. Unlike other approaches, we enqueue a process when {\it claim} begins to  provide mutual exclusion. Only the head of the wait queue can fully perform a {\it claim} and only a releasing process removes dequeues itself. Therefore, the algorithm controls most of the interaction by permitting only the head of the queue to execute {\it claim} and {\it release}. The main challenge is a concurrent {\it claim} and {\it release} without enqueued processes. That is, when {\it release} occurs while the next owner is actively executing {\it claim}. We examine these interactions in Section~\ref{sec:diagrammatic}.

Enqueuing first does permit safe non-busy locking, but our protocol will likely have slower best-case resolution. A typical fast locking mechanism would be similar to a {\it futux} --- an initial single fast user-space atomic operation may claim the lock before having to enqueue a thread in kernel space. Our protocol's best-case scenario is:
\begin{enumerate}
  \item Enqueue (based on Michael and Scott's~\cite{MichaelScott96} algorithm).
  \begin{enumerate}
    \item {\it Compare-and-swap} to set next node for tail in linked list.
    \item {\it Compare-and-swap} to set new tail in linked list.
  \end{enumerate}
  \item {\it Compare-and-swap} owner of the lock.
\end{enumerate}

Our best-case requires three atomic {\it CAS} operations minimum, in comparison to {\it futex} approaches best-path having one. A full experimental analysis under various contention scenarios will be required to better understand behaviour, but that is outside the scope of this paper. In Section~\ref{sec:resolution-cas} we provide an outline of the number of \textit{CAS} operations required under different scenarios. Our claim/release protocol does avoid busy waiting, and can resolve claim failure quickly, requiring only two {\it CAS} operations to enqueue and one {\it CAS} to change process state. As our protocol does not rely on busy waiting, it always runs ready code without wasting cycles through spinning.

One final behavioural consideration is potential starvation occurring as a process attempts to enqueue. If the runner thread of a process repeatedly loses the {\it CAS} race to enqueue (set the tail), it is possible a process will never be enqueued to the wait queue. Our assumption is that the coroutine runtime caps the number of simultaneously running system threads, approximately equal to the hardware processor count. An unsuccessful enqueue implies another thread modified the queue tail pointer, but this should only occur at most $k - 1$ times, where $k$ is the number of active runner threads. Non-executing processes must be retrieved from the run queue before they call {\it claim} and attempt to enqueue themselves. In practice, $k \leq 100$, so starvation is unlikely. 

\subsection{Diagrammatic Analysis}
\label{sec:diagrammatic}

It can be difficult to visualise why the protocol's interactions occur as they do. In this section, we provide a diagrammatic analysis to demonstrate the different interactions.

We use the following diagramming style. We represent line execution as a circle labelled with appropriate algorithm line number. The side label is the operation executed. We use dotted circles for modification of non-mutex properties, and red circles for modification of mutex shared data. An arrow between {\it claim} and {\it release} signifies a state change observed between the two processes. Green circles and lines indicate scheduling signals. Note that we present a particular permutation of operations for simplicity, but there are often slight variations. If general arrow direction (down-left or down-right) is maintained and no modification operations for {\it claim} moves above or below another in the release, operations can be reordering.

The \textit{claim} operation has three phases: resolution of a \textit{claimed/unclaimed} mutex, resolution with ongoing release, and resolution with a known completed release.

\subsubsection{Phase 1 --- Resolution of \textit{claimed/unclaimed} Mutex}

\begin{figure}
\begin{center}
\resizebox{0.75\textwidth}{!}{
\begin{tikzpicture}
    [
        op/.style={circle, dotted, draw=black, fill=white, minimum size=2mm},
        linked/.style={circle, double=red, draw=red, fill=white, minimum size=2mm},
        schedule/.style={circle, double=green, draw=green, minimum size=2mm}
    ]

    \tikzstyle{every node}=[font=\Large]

    \node (T1label) at (2, 0) {\bf (a) Unclaimed};
    \node at (0, -1) {\it Claim};
    \node[op, label=right:$state \gets ENGAGING$] (T1_1) at (0, -2) {1};
    \node[linked, label=right:$Enqueue$] (T1_2) at (0, -3) {2};
    \node[linked, label=right:$Peek$] (T1_3) at (0, -4) {3};
    \node[op, label=right:{$head$ = $pid$}] (T1_4) at (0, -5) {4};
    \node[linked, label=right:{\bf CAS} $owner \gets pid$] (T1_5) at (0, -6) {5};
    \node[op, label=right:$state \gets ACTIVE$] (T1_6) at (0, -7) {6};
    \node[op, label=right:{\bf return true}] (T1_7) at (0, -8) {7};
    \draw (T1_1) -- (T1_2) -- (T1_3) -- (T1_4) -- (T1_5) -- (T1_6) -- (T1_7);

    \node at (4, -1) {\it Release};

    \node (T2label) at (10, 0) {\bf (b) Claimed};
    \node at (8, -1) {\it Claim};
    \node[op, label=right:$state \gets ENGAGING$] (T2_1) at (8, -2) {1};
    \node[linked, label=right:$Enqueue$] (T2_2) at (8, -3) {2};
    \node[linked, label=right:$Peek$] (T2_3) at (8, -4) {3};
    \node[op, label=right:$head \neq pid$] (T2_4) at (8, -5) {4};
    \node[op, label=right:{{\bf CAS} $state \gets WAITING$}] (T2_5) at (8, -6) {32};
    \node[op, label=right:{\bf return false}] (T2_6) at (8, -7) {33};
    \node[schedule, label=right:{\it Yield}] (T2_7) at (8, -8) {};
    \draw (T2_1) -- (T2_2) -- (T2_3) -- (T2_4) -- (T2_5) -- (T2_6) -- (T2_7);

    \node at (12, -1) {\it Release};

    \node (T3label) at (18, 0) {\bf (c) Released-in-Time};
    \node at (16, -1) {\it Claim};
    \node[op, label=right:$state \gets ENGAGING$] (T3_1) at (16, -2) {1};
    \node[linked, label=right:$Enqueue$] (T3_2) at (16, -3) {2};
    \node[linked, label=right:$Peek$] (T3_3) at (16, -4) {3};
    \node[op, label=right:$head \neq pid$] (T3_4) at (16, -5) {4};
    \node[linked, label=right:{{\bf CAS} fails}] (T3_5) at (16, -11) {32};
    \node[op, label=right:$state \gets ACTIVE$] (T3_6) at (16, -12) {33};
    \node[op, label=right:{\bf return true}] (T3_7) at (16, -13) {34};
    \draw (T3_1) -- (T3_2) -- (T3_3) -- (T3_4) -- (T3_5) -- (T3_6) -- (T3_7);

    \node at (20, -1) {\it Release};
    \node[linked, label=right:$Dequeue$] (R3_1) at (20, -6) {1};
    \node[linked, label=right:$Peek$] (R3_2) at (20, -7) {2};
    \node[op, label=right:{$head \neq null$}] (R3_3) at (20, -8) {3};
    \node[linked, label=right:{{\bf CAS} $owner \gets head$}] (R3_4) at (20, -9) {5};
    \node[schedule, label=right:{{\sf SCHEDULE} $head$}] (R3_5) at (20, -10) {6};
    \draw (R3_1) -- (R3_2) -- (R3_3) -- (R3_4) -- (R3_5);

    \draw[draw=red, double, ->] (T3_2) -- (R3_2);
    \draw[draw=green, double, <-] (T3_5) -- (R3_5);

\end{tikzpicture}
}
\end{center}
\caption{Resolutions of (a) unclaimed and (b) claimed resource scenarios, and (c) just-in-time releasing.}
\label{fig:claimed-resolution}
\end{figure}
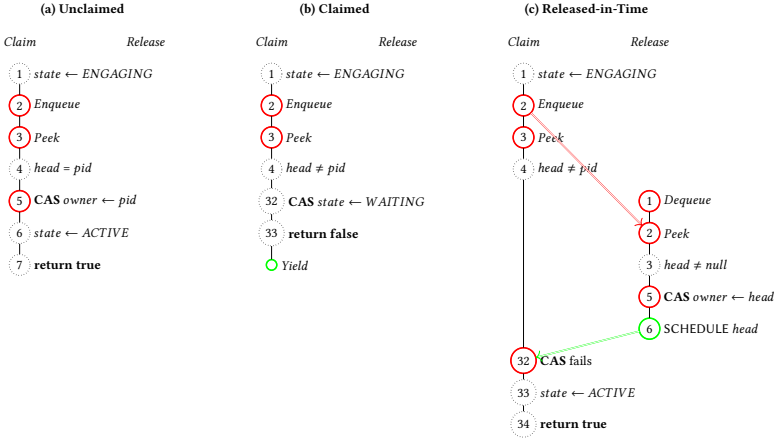

Figure~\ref{fig:claimed-resolution} presents the three scenarios for a \textit{claimed/unclaimed} mutex. An \textit{unclaimed} mutex (Figure~\ref{fig:claimed-resolution} (a)) means a process wins the enqueue race and there is no observed owner. When {\it peek} and {\it CAS} occur at the start of {\it claim}, any releasing process completed prior to the claiming process checking the shared data. \textit{claimed} resolution (Figure~\ref{fig:claimed-resolution} (b)) has the claiming process discover it is not the head of the queue and successfully sets its state to \textit{WAITING}. There is no inter-process interaction in either of these cases.

Finally, it is possible for a releasing process to notify the claiming process before waits (Figure~\ref{fig:claimed-resolution} (c)). Much of the behaviour is as \textit{claimed} resolution, but the final \textit{CAS} fails as the releasing process sends a {\it schedule} signal before the claiming process waits.

\subsubsection{Phase 2 --- Resolution with Concurrent Release}

\begin{figure}
\begin{center}
\resizebox{\textwidth}{!}{
\begin{tikzpicture}
    [
      op/.style={circle, dotted, draw=black, fill=white, minimum size=2mm},
      linked/.style={circle, draw=red, fill=white, minimum size=2mm},
      schedule/.style={circle, double=green, draw=green, minimum size=2mm}
    ]

    \tikzstyle{every node}=[font=\large]

    \node (T1label) at (2, 0) {\bf (a) Owner now \texttt{null}};

    \node at (0, -1) {\it Claim};
    \node[op, label=right:$state \gets ENGAGING$] (C1_1) at (0, -4) {1};
    \node[linked, label=right:$Enqueue$] (C1_2) at (0, -5) {2};
    \node[linked, label=right:$Peek$] (C1_3) at (0, -6) {3};
    \node[op, label=right:{$head = pid$}] (C1_4) at (0, -7) {4};
    \node[linked, label=right:{\bf CAS} fails] (C1_5) at (0, -8) {5};
    \node[linked, label=right:$local\_owner \gets owner$] (C1_6) at (0, -11) {9};
    \node[op, label=right:{$local\_owner = null$}] (C1_7) at (0, -12) {10};
    \node[linked, label=right:$owner \gets pid$] (C1_8) at (0, -13) {11};
    \node[op, label=right:$state \gets ACTIVE$] (C1_9) at (0, -14) {12};
    \node[op, label=right:{\bf return true}] (C1_10) at (0, -15) {13};
    \draw (C1_1) -- (C1_2) -- (C1_3) -- (C1_4) -- (C1_5) -- (C1_6) -- (C1_7) -- (C1_8) -- (C1_9) -- (C1_10);

    \node at (4, -1) {\it Release};
    \node[linked, label=right:$Dequeue$] (R1_1) at (4, -2) {1};
    \node[linked, label=right:$Peek$] (R1_2) at (4, -3) {2};
    \node[op, label=right:{$head = null$}] (R1_3) at (4, -9) {3};
    \node[linked, label=right:{\bf CAS} $owner \gets null$] (R1_4) at (4, -10) {4};
    \draw (R1_1) -- (R1_2) -- (R1_3) -- (R1_4);

    \draw[draw=red, double, <-] (C1_3) -- (R1_1);
    \draw[draw=red, double, <-] (C1_6) -- (R1_4);

    \node (T2label) at (10, 0) {\bf (b) Owner now \texttt{pid}, not scheduled};
    \node at (8, -1) {\it Claim};
    \node[op, label=right:$state \gets ENGAGING$] (C2_1) at (8, -3) {1};
    \node[linked, label=right:$Enqueue$] (C2_2) at (8, -4) {2};
    \node[linked, label=right:$Peek$] (C2_3) at (8, -6) {3};
    \node[op, label=right:{$head = pid$}] (C2_4) at (8, -7) {4};
    \node[linked, label=right:{\bf CAS} fails] (C2_5) at (8, -8) {5};
    \node[linked, label=right:$local\_owner \gets owner$] (C2_6) at (8, -11) {9};
    \node[op, label=right:{$local\_owner \neq null$}] (C2_7) at (8, -12) {10};
    \node[op, label=right:{$local\_owner = pid$}] (C2_8) at (8, -13) {14};
    \node[linked, label=right:{\textbf{CAS} $state \gets WAITING$}] (C2_9) at (8, -14) {14};
    \node[op, label=right:{\textbf return false}] (C2_10) at (8, -15) {15};
    \node[schedule, label=right:{\it Yield}] (C2_11) at (8, -16) {};
    \draw (C2_1) -- (C2_2) -- (C2_3) -- (C2_4) -- (C2_5) -- (C2_6) -- (C2_7) -- (C2_8) -- (C2_9) -- (C2_10) -- (C2_11);

    \node at (12, -1) {\it Release};
    \node[linked, label=right:$Dequeue$] (R2_1) at (12, -2) {1};
    \node[linked, label=right:$Peek$] (R2_2) at (12, -5) {2};
    \node[op, label=right:$head \neq null$] (R2_3) at (12, -9) {3};
    \node[linked, label=right:{\bf CAS} $owner \gets head$] (R2_4) at (12, -10) {5};
    \node[schedule, label=right:{\bf SCHEDULE}] (R2_5) at (12, -17) {6};
    \draw (R2_1) -- (R2_2) -- (R2_3) -- (R2_4) -- (R2_5);

    \draw[draw=red, double, <-] (C2_3) -- (R2_1);
    \draw[draw=red, double, ->] (C2_2) -- (R2_2);
    \draw[draw=red, double, <-] (C2_6) -- (R2_4);
    \draw[draw=green, double, <-] (C2_11) -- (R2_5);

    \node (T3label) at (18, 0) {\bf (c) Owner now \texttt{pid}, scheduled};
    \node at (16, -1) {\it Claim};
    \node[op, label=right:$state \gets ENGAGING$] (C3_1) at (16, -3) {1};
    \node[linked, label=right:$Enqueue$] (C3_2) at (16, -4) {2};
    \node[linked, label=right:$Peek$] (C3_3) at (16, -6) {3};
    \node[op, label=right:{$head = pid$}] (C3_4) at (16, -7) {4};
    \node[linked, label=right:{\bf CAS} fails] (C3_5) at (16, -8) {5};
    \node[linked, label=right:$local\_owner \gets owner$] (C3_6) at (16, -12) {9};
    \node[op, label=right:{$local\_owner \neq null$}] (C3_7) at (16, -13) {10};
    \node[op, label=right:{$local\_owner = pid$}] (C3_8) at (16, -14) {14};
    \node[linked, label=right:{\textbf{CAS} \textit{fails}}] (C3_9) at (16, -15) {14};
    \node[op, label=right:{$local\_owner = pid$}] (C3_10) at (16, -16) {16};
    \node[op, label=right:{$state \gets ACTIVE$}] (C3_11) at (16, -17) {17};
    \node[op, label=right:{\textbf{return true}}] (C3_12) at (16, -18) {18};
    \draw (C3_1) -- (C3_2) -- (C3_3) -- (C3_4) -- (C3_5) -- (C3_6) -- (C3_7) -- (C3_8) -- (C3_9) -- (C3_10) -- (C3_11) -- (C3_12);

    \node at (20, -1) {\it Release};
    \node[linked, label=right:$Dequeue$] (R3_1) at (20, -2) {1};
    \node[linked, label=right:$Peek$] (R3_2) at (20, -5) {2};
    \node[op, label=right:$head \neq null$] (R3_3) at (20, -9) {3};
    \node[linked, label=right:{\bf CAS} $owner \gets head$] (R3_4) at (20, -10) {5};
    \node[schedule, label=right:{\bf SCHEDULE}] (R3_5) at (20, -11) {6};
    \draw (R3_1) -- (R3_2) -- (R3_3) -- (R3_4) -- (R3_5);

    \draw[draw=red, double, <-] (C3_3) -- (R3_1);
    \draw[draw=red, double, ->] (C3_2) -- (R3_2);
    \draw[draw=red, double, <-] (C3_6) -- (R3_4);
    \draw[draw=green, double, <-] (C3_9) -- (R3_5);

    \node (T4label) at (26, 0) {\bf (d) Ownership Stealing};
    \node at (24, -1) {\it Claim};
    \node[op, label=right:$state \gets ENGAGING$] (C4_1) at (24, -2) {1};
    \node[linked, label=right:$Enqueue$] (C4_2) at (24, -3) {2};
    \node[linked, label=right:$Peek$] (C4_3) at (24, -5) {3};
    \node[op, label=right:{$head$ = $pid$}] (C4_4) at (24, -6) {4};
    \node[linked, label=right:{\bf CAS} fails] (C4_5) at (24, -7) {5};
    \node[linked, label=right:$local\_owner \gets owner$] (C4_6) at (24, -8) {9};
    \node[op, label=right:$local\_owner \neq null$] (C4_7) at (24, -9) {10};
    \node[op, label=right:$local\_owner \neq pid$] (C4_8) at (24, -10) {14};
    \node[op, label=right:$local\_owner \neq pid$] (C4_9) at (24, -11) {16};
    \node[linked, label=right:{\bf CAS} $owner \gets pid$] (C4_10) at (24, -12) {19};
    \node[op, label=right:$state \gets ACTIVE$] (C4_11) at (24, -13) {20};
    \node[op, label=right:{\bf return true}] (C4_12) at (24, -14) {21};
    \draw (C4_1) -- (C4_2) -- (C4_3) -- (C4_4) -- (C4_5) -- (C4_6) -- (C4_7) -- (C4_8) -- (C4_9) -- (C4_10) -- (C4_11) -- (C4_12);

    \node at (28, -1) {\it Release};
    \node[linked, label=right:$Dequeue$] (R4_1) at (28, -4) {1};
    \node[linked, label=right:$Peek$] (R4_2) at (28, -14) {2};
    \node[op, label=right:$head \neq null$] (R4_3) at (28, -15) {3};
    \node[linked, label=right:{\bf CAS} fails] (R4_4) at (28, -16) {5};
    \draw (R4_1) -- (R4_2) -- (R4_3) -- (R4_4);

    \draw[draw=red, double, <-] (C4_3) -- (R4_1);
    \draw[draw=red, double, ->] (C4_2) -- (R4_2);
    \draw[draw=red, double, ->] (C4_10) -- (R4_4);

\end{tikzpicture}
}
\end{center}
\caption{Resolution of {\it claim} concurrently with {\it release} where (a) owner became {\tt null}, (b) {\it release} transferred ownership but not scheduled, (c) {\it release} transferred ownership and scheduled, and (d) {\it claim} stole ownership.}
\label{fig:concurrent-release}
\end{figure}
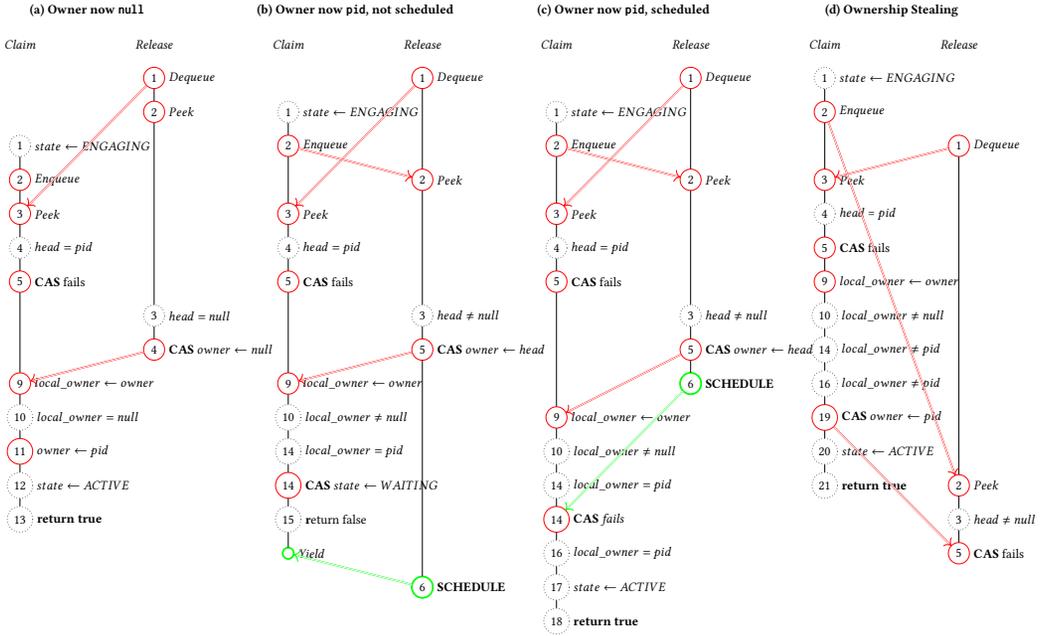

Figure~\ref{fig:concurrent-release} presents the scenarios where the claiming process knows a releasing process was active, but does not know if it has completed yet. These cases occur after the first {\it CAS} fails and the \texttt{owner} field is copied.

There are four potential scenarios. First, (Figure~\ref{fig:concurrent-release} (a)), the releasing process may have completed before the copy of \texttt{owner} and \texttt{owner} is now \texttt{null}. The claiming process can safely take ownership and return true on the successful claim.

Scenarios two and three (Figure~\ref{fig:concurrent-release} (b) and (c)) occur when the releasing process sets the claiming process as the owner. Reception of the schedule signal determines if the claiming process must wait (\textit{not received}, Figure~\ref{fig:concurrent-release} (b)), or continue (\textit{received}, Figure~\ref{fig:concurrent-release} (c)).

The final possibility (Figure~\ref{fig:concurrent-release} (d)) is when the copied owner value is the releasing process. The claiming process can attempt to steal ownership assuming the releasing process is still the owner. If it succeeds, the claiming process can end the claim successfully.

\paragraph{Why Must the Claiming Process Fail the Claim?}

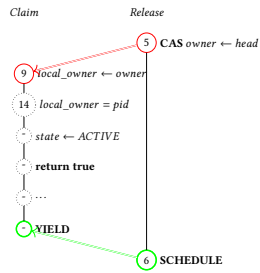
\begin{figure}
\begin{center}
\resizebox{0.25\textwidth}{!}{
\begin{tikzpicture}
    [
        op/.style={circle, dotted, draw=black, fill=white, minimum size=2mm},
        linked/.style={circle, draw=red, fill=white, minimum size=2mm},
        schedule/.style={circle, double=green, draw=green, minimum size=2mm}
    ]

    \tikzstyle{every node}=[font=\large]

    \node at (0, -1) {\it Claim};
    \node[linked, label=right:$local\_owner \gets owner$] (C1) at (0, -3) {9};
    \node[op, label=right:{$local\_owner = pid$}] (C2) at (0, -4) {14};
    \node[op, label=right:$state \gets ACTIVE$] (C3) at (0, -5) {-};
    \node[op, label=right:{\bf return true}] (C4) at (0, -6) {-};
    \node[op, label=right:{\it \dots}] (C5) at (0, -7) {-};
    \node[schedule, label=right:{\bf YIELD}] (C6) at (0, -8) {-};
    \draw (C1) -- (C2) -- (C3) -- (C4) -- (C5) -- (C6);

    \node at (4, -1) {\it Release};
    \node[linked, label=right:{\bf CAS} $owner \gets head$] (R1) at (4, -2) {5};
    \node[schedule, label=right:{\bf SCHEDULE}] (R2) at (4, -9) {6};
    \draw (R1) -- (R2);

    \draw[draw=red, double, <-] (C1) -- (R1);
    \draw[draw=green, double, <-] (C6) -- (R2);

\end{tikzpicture}
}
\end{center}
\caption{Spurious Scheduling After Claim Resolution.}
\label{fig:spurious}
\end{figure}

When the {\it claim} determines it is the owner (e.g., Figure~\ref{fig:concurrent-release} (b)), the assumption could be to allow {\it claim} to succeed and let the claiming process continue. However, {\it release} will schedule the claiming process --- the releasing process does not know if the process at the head of the wait queue is actively claiming or fully waiting. We can avoid scheduling an {\it active} process, but this is not the issue. Consider Figure~\ref{fig:spurious}.

Allowing {\it claim} to continue as succeeded does not consider the {\it schedule} signal. The claiming process could continue and then {\it yield} later. If the releasing process sends the {\it schedule} signal after this {\it yield}, the claiming process will be spuriously woken. ProcessJ shared channel access is sensitive to this as {\it claim} is for a channel, which contains a potential {\it yield} to await a communication partner.

A solution to avoid a descheduled successful claim is to busy wait on the state waiting for {\it scheduled}. Indeed, our model effectively does this (see Section~\ref{sec:verify}) for scheduler modelling simplification. However, making this explicitly a busy wait would be a spinlock, and thus we leave it to algorithm implementers to determine the best method for their runtime.

\paragraph{Why Does the Releasing Process Not Test the Claiming Process's State?}

A further consideration for a descheduled successful claim is whether the releasing process can determine if the schedule signal is necessary. The releasing process has observed a process in the wait queue, but it does not know whether that process is actively claiming or waiting. Any test of such state would be stale after retrieving it and could lead to incorrect behaviour. A particular problem occurs when the claiming process peeks at the queue and sees it is not the head. If the releasing process then dequeues itself, peeks at the queue, and checks the state of the head of the queue (the new claiming process), we have no observed state change to the claiming process. Consider the steps:
\begin{enumerate}
  \item The claiming process sets itself engaging.
  \item The claiming process enqueues itself.
  \item The claiming process peeks at the head of the wait queue. The claiming process has committed to claim-failure with one final {\it CAS} on its state.
  \item The releasing process dequeues itself.
  \item The releasing process peeks at the head and observes the claiming process.
  \item The releasing process checks the claiming processes state.
  \item Consider the two scenarios:
  \begin{itemize}
    \item The releasing process observes the claiming process is engaging.
    \item The claiming process sets itself to waiting. The releasing process still observes the claiming process is engaging.
  \end{itemize}
\end{enumerate}

Adding additional state (i.e., the claiming process knows it has claimed the resource and is continuing) will suffer from the same problem --- the releasing process may not have current knowledge to determine if it should schedule. Furthermore, an additional state would not resolve the underlying problem --- permitting the claiming process to proceed entails a spurious schedule signal. In theory (although unlikely in practice), a claiming process could loop around the claim again and have two (or more) spurious signals waiting.

For modelling purposes (to avoid divergence) we fail the claim which implies returning the new owner to the end of the run queue for future scheduling. Again, the simple solution is for the process to busy wait for its state to become \textit{scheduled}. Our model's implementation of the {\it schedule} operation demonstrates how the releasing process can manage a correct schedule signal. However, this does not mitigate the problem of a spurious signal.

\subsubsection{Phase 3 --- Release Completed}

\begin{figure}
\begin{center}
\resizebox{0.75\textwidth}{!}{
\begin{tikzpicture}
    [
        op/.style={circle, dotted, draw=black, fill=white, minimum size=2mm},
        linked/.style={circle, draw=red, fill=white, minimum size=2mm},
        schedule/.style={circle, double=green, draw=green, minimum size=2mm}
    ]

    \tikzstyle{every node}=[font=\large]

    \node (T1label) at (2, 0) {\bf (a) Release Complete, Owner \texttt{null}};
    \node at (0, -1) {\it Claim};
    \node[op, label=right:$state \gets ENGAGING$] (C1_1) at (0, -2) {1};
    \node[linked, label=right:$Enqueue$] (C1_2) at (0, -5) {2};
    \node[linked, label=right:$Peek$] (C1_3) at (0, -6) {3};
    \node[op, label=right:{$head = pid$}] (C1_4) at (0, -7) {4};
    \node[linked, label=right:{\bf CAS} fails] (C1_5) at (0, -8) {5};
    \node[linked, label=right:$local\_owner \gets owner$] (C1_6) at (0, -9) {9};
    \node[op, label=right:$local\_owner \neq null$] (C1_7) at (0, -12) {10};
    \node[op, label=right:$local\_owner \neq pid$] (C1_8) at (0, -13) {14};
    \node[op, label=right:$local\_owner \neq pid$] (C1_9) at (0, -14) {16};
    \node[linked, label=right:{\bf CAS} fails] (C1_10) at (0, -15) {19};
    \node[linked, label=right:{\bf CAS} $owner \gets pid$] (C1_11) at (0, -16) {20};
    \node[op, label=right:$state \gets ACTIVE$] (C1_12) at (0, -17) {21};
    \node[op, label=right:{\bf return true}] (C1_13) at (0, -18) {22};
    \draw (C1_1) -- (C1_2) -- (C1_3) -- (C1_4) -- (C1_5) -- (C1_6) -- (C1_7) -- (C1_8) -- (C1_9) -- (C1_10) -- (C1_11) -- (C1_12) -- (C1_13);
    
    \node at (4, -1) {\it Release};
    \node[linked, label=right:$Dequeue$] (R1_1) at (4, -3) {1};
    \node[linked, label=right:$Peek$] (R1_2) at (4, -4) {2};
    \node[op, label=right:{$head = null$}] (R1_3) at (4, -10) {3};
    \node[linked, label=right:{\bf CAS} $owner \gets null$] (R1_4) at (4, -11) {4};
    \draw (R1_1) -- (R1_2) -- (R1_3) -- (R1_4);

    \draw[draw=red, double, <-] (C1_3) -- (R1_1);
    \draw[draw=red, double, <-] (C1_10) -- (R1_4);
    \draw[draw=red, double, <-] (C1_11) -- (R1_4);

    \node (T2label) at (10, 0) {\bf (b) Release complete, owner \texttt{pid}, not scheduled};
    \node at (8, -1) {\it Claim};
    \node[op, label=right:{$state \gets ENGAGING$}] (C2_1) at (8, -2) {1};
    \node[linked, label=right:$Enqueue$] (C2_2) at (8, -5) {2};
    \node[linked, label=right:$Peek$] (C2_3) at (8, -6) {3};
    \node[op, label=right:{$head = pid$}] (C2_4) at (8, -7) {4};
    \node[linked, label=right:{\bf CAS} fails] (C2_5) at (8, -8) {5};
    \node[linked, label=right:{$local\_owner \gets owner$}] (C2_6) at (8, -9) {9};
    \node[op, label=right:{$local\_owner \neq null$}] (C2_7) at (8, -12) {10};
    \node[op, label=right:{$local\_owner \neq pid$}] (C2_8) at (8, -13) {14};
    \node[op, label=right:{$local\_owner \neq pid$}] (C2_9) at (8, -14) {16};
    \node[linked, label=right:{{\bf CAS} fails}] (C2_10) at (8, -15) {19};
    \node[linked, label=right:{{\bf CAS} fails}] (C2_11) at (8, -16) {22};
    \node[linked, label=right:{{\bf CAS} $state \gets WAITING$}] (C2_12) at (8, -17) {25};
    \node[op, label=right:{\textbf{return false}}] (C2_13) at (8, -18) {26};
    \node[schedule, label=right:{\textit{Yield}}] (C2_14) at (8, -19) {};
    \draw (C2_1) -- (C2_2) -- (C2_3) -- (C2_4) -- (C2_5) -- (C2_6) -- (C2_7) -- (C2_8) -- (C2_9) -- (C2_10) -- (C2_11) -- (C2_12) -- (C2_13) -- (C2_14);
    
    \node at (12, -1) {\it Release};
    \node[linked, label=right:$Dequeue$] (R2_1) at (12, -3) {1};
    \node[linked, label=right:$Peek$] (R2_2) at (12, -4) {2};
    \node[op, label=right:{$head \neq null$}] (R2_3) at (12, -10) {3};
    \node[linked, label=right:{{\bf CAS} $owner \gets head$}] (R2_4) at (12, -11) {5};
    \node[linked, label=right:{\textbf{SCHEDULE}}] (R2_5) at (12, -20) {6};
    \draw (R2_1) -- (R2_2) -- (R2_3) -- (R2_4) -- (R2_5);

    \draw[draw=red, double, <-] (C2_3) -- (R2_1);
    \draw[draw=red, double, <-] (C2_10) -- (R2_4);
    \draw[draw=red, double, <-] (C2_11) -- (R2_4);
    \draw[draw=green, double, <-] (C2_14) -- (R2_5);

    \node (T3label) at (18, 0) {\bf (c) Release complete, owner \texttt{pid}, scheduled};
    \node at (16, -1) {\it Claim};
    \node[op, label=right:$state \gets engaging$] (C3_1) at (16, -2) {1};
    \node[linked, label=right:$Enqueue$] (C3_2) at (16, -3) {2};
    \node[linked, label=right:$Peek$] (C3_3) at (16, -5) {3};
    \node[op, label=right:{$head$ = $pid$}] (C3_4) at (16, -6) {4};
    \node[linked, label=right:{\bf CAS} fails] (C3_5) at (16, -7) {5};
    \node[linked, label=right:$local\_owner \gets owner$] (C3_6) at (16, -8) {9};
    \node[op, label=right:$local\_owner \neq null$] (C3_7) at (16, -9) {10};
    \node[op, label=right:$local\_owner \neq pid$] (C3_8) at (16, -10) {14};
    \node[op, label=right:$local\_owner \neq pid$] (C3_9) at (16, -11) {16};
    \node[linked, label=right:{\bf CAS} fails] (C3_10) at (16, -16) {19};
    \node[linked, label=right:{\bf CAS} fails] (C3_11) at (16, -17) {22};
    \node[linked, label=right:{\bf CAS} fails] (C3_12) at (16, -18) {25};
    \node[op, label=right:{$state \gets ACTIVE$}] (C3_13) at (16, -19) {28};
    \node[op, label=right:{{\bf return true}}] (C3_14) at (16, -20) {29};
    \draw (C3_1) -- (C3_2) -- (C3_3) -- (C3_4) -- (C3_5) -- (C3_6) -- (C3_7) -- (C3_8) -- (C3_9) -- (C3_10) -- (C3_11) -- (C3_12) -- (C3_13) -- (C3_14);

    \node at (20, -1) {\it Release};
    \node[linked, label=right:$Dequeue$] (R3_1) at (20, -4) {1};
    \node[linked, label=right:$Peek$] (R3_2) at (20, -12) {2};
    \node[op, label=right:$head \neq null$] (R3_3) at (20, -13) {3};
    \node[linked, label=right:{\bf CAS} $owner \gets head$] (R3_4) at (20, -14) {5};
    \node[schedule, label=right:{\sf SCHEDULE} $pid$] (R3_5) at (20, -15) {6};
    \draw (R3_1) -- (R3_2) -- (R3_3) -- (R3_4) -- (R3_5);

    \draw[draw=red, double, <-] (C3_3) -- (R3_1);
    \draw[draw=red, double, ->] (C3_2) -- (R3_2);
    \draw[draw=red, double, <-] (C3_10) -- (R3_4);
    \draw[draw=red, double, <-] (C3_11) -- (R3_4);
    \draw[draw=green, double, <-] (C3_12) -- (R3_5);
    
\end{tikzpicture}
}
\end{center}
\caption{Resolution of {\it claim} after {\it release} completion, with (a) owner now \texttt{null}, (b) owner now \texttt{pid} but not scheduled, and (c) owner now \texttt{pid} and scheduled.}
\label{fig:release-completed}
\end{figure}
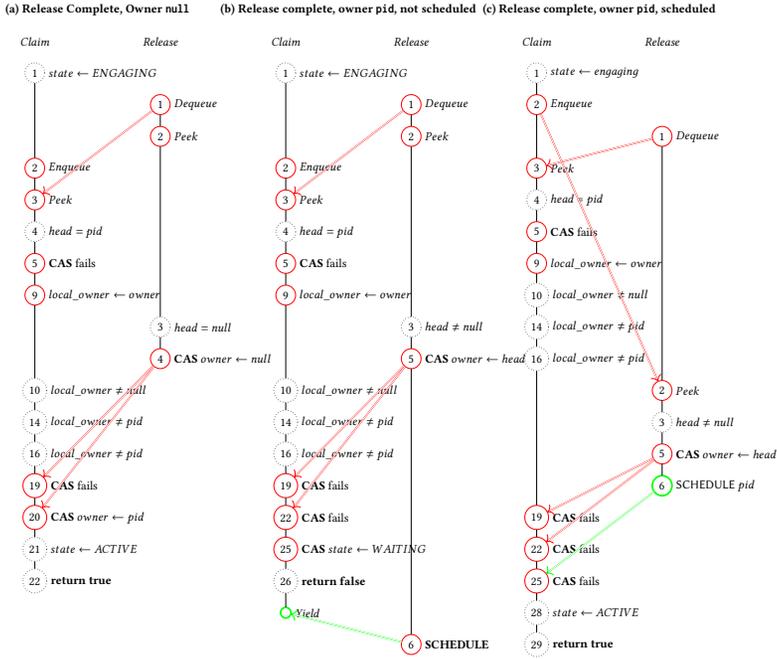

If the claiming process fails to steal ownership, then it knows the releasing process has completed. Figure~\ref{fig:release-completed} presents the final three scenarios to resolve.

First, we check if \texttt{owner} is \texttt{null} (Figure~\ref{fig:release-completed} (a)). Otherwise, we know the releasing process set the claiming process as the owner, and we must determine {\it claim} waits for the schedule signal (Figure~\ref{fig:release-completed} (b)) or not (Figure~\ref{fig:release-completed} (c)).

\subsubsection{Releasing Behaviour}

Releasing behaviour is simpler and covered across Figures~\ref{fig:claimed-resolution}, \ref{fig:concurrent-release}, and \ref{fig:release-completed}. The one consideration is {\it release} only reacts to a successful {\it CAS} setting a new process as {\tt owner}. {\it release} is then responsible for scheduling. In other cases, {\it release} may succeed or fail to set the \texttt{owner}, but this does not impact {\it release} success.
\section{CSP Model}
\label{sec:verify}

In the previous section, we presented our lock-free claim/release protocol. The next step is to translate both the {\it claim} and the {\it release} operations to CSP.

CSP cannot directly represent imperative code, so we undertake a conversion process with the following steps:
\begin{enumerate}
    \item Create globally accessible processes to represent shared atomic state.
    \item Replace method calls to the queue with event pairs to call and return --- for example, {\it enqueue} and {\it end\_enqueue}.
    \item Add events to indicate the start and end of the claim and release calls.
    \item Add {\sf SKIP} events to represent the completion of an operation.
    \item Store return values before use --- for example storing the return of {\it CAS} operations in temporary {\it succ} variables.
\end{enumerate}

Our model defines the following data types:
\begin{description}
    \item[PID] --- the set of all process identifiers, including {\it NULL}.
    \item[NonNullPID] --- a subtype of {\it PID}, excluding {\it NULL}.
    \item[Operations] --- {\it load | store}, representing variable interaction operations.
    \item[STATE] --- {\it WAITING | ENGAGING | SCHEDULED | ACTIVE}; the possible states of a process.
\end{description}

\subsection{Shared State}

CSP provides no globally accessible shared state only globally accessible events. Our claim/release protocol relies on the operations sharing {\it owner} and process {\it state} atomic values. In CSP, we represent a shared variable as a process that we can {\it get} or {\it set} the value of (examples are available in our previous work~\cite{ChalmersPedersen}). To support {\it CAS} operations, we extend this idea. A compare-and-swap operation requires three parameters:
\begin{itemize}
    \item The {\it concurrent object} manipulated. Individual {\it state} variables are further differentiated via process identifiers. {\it owner} is a single value, so we have a dedicated channel.
    \item The current {\it expected} value of the concurrent object.
    \item The {\it new value} for the concurrent object.
\end{itemize}

A successful modification occurs when $expected = current$. {\it CAS} returns true or false based on the outcome success.

In CSP, a channel can both send and receive in a single operation. Thus, we implement {\it CAS} in the general form $cas?expected?newVal!succ$, where $succ$ represents the modification outcome. We first define an {\it atomic variable} process:
\begin{align*}
AT&OMIC\_VARIABLE(get, set, cas, val) = \\
&\quad get!val \then ATOMIC\_VARIABLE(get, set, cas, val) \\
&\extchoice \ set?newVal \then ATOMIC\_VARIABLE(get, set, cas, newVal) \\
&\extchoice \ cas?expected?newVal!(expected = val) \then \\
&\qquad {\sf if } \ (expected = val) \ {\sf then} \ ATOMIC\_VARIABLE(get, set, cas, newVal)\\
&\qquad {\sf else} \ ATOMIC\_VARIABLE(get, set, cas, val)
\end{align*}

\noindent
and define {\it OWNER} as an {\it ATOMIC\_VARIABLE}:
\begin{align*}
OWNER = ATOMIC\_VARIABLE(owner.load, owner.store, casOwner, NULL)
\end{align*}

\noindent
with the {\it owner} and {\it owner\_cas} channels, with appropriate aliases, defined as:

\begin{tabbing}
==\===\===\kill
\>{\sf channel} {\it owner : Operations.PID}\\
\>{\sf channel} {\it owner\_cas : PID.PID.Bool}\\
\>{\it getOwner = owner.load}\\
\>{\it setOwner = owner.store}\\
\>{\it casOwner = owner\_cas}
\end{tabbing}

The {\it state} variables are similar, but we create one for each process using the process identifier to distinguish between different {\it state} variables.

\begin{tabbing}
==\===\===\kill
\>{\sf channel} {\it state : Operations.NonNullPID.STATE}\\
\>{\sf channel} {\it cas\_state : NonNullPID.STATE.STATE.Bool}
\end{tabbing}
\begin{align*}
ST&ATE\_VAR = \\
&\big|\big|\big| \ p \in NonNullPID \bullet ATOMIC\_VARIABLE(state.load.p, state.store.p, cas\_state.p, ACTIVE)
\end{align*}

\subsubsection{Scheduling}

We also provide the following processes for interaction with the scheduling system.
\begin{align*}
YIELD(pid) = \ &getState.pid.SCHEDULED \then setState.pid.ACTIVE \then {\sf SKIP}\\
SCHEDULE(pid) = \ &casState.pid!ENGAGING!SCHEDULED?succ \then \\
&\qquad{\sf if} \ ({\sf not} \ succ) \ {\sf then} \ casState.pid!WAITING!SCHEDULED?\_ \then \ {\sf SKIP}\\
&\qquad{\sf else \ SKIP}
\end{align*}
 
{\it YIELD} waits for {\it state} to be {\it SCHEDULED} and then sets it to {\it ACTIVE}. {\it SCHEDULE} attempts to set {\it state} to {\it SCHEDULED}. That is, a process can immediately continue after yielding when it is {\it SCHEDULED}, and when we schedule we set a process's state to {\it SCHEDULED}, ending any yielding. Thus, our processes act as if there is always a runner available when they are {\it SCHEDULED} --- there is no resource contention. {\it YIELD} can also be implemented as a busy wait for {\it state} to become {\it SCHEDULED}.

\subsection{Concurrent Queue}

The {\tt waitQueue} is a lock-free queue permitting multiple concurrent {\it enqueue} and {\it dequeue} operations. We have previously defined a simple specification for a concurrent queue~\cite{software4030015} that can be plugged into our model for verification purposes. We model the concurrent queue with internal linearization points, which we will discuss further in Section~\ref{sec:spec}. First, we undertake step 2 of our modification process --- converting method calls for the queue into call and return pairs:

\begin{tabbing}
==\===\===\kill
\>{\sf channel} {\it enqueue} : {\it NonNullPID.NonNullPID} \\
\>{\sf channel} {\it end\_enqueue} : {\it NonNullPID} \\
\>{\sf channel} {\it dequeue} : {\it NonNullPID} \\
\>{\sf channel} {\it peek} : {\it NonNullPID} \\
\>{\sf channel} {\it return} : {\it NonNullPID.PID}
\end{tabbing}

Both {\it dequeue} and {\it peek} use the {\it return} channel to end a call. We are safe to do so as each call and return event contains the process identifier of the caller and processes do not interact directly. Internally, operations commit at some point between the call and return ({\it begin} and {\it end}) events. We model the commit point via internal {\it linearization} events, defined as:

\begin{tabbing}
==\===\===\===\kill
\>{\sf channel} {\it lin\_enqueue} : {\it NonNullPID.NonNullPID} \\
\>{\sf channel} {\it lin\_dequeue} : {\it NonNullPID.PID} \\
\>{\sf channel} {\it lin\_peek} : {\it NonNullPID.PID} 
\end{tabbing}

We can now define {\it QUEUE\_SPEC} as accepting {\it begin}, {\it linearization}, or {\it end} events at any time, but only committing changes to queue state on {\it linearization} events. For brevity, we present the {\it enqueue} operation example below. The full CSP model is available in Appendix~\ref{app:conqueue}.
\begin{align*}
QUEU&E\_SPEC(q) = \ {\sf length}(q) \leq max \ \& (\\
&\quad (enqueue?\_?\_ \then QUEUE\_SPEC(q))\\
&\extchoice \ \ (lin\_enqueue?proc?v \then QUEUE\_SPEC(q ^\frown \langle v \rangle))\\
&\extchoice \ \ (end\_enqueue?\_ \then QUEUE\_SPEC(q))\\
&\dots
\end{align*}

We guard all operations --- {\sf length}({\it q}) $\leq$ {\it max} --- to manage the maximum queue size for state space purposes. The use of {\it \_} signifies values we ignore. For example, for {\it enqueue}, the queue does not retain the caller or the value being enqueued. Thus, we can use {\it enqueue?\_?\_}. Only at linearization do we require the value being enqueued.

$QUEUE\_SPEC$ alone allows {\it begin}, {\it linearization}, and {\it end} events to occur at anytime. For specification purposes, we restrict the calls by composing {\it USER} processes with the {\it QUEUE\_SPEC} (where $users \subseteq NonNullPID$):
\begin{align*}
USER(id) &= \ (enqueue.id?val \then lin\_enqueue.id.va \then end\_enqueue.id \then USER(id))\\
&\ \extchoice \ \dots
\end{align*}

The {\it QUEUE} specification is a set of interleaving {\it USER} processes synchronising with {\it QUEUE\_SPEC} via all the {\it begin}, {\it linearization}, and {\it end} events ({\it SYNC}), hiding the {\it linearization} events ({\it HIDE}).
\begin{align*}
QUEUE(users) = \bigl( \ \big|\big|\big| \ id \in users \bullet USER(id) \underset{SYNC}{\big|\big|} QUEUE\_SPEC(\langle \rangle)\ \bigr) \ \backslash \ HIDE
\end{align*}

{\it USER} processes control {\it begin}, {\it end}, and {\it linearization} events. We attach $QUEUE$ to the protocol, knowing that when an operation begins (e.g., {\it enqueue}), the appropriate {\it USER} process will continue, ready to accept the hidden {\it linearization} event. The operation only ends after the {\it linearization} event. Thus, externally, a queue user will invoke {\it enqueue} then {\it end\_enqueue}, with its process identifier. It can start an operation ({\it enqueue}) at any point, but cannot complete it ({\it end\_enqueue}) until the internal {\it USER} with the same process identifier has performed the hidden {\it linearization} event {\it lin\_enqueue}, thus committing the value to the queue.

We encapsulate the complete set of processes and channels for the queue within a module, exposing only the {\it QUEUE} process.

\subsection{Start and End Events}

{\it claim} and {\it release} operations also require {\it begin} and {\it end} events to signify operation lifetime. We use these events for specification checking:

\begin{tabbing}
==\===\===\kill
\>{\sf channel} {\it begin\_claim : NonNullPID} \\
\>{\sf channel} {\it end\_claim : NonNullPID} \\
\>{\sf channel} {\it begin\_release : NonNullPID} \\
\>{\sf channel} {\it end\_release : NonNullPID}
\end{tabbing}

We explain in Section~\ref{sec:spec} how we use these events to verify our implementation model. Effectively, we expect to demonstrate that there exists a trace:
\begin{align*}
\trace{\dots, begin\_claim.x, \dots, end\_claim.x, \dots, begin\_release.x, \dots, end\_release.x, \dots}
\end{align*}

\noindent
that ensures only the process with identifier $x$ owns the resource between $end\_claim.x$ and $begin\_release.x$ --- that is no other $end\_claim$ or $begin\_release$ occurs between these events.

\subsection{Operation Completion}

Both {\it claim} and {\it release} will be represented as processes ({\it CLAIM} and {\it RELEASE}). CSP process definitions must end with a process, we use {\sf SKIP} to signify successful operation completion. In Algorithm~\ref{alg:claim}, the operation returns true or false to signify the claiming outcome. For verification purposes, we incorporate yielding directly into {\it claim}. As an example, the following line in the claim operation:

\begin{algorithmic}
\State \textbf{return false}
\end{algorithmic}

\noindent
will become:
\begin{align*}
&YIELD(pid); \ {\sf SKIP}
\end{align*}

Taking this approach does have some considerations on the validity of our verified model, which we will discuss further in Section~\ref{sec:lin-analysis}.

\subsection{Temporary Variables}

When modelling the {\it CAS} operation, we noted a CSP channel's ability to both input and output values in a single event. Atomic variable processes output true or false from a {\it CAS}. In CSP, we must store this value in a temporary variable for the subsequent {\it if} statement. For example, the algorithm code:

\begin{algorithmic}
\If {\Call{CAS}{$owner$, \textbf{null}, $pid$}}
\State $\dots$
\EndIf
\end{algorithmic}

\noindent
is implemented as:
\begin{align*}
&casOwner!null!pid?succ \then \ {\sf if} \ (succ) \ {\sf then}\\
&\dots
\end{align*}

\subsection{Model Implementation}

With the steps taken above, we can replicate the protocol directly in CSP. The full CSP models for {\it CLAIM} and {\it RELEASE} are available in Appendix~\ref{app:models}. To model protocol usage we attach user processes. We define a {\it USER} process to repeatedly perform claim and release as:
\begin{align*}
USER&(pid) = CLAIM(pid); RELEASE(pid); USER(pid)
\end{align*}

\noindent
and interleave a set of {\it USER} processes that synchronise on the memory objects.
\begin{align*}
SYSTEM = \Bigl( \ QUEUE(users) \ \big|\big|\big| \ {\it STATE\_VAR} \ \big|\big|\big| \ OWNER \ \Bigr) \ \ \mathclap{\underset{\it DATA}{\big|\big|}} \ \ \Bigl( \ \big|\big|\big| \ u \in users \ \bullet USER(u) \ \Bigr)
\end{align*}

\noindent
where {\it DATA} is the set of all events interacting with the variable processes, including the queue. For model checking purposes, we also hide the {\it DATA} events, exposing only the {\it begin} and {\it end} events for claiming and releasing.

With the model in place, we can discuss how we derived an appropriate mutex specification and how we test properties of interest. This is presented in the next section. 
\section{Specification}
\label{sec:spec}

Our claim/release protocol is a mutex, and modelling mutex implementations with CSP exists~\cite{WelchMartin00,ChalmersPedersen,Lowe2019}. However, we are not implementing mutex behaviour, but specifying the abstract behaviour of a mutex during process interaction. Our claim/release protocol does not block the cooperative scheduler, and so we cannot use blocking events to model the mutex. As an example, we have previously modelled a mutex as follows~\cite{ChalmersPedersen}:
\begin{align*}
MUTEX(claim, release) = claim?pid \then release.pid \then MUTEX(claim, release)
\end{align*}

This is insufficient as it prevents multiple concurrent {\it claim} operations and prevents concurrent {\it claim} and {\it release} --- we have a blocking claim call. $MUTEX$ enables mutual exclusion for implementation modelling, but does so in a blocking manner. We could extend {\it MUTEX} to include a queue of waiting processes (ignoring the case when $Q$ is empty) with a $continue$ event allowing only the queue head to enter the critical section:
\begin{align*}
MUTEX(claim, release, \langle h \rangle ^\frown Q) &= \ claim?pid \then MUTEX(claim, release, \langle h \rangle ^\frown Q ^\frown \langle pid \rangle)\\
&\ \extchoice \ \ continue.h \then MUTEX(claim, release, \langle h \rangle ^\frown Q ^\frown \langle pid \rangle)\\
&\ \extchoice \ \ release.h \then MUTEX(claim, release, Q)
\end{align*}

However, this would also be insufficient as this model allows immediate completion of {\it claim} and {\it release} (they are atomic), prevents operation overlap ({\it claim} and {\it release} cannot occur concurrently), and queue contains claimers in explicit call order. Although it is possible for claimers to be enqueued in call order, it is not guaranteed. Interleaving processes could commit enqueues in unknown orders due to overlapping enqueuing operations (see Figure~\ref{fig:concqueue} for an example).

Therefore, much like the queue specification in Section~\ref{sec:verify}, we require a specification enabling multiple concurrent {\it claim} operations. We can do this by considering {\it linearizability} in more detail.

\subsection{Linearization}

Linearization~\cite{HerlihyWing90} is a correctness property for object state during concurrent operations. Effectively, an operation has a start (call) and an end (return). At some point between start and end, the operation commits --- termed the {\it linearization point}. Even with overlapping operations, we can understand correct ordering by considering the order of linearization points.

In our previous work~\cite{software4030015}, we modelled a concurrent queue by creating a specification using CSP events as operation calls, returns, and linearization points. From this, we produced a concurrent queue specification that was both lightweight and reusable. In this paper we reuse that concurrent queue specification (see Section~\ref{sec:verify}). We will also adopt the linearization modelling technique to create a specification for the mutex.

\subsection{Linearizability and Stability}
\label{sec:lin-stable}

We root our concurrent data object modelling technique in linearizability. We consider any concurrent, non-synchronising data object to expose its operations through begin and end events that signify the call and return of the operation by a thread. Therefore, we have some thread, $T_i$, interacting with the concurrent object.

If the concurrent data object has {\tt synchronized} methods, we can take a simplified view of modelling. In CSP, we create a process that represents the {\tt synchronized} object, accepting {\it begin} then {\it end} events for each operation call. For example, consider the specification of the concurrent {\tt synchronized} object, $SOBJ$, with operations $op_1, op_2, \dots, op_n$:
\begin{align*}
SOBJ(state) &= \ (begin\_op_1?tid?param \then \dots \then end\_op_1!tid!result \then SOBJ(state'))\\
&\ \extchoice \ \ (begin\_op_2?tid?param \then \dots \then end\_op_2!tid!result \then SOBJ(state'))\\
&\ \extchoice \ \ \dots\\
&\ \extchoice \ \ (begin\_op_n?tid?param \then \dots \then end\_op_n!tid!result \then SOBJ(state'))
\end{align*}

\noindent
where $state$ represents the object state, $state'$ the state after an operation, $tid$ the identifier of the thread invoking the operation (e.g., $T_i$), $param$ the parameters for the operation, and $result$ the result returned from the operation.

With this specification, $SOBJ$ will produce traces where {\it begin} events immediately precede the partner {\it end} events, such as:
\begin{align*}
\trace{begin\_op_i.T_p.par, end\_op_i.T_p.res, begin\_op_j.T_q.par, end\_op_j.T_q.res, \dots}
\end{align*}

\begin{figure}
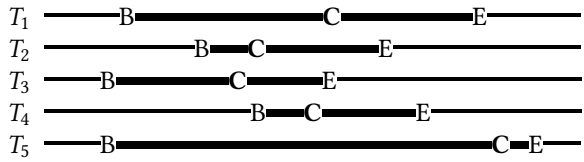

\begin{center}
\noindent
$T_1$ \ \rule[1mm]{1cm}{0.5mm}B\rule[0.5mm]{2.5cm}{1mm}{\bf C}\rule[0.5mm]{1.75cm}{1mm}E\rule[1mm]{1.25cm}{0.5mm}\\
$T_2$ \ \rule[1mm]{2cm}{0.5mm}B\rule[0.5mm]{0.5cm}{1mm}{\bf C}\rule[0.5mm]{1.5cm}{1mm}E\rule[1mm]{2.5cm}{0.5mm}\\
$T_3$ \ \rule[1mm]{0.75cm}{0.5mm}B\rule[0.5mm]{1.5cm}{1mm}{\bf C}\rule[0.5mm]{1cm}{1mm}E\rule[1mm]{3.25cm}{0.5mm}\\
$T_4$ \ \rule[1mm]{2.75cm}{0.5mm}B\rule[0.5mm]{0.5cm}{1mm}{\bf C}\rule[0.5mm]{1.25cm}{1mm}E\rule[1mm]{2cm}{0.5mm}\\
$T_5$ \ \rule[1mm]{0.75cm}{0.5mm}B\rule[0.5mm]{5cm}{1mm}{\bf C}\rule[0.5mm]{.25cm}{1mm}E\rule[1mm]{0.5cm}{0.5mm}\\
\end{center}
\caption{Example of different commit/linearization times.}
\label{fig:concqueue}
\end{figure}

With lock-free access (with no {\tt synchronized} methods), other traces are possible as multiple threads can invoke object operations concurrently. This is true for any non-atomic shared data object in a concurrent system, including mutexes. Invoking {\it claim} does not atomically provide the lock to the caller --- some internal code will run prior to claim commitment. Therefore, like a concurrent queue, between the {\it begin} and {\it end} of {\it claim} some form of commit point occurs. Linearization~\cite{HerlihyWing90} considers such operations, using linearization points to signify when an operation commits during the operation invocation. Figure~\ref{fig:concqueue} illustrates some behaviour a shared object can express from {\it begin} ({\bf B}), {\it commit} ({\bf C}), and {\it end} ({\bf E}) events.

Note that commit events can happen in a different order to {\it begin} events, and {\it end} events likewise in a different order to commits. A model specifying correct overlapping behaviour is our aim. First we consider how different object types --- {\tt synchronized}, concurrent, and a mutex --- behave under such conditions in a CSP model.

\subsubsection{Concurrent Object Lifelines}

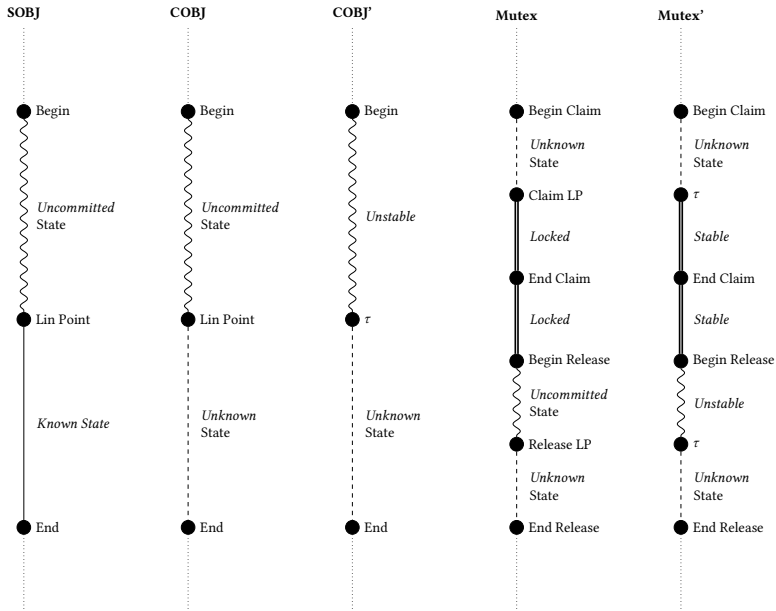
\begin{figure}
\begin{center}
\resizebox{0.75\textwidth}{!}{
\begin{tikzpicture}
    [event/.style={circle, draw=black, fill=black, minimum size=2mm}]
    \node (sbegin) at (0, 12) [event, label=right:Begin] {};
    \node (slp) at (0, 7) [event, label=right:Lin Point] {} ;
    \node (send) at (0, 2) [event, label=right:End] {};
    \draw[dotted] (0, 14) -- node[above, outer sep=10mm] {\bf SOBJ} (sbegin);
    \draw [decorate, decoration=snake] (sbegin) -- node[right, outer sep=2mm, align=left] {\it Uncommitted\\State} (slp);
    \draw (slp) -- node[right, outer sep=2mm] {\it Known State} (send);
    \draw[dotted] (send) -- (0, 0);

    \node (cbegin) at (4, 12) [event, label=right:Begin] {};
    \node (clp) at (4, 7) [event, label=right:Lin Point] {};
    \node (cend) at (4, 2) [event, label=right:End] {};
    \draw[dotted] (4, 14) -- node[above, outer sep=10mm] {\bf COBJ} (cbegin);
    \draw [decorate, decoration=snake] (cbegin) -- node[right, outer sep=2mm, align=left] {\it Uncommitted\\State} (clp);
    \draw [dashed] (clp) -- node[right, outer sep=2mm, align=left] {\it Unknown\\State} (cend);
    \draw[dotted] (cend) -- (4, 0);

    \node (hcbegin) at (8, 12) [event, label=right:Begin] {};
    \node (hclp) at (8, 7) [event, label=right:$\tau$] {};
    \node (hcend) at (8, 2) [event, label=right:End] {};
    \draw[dotted] (8, 14) -- node[above, outer sep=10mm] {\bf COBJ'} (hcbegin);
    \draw [decorate, decoration=snake] (hcbegin) -- node[right, outer sep=2mm, align=left] {\it Unstable} (hclp);
    \draw [dashed] (hclp) -- node[right, outer sep=2mm, align=left] {\it Unknown\\State} (hcend);
    \draw[dotted] (hcend) -- (8, 0);

    \node (mclaim) at (12, 12) [event, label=right:Begin Claim] {};
    \node (mclaim lp) at (12, 10) [event, label=right:Claim LP] {};
    \node (end mclaim) at (12, 8) [event, label=right:End Claim] {};
    \node (mrelease) at (12, 6) [event, label=right:Begin Release] {};
    \node (mrelease lp) at (12, 4) [event, label=right:Release LP] {};
    \node (end mrelease) at (12, 2) [event, label=right:End Release] {};
    \draw[dotted] (12, 14) -- node[above, outer sep=10mm] {\bf Mutex} (mclaim);
    \draw [dashed] (mclaim) -- node[right, outer sep=2mm, align=left] {\it Unknown\\State} (mclaim lp);
    \draw[very thick, double] (mclaim lp) -- node[right, outer sep=2mm, align=left] {\it Locked} (end mclaim);
    \draw[very thick, double] (end mclaim) -- node[right, outer sep=2mm, align=left] {\it Locked} (mrelease);
    \draw [decorate, decoration=snake] (mrelease) -- node[right, outer sep=2mm, align=left] {\it Uncommitted\\State} (mrelease lp);
    \draw [dashed] (mrelease lp) -- node[right, outer sep=2mm, align=left] {\it Unknown\\State} (end mrelease);
    \draw[dotted] (end mrelease) -- (12, 0);

    \node (hclaim) at (16, 12) [event, label=right:Begin Claim] {};
    \node (hclaim lp) at (16, 10) [event, label=right:$\tau$] {};
    \node (end hclaim) at (16, 8) [event, label=right:End Claim] {};
    \node (hrelease) at (16, 6) [event, label=right:Begin Release] {};
    \node (hrelease lp) at (16, 4) [event, label=right:$\tau$] {};
    \node (end hrelease) at (16, 2) [event, label=right:End Release] {};
    \draw[dotted] (16, 14) -- node[above, outer sep=10mm] {\bf Mutex'} (hclaim);
    \draw [dashed] (hclaim) -- node[right, outer sep=2mm, align=left] {\it Unknown\\State} (hclaim lp);
    \draw[very thick, double] (hclaim lp) -- node [right, outer sep=2mm]{\it Stable} (end hclaim);
    \draw[very thick, double] (end hclaim) -- node [right, outer sep=2mm]{\it Stable} (hrelease);
    \draw [decorate, decoration=snake] (hrelease) -- node [right, outer sep=2mm]{\it Unstable} (hrelease lp);
    \draw [dashed] (hrelease lp) -- node[right, outer sep=2mm, align=left] {\it Unknown\\State} (end hrelease);
    \draw[dotted] (end hrelease) -- (16, 0);
\end{tikzpicture}
}
\end{center}
\caption{Operational Lifetimes of a Lock Protected Object, Concurrent Object, and Mutex.}
\label{fig:lifetime}
\end{figure}

Figure~\ref{fig:lifetime} presents five concurrent object lifelines. $SOBJ$ is an object with {\tt synchronized} operations. After an operation call, $SOBJ$ enters an uncommitted operation state. That is, there are still internal behaviours that will modify the state of $SOBJ$. At the commitment point (we use linearization point for consistency across lifelines), we know the object's state again. Although the operation has not ended, we know internally the state change has occurred. The calling thread has locked $SOBJ$ so no other operation can be performed. Therefore, we know $SOBJ$'s. We do have uncertainty regarding when the operation will commit. However, $SOBJ$ creates the correct sequential histories Herilhy and Wing~\cite{HerlihyWing90} describe for linearization.

$COBJ$ has no {\tt synchronized} methods, allowing concurrent overlapping operations from multiple threads. $COBJ$ will behave differently to $SOBJ$. On operation invocation, $COBJ$, like $SOBJ$, enters an uncommitted state. However, after the linearization point, $COBJ$'s state is unknown as we do not know if other operations are in progress. Although linearization points demonstrate that correct sequential behaviour is possible (i.e., in some regard $SOBJ$ provides a partial specification of $COBJ$), any individual thread is uncertain of $COBJ$'s state.

In our previous work~\cite{software4030015}, we compared linearization points between equivalent $SOBJ$ and $COBJ$ style specifications to demonstrate linearizability. For full specification, we hid linearization points to verify correct behaviour of an implementation. This is represented by $COBJ'$ in Figure~\ref{fig:lifetime}. We replace {\it Lin Point} with $\tau$ (the hidden event).

When considering $COBJ'$ within the stable failures model of CSP, an operation call makes $COBJ'$ become $\tau$-enabled and entering an unstable state. This unstable state remains until at least the hidden {\it Lin Point} event occurs. After this point, $COBJ'$ may still be in an unstable state if other operations are in progress, or it may return to a stable state if no other operations are in progress.

When considering a mutex, we observe a different behaviour pattern due to the relationship between the {\it claim} and {\it release} operations. When we begin a {\it claim}, we do not know the state of the mutex. Although it could be argued that when {\it claim} returns false we know that some process owns the lock, this is not true at the return point itself. On return, a {\it release} could have started, placing the mutex into an uncommitted state. Rather, we consider {\it claim} to end when the thread owns the mutex. We discuss this consideration further in Section~\ref{sec:lin-analysis}.

At {\it Claim LP}, the mutex enters a known {\it locked} state, which remains until we {\it release} is called. Although other threads may call {\it claim} (and be added to the queue), we know that no other thread will emit {\it Claim LP}, {\it End Claim}, {\it Begin Release}, or {\it Release LP}. Any previous releasing thread may emit the {\it End Release} event due to scheduling and concurrent {\it claim} and {\it release} operations. After {\it Begin Release}, we enter the only known uncommitted state. At this point, {\it release} tries to complete, potentially with a concurrent claimer. After {\it Release LP}, the mutex enters an unknown state as we do not know if another thread is active within it.

We can again hide the internal linearization points of the {\it claim} and {\it release} operations for \textit{Mutex}, providing {\it Mutex'}. Considering the stable failures model again, when the lock commits (via the hidden claim linearization point), the mutex enters a steady state. Although other threads can begin claiming, the mutex will not stop offering any events --- {\it Mutex'} is not $\tau$-enabled after {\it Claim LP}. This stable state remains until {\it Begin Release}, at which point we enter a known unstable state until {\it release} commits. At that point, there is no stability certainty of the mutex as we do not know the claim state.

Understanding these behaviours of general lock-free objects, and specifically the mutex, enables deeper understanding of specification behavioural needs, as well as the specification's correctness. Indeed, we can demonstrate that a mutex is linearizable by defining a simple mutex specification to check against much like $SOBJ$ specifies a correct linearizable behaviour of $COBJ$. We can also demonstrate that our mutex specification does indeed lock when required by checking the stability of the mutex process model at that point.

Therefore, we develop the following three specifications:
\begin{enumerate}
    \item A general specification of a mutex, allowing multiple concurrent claimers, using our linearization point approach to manage internal commitment.
    \item A simplified specification of a mutex that demonstrates the linearizability of the mutex via an equivalent {\tt synchronized} object approach.
    \item A specification that considers process stability to demonstrate mutex locking.
\end{enumerate}

\subsection{General Mutex Specification}

We take the same approach for specifying a mutex as we did for a concurrent queue in our previous work~\cite{software4030015}. We consider that between each {\it begin} and {\it end} event, there is a {\it linearization point} event denoting operation commitment. Implementation linearization points are unknown --- we work to the assumption that it happens between {\it begin} and {\it end}.

First, we add two additional channels, {\it lin\_claim} and {\it lin\_release}, to signify linearization points within {\it claim} and {\it release} respectively. These events are hidden for model checking.

\begin{tabbing}
==\===\===\kill
\>{\sf channel} {\it lin\_claim : users}\\
\>{\sf channel} {\it lin\_release : users}
\end{tabbing}

A mutex (defined as {\it MUTEX\_SPEC}) can always accept any available {\it begin\_claim} event. It can likewise always accept any available {\it end\_release} event. These two events signify the beginning and end of the claim/release, and can overlap with other {\it begin\_claim} and {\it end\_release} events as Figure~\ref{fig:concqueue} demonstrated. Between these two events, a successful lock and release transaction must occur.

At some point, a {\it lin\_claim.user} event can occur, and $user$ becomes the mutex owner. The events {\it end\_claim.user}, {\it begin\_release.user}, and {\it lin\_release.user} will progress the claim and release transaction. The owner resets to $NULL$ after {\it lin\_release.user}, allowing another {\it lin\_claim} event to occur.
\begin{align*}
MUTEX\_SPEC(user) &= \ begin\_claim?\_ \then MUTEX\_SPEC(user)\\
&\ \extchoice \ \ user = NULL \ \& \ lin\_claim?proc \then MUTEX\_SPEC(proc)\\
&\ \extchoice \ \ user \neq NULL \ \& \ end\_claim.user \then MUTEX\_SPEC(user)\\
&\ \extchoice \ \ user \neq NULL \ \& \ begin\_release.user \then MUTEX\_SPEC(user)\\
&\ \extchoice \ \ user \neq NULL \ \& \ lin\_release.user \then MUTEX\_SPEC(NULL)\\
&\ \extchoice \ \ end\_release \then MUTEX\_SPEC(user)
\end{align*}

{\it MUTEX\_SPEC} does not restrict the behaviour of the mutex for specification, and we again attach controlling {\it USER} processes to {\it MUTEX\_SPEC}:
\begin{align*}
USER(pid) = \ &begin\_claim.pid \then lin\_claim.pid \then end\_claim.pid \then\\
&begin\_release.pid \then lin\_release.pid \then end\_release.pid \then USER(pid)
\end{align*}

We then interleave a set of {\it USER} processes in parallel with {\it MUTEX\_SPEC}, synchronising on all events.
\begin{align*}
SPEC = \bigl(\big|\big|\big| \ u \in users \bullet USER(u) \underset{\alpha SPEC}{\big|\big|} MUTEX\_SPEC(NULL)\bigr) \ \backslash \ \{ | lin\_claim, lin\_release | \}
\end{align*}

The {\it linearization point} events are hidden for comparison with the implementation. We have encapsulated this complete specification within a module to simplify use.

\subsection{Linearizable Mutex Specification}
\label{sec:lin_mutex}

The mutex specification internally acts as a linearizable mutex via internal linearization points. Our argument is that any implementation meeting this specification must behave internally as a linearizable mutex. If the observed external events ({\it begin} and {\it end}) of the specification and implementation are equivalent in the {\it stable-failures} model, we claim the claim/release protocol behaves externally as a linearizable mutex.

We can check for linearizability of our specification by observing the hidden {\it linearization point} events, and then arguing that the {\it begin} and {\it end} events will happen immediately before and after the {\it linearization point}. Therefore, a sequential history (of unstable-stable state pairs) is possible.

\begin{figure}
\begin{center}
\resizebox{0.75\textwidth}{!}{
\begin{tikzpicture}
    [
        begin/.style={circle, draw=black, fill=white, minimum size=2mm},
        end/.style={circle, draw=black, fill=black, minimum size=2mm},
        lin/.style={circle, draw=black, fill=white, double=black, line width=1pt, minimum size=2mm},
        transform/.style={single arrow, draw=black, fill=white}
    ]

    \node (T1label) at (0, 20) {$T_1$};
    \node[begin] (T1begin) at (0, 18) {};
    \node[lin] (T1commit) at (0, 12) {};
    \node[end] (T1end) at (0, 10) {};
    \draw[dotted] (T1label) -- (T1begin);
    \draw[decorate, decoration=snake] (T1begin) -- (T1commit);
    \draw[dashed] (T1commit) -- (T1end);
    \draw[dotted] (T1end) -- (0, 0);

    \node (T2label) at (1, 20) {$T_2$};
    \node[begin] (T2begin) at (1, 16) {};
    \node[lin] (T2commit) at (1, 4) {};
    \node[end] (T2end) at (1, 2) {};
    \draw[dotted] (T2label) -- (T2begin);
    \draw[decorate,decoration=snake] (T2begin) -- (T2commit);
    \draw[dashed] (T2commit) -- (T2end);
    \draw[dotted] (T2end) -- (1, 0);

    \node (T3label) at (2, 20) {$T_3$};
    \node[begin] (T3begin) at (2, 14) {};
    \node[lin] (T3commit) at (2, 8) {};
    \node[end] (T3end) at (2, 6) {};
    \draw[dotted] (T3label) -- (T3begin);
    \draw[decorate,decoration=snake] (T3begin) -- (T3commit);
    \draw[dashed] (T3commit) -- (T3end);
    \draw[dotted] (T3end) -- (2, 0);

    \node[transform] at (3, 10) {Lin};

    \node (LinLabel) at (4, 20) {Lin Points};
    \node[lin, label=right:$T_1$] (T1lin) at (4, 12) {};
    \node[lin, label=right:$T_3$] (T3lin) at (4, 8) {};
    \node[lin, label=right:$T_2$] (T2lin) at (4, 4) {};
    \draw[dotted] (LinLabel) -- (T1lin);
    \draw (T1lin) -- (T3lin);
    \draw (T3lin) -- (T2lin);
    \draw[dotted] (T2lin) -- (4, 0);

    \node[transform] at (5, 10) {Seq};

    \node (LinearizeLabel) at (6, 20) {Seq};
    \node[begin] (T1LPbegin) at (6, 13) {};
    \node[lin, label=right:$T_1$] (T1LP) at (6, 12) {};§
    \node[end] (T1LPend) at (6, 11) {};
    \node[begin] (T3LPbegin) at (6, 9) {};
    \node[lin, label=right:$T_3$] (T3LP) at (6, 8) {};
    \node[end] (T3LPend) at (6, 7) {};
    \node[begin] (T2LPbegin) at (6, 5) {};
    \node[lin, label=right:$T_2$] (T2LP) at (6, 4) {};
    \node[end] (T2LPend) at (6, 3) {};
    \draw[dotted] (LinearizeLabel) -- (T1LPbegin);
    \draw[decorate,decoration=snake] (T1LPbegin) -- (T1LP);
    \draw (T1LP) -- (T1LPend);
    \draw (T1LPend) -- (T3LPbegin);
    \draw[decorate,decoration=snake] (T3LPbegin) -- (T3LP);
    \draw (T3LP) -- (T3LPend);
    \draw (T3LPend) -- (T2LPbegin);
    \draw[decorate,decoration=snake] (T2LPbegin) -- (T2LP);
    \draw (T2LP) -- (T2LPend);
    \draw[dotted] (T2LPend) -- (6, 0);

    \node (T1Mutex) at (10, 20) {$T_1$};
    \node[begin, label=right:Claim] (T1claim) at (10, 19) {};
    \node[lin] (T1linclaim) at (10, 6) {};
    \node[end] (T1endclaim) at (10, 5) {};
    \node[begin, label=right:Release] (T1release) at (10, 4) {};
    \node[lin] (T1linrelease) at (10, 3) {};
    \node[end] (T1endrelease) at (10, 2) {};
    \draw[dotted] (T1Mutex) -- (T1claim);
    \draw[dashed] (T1claim) -- (T1linclaim);
    \draw[very thick, double] (T1linclaim) -- (T1endclaim);
    \draw[very thick, double] (T1endclaim) -- (T1release);
    \draw[decorate, decoration=snake] (T1release) -- (T1linrelease);
    \draw[dashed] (T1linrelease) -- (T1endrelease);
    \draw[dotted] (T1endrelease) -- (10, 0);
    
    \node (T2Mutex) at (12, 20) {$T_2$};
    \node[begin, label=right:Claim] (T2claim) at (12, 18) {};
    \node[lin] (T2linclaim) at (12, 16) {};
    \node[end] (T2endclaim) at (12, 15) {};
    \node[begin, label=right:Release] (T2release) at (12, 14) {};
    \node[lin] (T2linrelease) at (12, 13) {};
    \node[end] (T2endrelease) at (12, 10) {};
    \draw[dotted] (T2Mutex) -- (T2claim);
    \draw[dashed] (T2claim) -- (T2linclaim);
    \draw[very thick, double] (T2linclaim) -- (T2endclaim);
    \draw[very thick, double] (T2endclaim) -- (T2release);
    \draw[decorate, decoration=snake] (T2release) -- (T2linrelease);
    \draw[dashed] (T2linrelease) -- (T2endrelease);
    \draw[dotted] (T2endrelease) -- (12, 0);

    \node (T3Mutex) at (14, 20) {$T_3$};
    \node[begin, label=right:Claim] (T3claim) at (14, 17) {};
    \node[lin] (T3linclaim) at (14, 12) {};
    \node[end] (T3endclaim) at (14, 11) {};
    \node[begin, label=right:Release] (T3release) at (14, 9) {};
    \node[lin] (T3linrelease) at (14, 8) {};
    \node[end] (T3endrelease) at (14, 1) {};
    \draw[dotted] (T3Mutex) -- (T3claim);
    \draw[dashed] (T3claim) -- (T3linclaim);
    \draw[very thick, double] (T3linclaim) -- (T3endclaim);
    \draw[very thick, double] (T3endclaim) -- (T3release);
    \draw[decorate, decoration=snake] (T3release) -- (T3linrelease);
    \draw[dashed] (T3linrelease) -- (T3endrelease);
    \draw[dotted] (T3endrelease) -- (14, 0);

    \node[transform] at (15.5, 10) {Mutex};

    \node (Mutex) at (17, 20) {Mutex};
    \node[lin, label=right:$T_2$ Lock] (T2lock) at (17, 16) {};
    \node[lin, label=right:$T_2$ Unlock] (T2unlock) at (17, 13) {};
    \node[lin, label=right:$T_3$ Lock] (T3lock) at (17, 12) {};
    \node[lin, label=right:$T_3$ Unlock] (T3unlock) at (17, 8) {};
    \node[lin, label=right:$T_1$ Lock] (T1lock) at (17, 6) {};
    \node[lin, label=right:$T_1$ Unlock] (T1unlock) at (17, 3) {};
    \draw[dotted] (Mutex) -- (T2lock);
    \draw[very thick, double] (T2lock) -- (T2unlock);
    \draw (T2unlock) -- (T3lock);
    \draw[very thick, double] (T3lock) -- (T3unlock);
    \draw (T3unlock) -- (T1lock);
    \draw[very thick, double] (T1lock) -- (T1unlock);
    \draw[dotted] (T1unlock) -- (17, 0);

    \node[begin, label=right:Begin] at (4, -2) {}; 
    \node[lin, label=right:Commit] at (8, -2) {};
    \node[end, label=right:End] at (12, -2) {};

\end{tikzpicture}
}
\end{center}
\caption{Transforming Concurrent Behaviours into Linearizable Specifications.}
\label{fig:linearizable}
\end{figure}
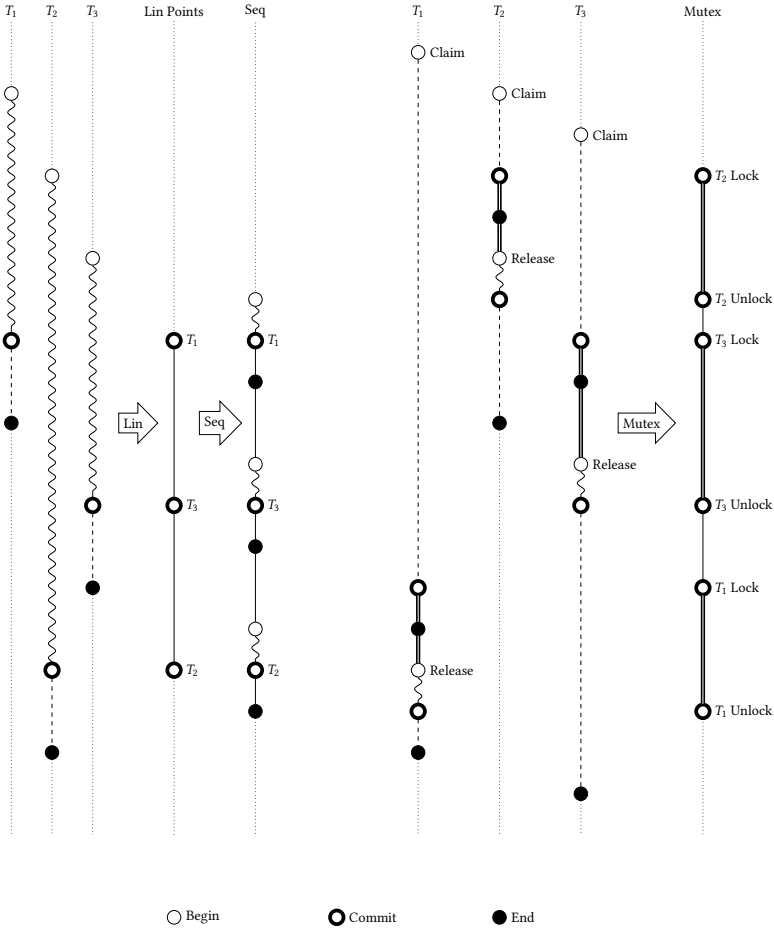

Figure~\ref{fig:linearizable} presents how we can test for linearizable behaviour in FDR by observing linearization point events. On the left of Figure~\ref{fig:linearizable}, three threads ($T_1, T_2, T_3$) are interacting with some concurrent object. The threads have the same triple-event signature --- {\it begin}, {\it commit}, and {\it end} --- as used in Figure~\ref{fig:concqueue}. The operations are also overlapping. The three linearization points occur in some order (here, $T_1, T_3, T_2$). We know linearization points are when object modification occurs. As such, rather than hide linearization points as we do for main specification checking, we can hide the {\it begin} and {\it end} events, observing state change events. This transformation, {\it Lin}, provides the {\it Lin Points} timeline.

Linearization includes reordering \textit{begin} and \textit{end} events into a valid sequential history of the concurrent object. Linearization point analysis makes this trivial, as we know that any ordering of linearization points produces a correct sequential history. We can ignore when {\it begin} and {\it end} occur in this scenario --- their order does not matter. Indeed, we can argue that a sequential history of {\it begin}, {\it commit}, and {\it end} events occurs when we move the {\it begin} and {\it end} events immediately preceding and proceeding the \textit{linearization point} event. The second transformation --- {\it Seq} --- demonstrates this, leading to timeline {\it Seq}. It is a trivial observation that considering only the concurrent object's interaction, {\it Seq} provides the same result as $T_1, T_2$ and $T_3$ executing concurrently in Figure~\ref{fig:linearizable}. The commit order is the same for both.

We can apply the same approach for claim/release behaviour. Figure~\ref{fig:linearizable} also presents three threads interacting with the claim/release protocol. The claim/release has two commit points (one for {\it claim} and one for {\it release}), and thus the transformation is different. Each thread will have a {\it lock}-{\it unlock} chain, between which the thread has exclusive access to the resource controlled. This linearized version of the mutex is the standard CSP mutex implementation (see our example at the start of this section). That is, we can define a {\it linearizable} mutex as:
\begin{align*}
LIN\_MUTEX = lock?pid \then unlock.pid \then LIN\_MUTEX
\end{align*}

If our specification provides linearizable behaviour, we can test by swapping visible and hidden events. The main specification hides linearization events. If we hide the {\it begin} and {\it end} events (set {\it BEGIN\_END}) and expose the {\it linearization} events, and rename {\it lin\_claim} and {\it lin\_release} to {\it lock} and {\it unlock}, we have a specification that should match $LIN\_MUTEX$.
\begin{align*}
SPEC' &= \Bigl( \ \big|\big|\big| \ u \in users \bullet USER(u) \underset{\alpha SPEC}{||} \ MUTEX\_SPEC(NULL) \ \Bigr) \ \backslash \ BEGIN\_END\\
LIN\_SPEC &= SPEC' \ \llbracket lin\_claim, lin\_release \backslash lock, unlock \rrbracket
\end{align*}

We can test that {\it LIN\_SPEC} behaves as {\it LIN\_MUTEX}, and thus argue that our implementation is also linearizable.
\begin{align*}
LIN\_MUTEX \refinedby{FD} LIN\_SPEC \land SPEC \refinedby{FD} IMPL
\end{align*}

If both pass, we have demonstrated that $SPEC$ behaves as a linearizable mutex (the internal behaviour of $SPEC$ is equivalent to $LIN\_MUTEX$). We will have also demonstrated that our implementation behaves as $SPEC$ (the external behaviours of $SPEC$ and $IMPL$ are equivalent). Thus, our argument is that given the internal behaviour of $SPEC$ is linearizable, the external behaviour emitted must demonstrate linearizable behaviour. Then, as $IMPL$ emits the same external behaviour as $SPEC$, $IMPL$ must also emit external behaviour that is linearizable.

These assertions are not transitive as $LIN\_SPEC$ exposed different external events to $SPEC$. Our argument is not for equivalence between $LIN\_MUTEX$ and $IMPL$. It is that $SPEC$ emits linearizable behaviour, and given this our $IMPL$ will do likewise.

\subsection{Ensuring Locking via Stability Testing}

With $LIN\_MUTEX \refinedby{FD} LIN\_SPEC$ we have not verified the stable state concept presented in Figure~\ref{fig:linearizable} that meets locking conditions. Indeed, we have stated that $LIN\_MUTEX$ does not specify our implementation as it considers a modified $SPEC$, and also argued that a specification such as $LIN\_MUTEX$ only enables locking for implementation purposes. We have not demonstrated that the implementation has {\it lock-unlock} behaviour.

Unlike a general concurrent lock free object, the mutex will enter a stable state when {\it lin\_claim} occurs. A general concurrent lock free object can accept multiple {\it begin} events and therefore can be in a constant state of instability with some process expecting an {\it end} event to occur after the internal (hidden) {\it linearization point} event. Given the mutex enters a known stable state, we can verify that the lock does indeed ensure mutual exclusion.

After {\it lin\_claim}, the mutex enters a stable state --- it becomes locked. This stable state remains until {\it begin\_release} occurs. We want to know that no $\tau$ events can occur between these events in our specification, or more simply after {\it claim} completes and before {\it release} starts.

FDR provides a number of built-in functions to support verification. One such function is {\tt chase}, which provides a form of state compression by modifying a process to prioritise entering stable states. For any state, $s$ of a process, if $\tau$ will lead to a state $s'$, then {\tt chase(s) = chase(s')}. If the state is not $\tau$ enabled, {\tt chase(s) = s}. Thus, {\tt chase} prioritises moving into a stable state, by transforming an unstable state to a stable one, and removing the $\tau$ event. This modification is non-semantics preserving, leading to a process with different observed behaviour.

As an example, consider the following process, and its transformation under {\tt chase}:
\begin{align*}
P &= (a \then ((b \then c \then {\sf SKIP}) \ \extchoice \ (d \then {\sf SKIP}))) \ \backslash \ \{ b \}\\
{\tt chase}(P) &= a \then c \then {\sf SKIP}
\end{align*}

{\tt chase(P)} transforms $P$ such that the hidden event $b$ is prioritised, combining the unstable state between $a$ and $c$ caused by the hidden event $b$ with the stable state where $c$ is ready. We have effectively removed the unstable state, and prioritised the unstable pathway in the process execution.

We can use this approach for checking if our lock is unstable when it should be locked. First, we create a new specification to test our {\it mutex} specification against, $TauPriorSpec = chase(SPEC)$. On the first $begin\_claim$, $TauPriorSpec$ will immediately transition into a state where the partner $end\_claim$ is available. No other $end\_claim$ will become available until the first one occurs and transitions through {\it locked} to {\it unlocked}. Likewise, on $begin\_release$, $TauPriorSpec$ will immediately transition, either to a state where only $end\_release$ is available (as no other $begin\_claim$ has occurred), or one where an $end\_claim$ is also available to remove the instability of a waiting $begin\_claim$ event. In other words both the $end\_release$ and an $end\_claim$ will be available.

It should be obvious that $traces(SPEC) \not\subset traces(TauPriorSpec)$. If a $begin\_claim$ occurs with no mutex owner, the {\it claim} immediately resolves to being ready to complete. This is not the same as enforcing a $begin\_claim \then end\_claim$ sequence. Rather, we have moved the process into a state where $end\_claim$ is immediately available if the mutex is unlocked. As such, the following should be true:
\begin{align*}
\lnot (TauPriorSpec \refinedby{T} SPEC)
\end{align*}

$SPEC$ has different behaviour to $TauPriorSpec$ --- $SPEC$ can perform behaviours $TauPriorSpec$ does not allow. This is understandable. {\tt chase} is non-semantics preserving and multiple $begin\_claim$ events can occur before one moves into a resolution point (via a $lin\_claim$ event). However, we are only concerned about stability between {\it end\_claim} and {\it begin\_release}. Therefore, we can hide the {\it begin\_claim} and {\it end\_release} events (set $HIDE$ below) and compare $TauPriorSpec$ and $SPEC$ in this condition. That is, we test that between {\it end\_claim} and {\it begin\_release}, both $TauPriorSpec$ and $SPEC$ have the same stable state behaviour, knowing that $TauPriorSpec$ {\it only} contains stable states. Our test for correct lock behaviour is therefore:
\begin{align*}
TauPriorSpec \ \backslash \ HIDE \refinedby{F} SPEC \ \backslash \ HIDE
\end{align*}

However, we can also directly check with the implementation model to reassure ourselves that correct locking occurs in the implementation. $IMPL$ will not have constant stable states between $end\_claim$ and $begin\_release$ as we hide internal behaviour (such as enqueuing) of $IMPL$, but our concern is whether the locking events are stable. Therefore, we check that:
\begin{align*}
TauPriorSpec \ \backslash \ HIDE \refinedby{F} IMPL \ \backslash \ HIDE
\end{align*}

\noindent
to demonstrate the implementation has the same behaviour as the stabilised specification.

\subsection{Safe States}

We must also ensure correct state on mutex ownership. After a successful $CLAIM$, the claiming process must be in the \textit{ACTIVE} state. Indeed, it must also be so prior to claiming (and thus after releasing). We can add a simple $CHECK\_STATE$ process that diverges if a process's state is not \textit{ACTIVE}, using it in a $USER\_SAFE$ process.
\begin{align*}
CHECK\_STATE(pid, s) = \ &getState.pid?st \then \ {\sf if} \ st == s \ {\sf then \ SKIP \ else \ DIV}\\
USER\_SAFE(pid) = \ &CHECK\_STATE(pid, ACTIVE); CLAIM(pid); \\
&CHECK\_STATE(pid, ACTIVE); RELEASE(pid); USER\_SAFE(pid)
\end{align*}

$SYSTEM\_SAFE$ uses a set of $USER\_SAFE$ processes instead of $USER$, and check $SYSTEM\_SAFE$ is divergence free. If it is, we know that on a successful $claim$ a process is always in the $ACTIVE$ state.

\subsection{Fairness}

One final property we can test for is fairness. Our mutex should be fair, insofar that the order processes are enqueued to the wait queue is the order that processes will claim the resource. We can test this ordering by examining the behaviour between {\it begin\_claim}, {\it lin\_enqueue} (the commit point of enqueue), and {\it end\_claim}. In simple terms, we want to demonstrate that, given the trace:
\begin{align*}
\trace{\dots, begin\_claim.p_i, \dots, lin\_enqueue.p_i.p_i, \dots, end\_claim.p_i, \dots}
\end{align*}

\noindent
for any process $p_i$, if $lin\_enqueue.p_i.p_i$ precedes $lin\_enqueue.p_j.p_j$, then $end\_claim.p_i$ also precedes $end\_claim.p_j$.

A simple process to demonstrate fairness will maintain a set of started claimers (only started claimers can enqueue, and must enqueue) and a sequence of committed enqueues, managing these based on $begin\_claim$, $lin\_enqueue$, and $end\_claim$ events:
\begin{align*}
FAIR&(claims, waiting) =\\
&\quad begin\_claim?pid \in NonNullPID - claims \then FAIR(claims \cup \{ pid \}, waiting)\\
&\extchoice \ pid \in claims - waiting \ \bullet lin\_enqueue.pid.pid \then FAIR(claims, waiting ^\frown \langle pid \rangle)\\å
&\extchoice \ end\_claim.head(waiting) \then FAIR(claims - \{ head(waiting) \}, tail(waiting))
\end{align*}

We can then demonstrate that an implementation model $FAIR\_IMPL$ --- which only exposes $begin\_claim, lin\_enqueue$ and $end\_claim$ --- meets this specification. That is:
\begin{align*}
FAIR(\{ \}, \langle \rangle) \refinedby{FD} FAIR\_IMPL
\end{align*}

\noindent
should hold if the implementation ensures fairness.
\section{Results and Discussion}
\label{sec:discussion}

In this section we discuss the results from the FDR analysis. We also discuss aspects of our implementation and modelling approach, including scalability of our results, linearization analysis, and claim resolution costs. Finally, we discuss the use of FDR as an iterative lock-free algorithm development tool.

\subsection{Results}

In Section~\ref{sec:spec} we presented five different checks to undertake with FDR: \textit{specification refinement}, \textit{linearizability of the specification}, \textit{stability}, \textit{state safety}, and \textit{fairness}. Furthermore, we test for \textit{deadlock freedom}, \textit{divergence freedom}, and \textit{determinism}. The results of our tests are below.

\begin{table}[H]
\begin{tabular}{|llll|}
\hline
(1) & $IMPL$ {\bf deadlock free} & $1 \leq N \leq 4$ users & \\
(2) & $IMPL$ {\bf divergence free} & $1 \leq N \leq 4$ users & \\
(3a) & $SPEC$ {\bf not deterministic} & $2 \leq N \leq 4$ users & \\
(3b) & $IMPL$ {\bf not deterministic} & $2 \leq N \leq 4$ users & \\
(4) & $SPEC \refinedby{FD} IMPL \refinedby{FD} SPEC$ & $1 \leq N \leq 4$ users & \\
(5) & $LIN\_MUTEX \refinedby{FD} LIN\_SPEC$ & $1 \leq N \leq 4$ users & {\bf Linearizable} \\
(6a) & $TauPriorSpec \refinedby{FD} SPEC$ & $1 \leq N \leq 4$ users & {\bf Lockable}\\
(6b) & $TauPriorSpec \refinedby{FD} IMPL$ & $1 \leq N \leq 4$ users & {\bf Lockable}\\
(7) & $SYSTEM\_SAFE$ \textbf{divergence free} & $1 \leq N \leq 4$ users & \textbf{Safe} \\
(8) & $FAIR \refinedby{FD} FAIR\_IMPL$ & $1 \leq N \leq 4$ users & {\bf Fair} \\
\hline
\end{tabular}
\end{table}

In the following sections we discuss the implication of these results.

\subsubsection{Deadlock Free}

Deadlock freedom is a standard safety condition for any concurrent system. Our $IMPL$ being deadlock free means that there is no condition where the protocol deadlocks, implying that processes transition through the claim/release cycle without getting into a state where they can no longer progress. We have demonstrated correct liveness of the protocol.

\subsubsection{Divergence Free}

Divergence indicates internal hidden loops with no progress, such as would be seen if the claim/release protocol featured busy (spinning) waits. Our protocol is divergence free, and therefore we can state it has no busy waits. Indeed, this is somewhat obvious from the implementation model.

\subsubsection{Not Deterministic}

One failed FDR property is determinism when there is greater than one $USER$ process. Both the $SPEC$ and the $IMPL$ are non-deterministic. However, this is not a failed outcome, and indeed it demonstrates a key property we have argued for --- the order of \textit{begin\_claim} events does not predict the order of \textit{end\_claim} events. The determinism analysis confirms this. The failed cases are where the visible \textit{begin\_claim} events do not lead to the same visible \textit{end\_claim} events. Thus, our specification and protocol being non-deterministic is a desired outcome.

\subsubsection{$SPEC \refinedby{FD} IMPL$}

Our main result is that the protocol behaves as a mutex within the \textit{stable failures} model, and as our models are \textit{divergence free}, this means our implementation behaves as specified in the \textit{failures/divergences} model. Furthermore, we have also reversed the check, such that $IMPL \refinedby{FD} SPEC$ is also demonstrated, demonstrating equivalence between $IMPL$ and $SPEC$.

Meeting this test provides us with proof that our protocol does behave as the specified mutex. As such, we have not only verified our protocol is correct, but we can also use the specification as a lightweight drop-in replacement for future modelling. This is important to our ongoing work developing a lock-free runtime for ProcessJ.

\subsubsection{Linearizable Specification}

A question we posed is whether the specification provides linearizable behaviour. Although not a requirement for correctness, we will discuss linearization and the outcomes of this property further in Section~\ref{sec:lin-analysis}.

As $LIN\_MUTEX \refinedby{FD} LIN\_SPEC$ --- which tests that the internal behaviour of the specification behaves as a linearizable mutex --- we argue that the specification must emit the external behaviour of a linearizable mutex. There are considerations for this claim, which we discuss in Section~\ref{sec:lin-analysis}, but our argument of linearizability stands.

\subsubsection{Lockable}

In Section~\ref{sec:spec}, and specifically Section~\ref{sec:lin-stable}, we argued that if our specification is lockable, it should enter a stable state between $lin\_claim$ and $begin\_release$. Our test, via $TauPriorSpec$, tests between $end\_claim$ and $begin\_release$ which is sufficient for our purposes. Our aim with $TauPriorSpec$ is to demonstrate that both $SPEC$ and $IMPL$ behave in that stable manner. Passing this test means that once $end\_claim$ becomes available, it continues to do so. Once $end\_claim$ occurs, $begin\_release$ becomes available and continues to do so. There are no $\tau$ events that lead to either event being unavailable. As $end\_claim$ becomes available after $lin\_claim$, we have demonstrated that the owning process can always continue progress, and no other process can enter the critical section as defined in $SPEC$.

\subsubsection{Safe}

We also want to ensure that the state changes within the protocol ensure a process is \textit{ACTIVE} when it has successfully claimed. This is a safety property of the protocol --- a process not \textit{ACTIVE} should not be executing within the critical section.

Our successful test demonstrates that the protocol ensures a process is \textit{ACTIVE} after a successful \textit{claim}, and also after a successful \textit{release}. As such, our protocol does ensure safe behaviour.

\subsubsection{Fair}

Our final property of interest is fairness. We have argued that our protocol is fair, insofar that processes enter the critical section of the mutex in the order they enter the internal wait queue. Our $FAIR$ oracle tests that only claiming processes can be enqueued, and that $end\_claim$ events occur in the same process order as $lin\_enqueue$ events. Therefore, we have demonstrated that our protocol ensures fairness based on enqueue order, and that no process can jump the queue. All processes must start claiming, then enter the queue, and then $end\_claim$ in the enqueue order.

As we have kept fairness outside the specification, our mutex specification can be reused in other modelling studies. Indeed, we can create different fairness policies (e.g., priority queues) and model those instead using an appropriate fairness oracle. This approach to fairness checking is therefore versatile enough to support various modelling scenarios.

\subsection{Verification Scalability}

Our FDR tests are limited to a maximum of four user processes. This is due to the state-space growing, especially with possible queue permutations. However, we argue that our protocol does scale safely beyond four processes.

Appendix~\ref{app:proof} sketches an inductive proof based on mutex state including queue length. Our proof demonstrates that an invariant $INV(n)$-structure ensures correct exclusive ownership and fairness for all $n$. The machinery adapts ownership into a one of \textit{locked, unlocked}, \textit{claiming} or \textit{releasing}, using a $Qstate$ property for each $n$ to demonstrate arbitrary processes ensure correct queue structure, ownership, and fairness.

With this proof, we argue that the protocol will work for any number of claiming processes, given the operating conditions and subsequent assumptions described in Section~\ref{sec:runtime}.

\subsection{Linearization Analysis and Limit of Verification}
\label{sec:lin-analysis}

Our approach to linearizability checking simplifies proof obligation checking to a refinement check in the stable failures model of CSP. By using an event triple (\textit{begin-linearize-end}) and hiding the \textit{linearize} events, we avoid explicit linearization-point reasoning.

An unanswered question is where the linearization points are within our claim/release protocol. We have argued that we do not need to define these as the specification we verify against contains linearization points internally. However, it is worth considering some specification and modelling design choices we have made around linearization which impact the validity of our verification.

The challenge of linearization point determination is specifying what the commit point would be for the operation. We could consider the two individual {\it linearization points} for {\it claim} and {\it release}, with the most likely candidate for {\it claim} being when the claiming process becomes the {\it owner}. For {\it release}, setting the {\it owner} to something different --- {\tt null} or another claiming process --- is the best candidate. However, this view of {\it claim} raises some limitations regarding our verification.

Consider the point of {\it claim} failure (the {\it claimed resource} scenario). Our arguments in Section~\ref{sec:lin_mutex} was locking persisted between the $lin\_claim$ and $begin\_release$ events. The mutex becomes locked due to entering a specification stable state. We have demonstrated between the $end\_claim$ and $begin\_release$ events the mutex specification is indeed stable, and the implementation meets this behaviour. However, in our model, {\it yield} is within the algorithm, occurring before $end\_claim$. Algorithm~\ref{alg:claim} rather returns true of false depending on claim success.

\begin{figure}
\begin{center}
\resizebox{0.75\textwidth}{!}{
\begin{tikzpicture}
    [
        begin/.style={circle, draw=black, fill=white, minimum size=2mm},
        end/.style={circle, draw=black, fill=black, minimum size=2mm},
        yield/.style={circle, dotted, draw=black, fill=white, minimum size=2mm},
        lin/.style={circle, draw=black, fill=white, double=black, line width=1pt, minimum size=2mm},
        transform/.style={single arrow, draw=black, fill=white}
    ]

    \node (S1Label) at (2, 21) {\bf Internal Yield};

    \node (T1Mutex) at (0, 20) {$T_1$};
    \node[begin, label=right:Claim] (T1claim) at (0, 19) {};
    \node[yield, label=right:Yield] (T1yield) at (0, 15) {};
    \node[lin] (T1linclaim) at (0, 11) {};
    \node[end] (T1endclaim) at (0, 10) {};
    \node[begin, label=right:Release] (T1release) at (0, 9) {};
    \node[lin] (T1linrelease) at (0, 8) {};
    \node[end] (T1endrelease) at (0, 7) {};
    \draw[dotted] (T1Mutex) -- (T1claim);
    \draw[dashed] (T1claim) -- (T1yield);
    \draw[dashed] (T1yield) -- (T1linclaim);
    \draw[very thick, double] (T1linclaim) -- (T1endclaim);
    \draw[very thick, double] (T1endclaim) -- (T1release);
    \draw[decorate, decoration=snake] (T1release) -- (T1linrelease);
    \draw[dashed] (T1linrelease) -- (T1endrelease);
    \draw[dotted] (T1endrelease) -- (0, 6);

    \node (T2Mutex) at (4, 20) {$T_2$};
    \node[begin, label=right:Claim] (T2claim) at (4, 18) {};
    \node[lin] (T2linclaim) at (4, 17) {};
    \node[end] (T2endclaim) at (4, 16) {};
    \node[begin, label=right:Release] (T2release) at (4, 13) {};
    \node[lin] (T2linrelease) at (4, 12) {};
    \node[end] (T2endrelease) at (4, 11) {};
    \draw[dotted] (T2Mutex) -- (T2claim);
    \draw[dashed] (T2claim) -- (T2linclaim);
    \draw[very thick, double] (T2linclaim) -- (T2endclaim);
    \draw[very thick, double] (T2endclaim) -- (T2release);
    \draw[decorate, decoration=snake] (T2release) -- (T2linrelease);
    \draw[dashed] (T2linrelease) -- (T2endrelease);
    \draw[dotted] (T2endrelease) -- (4, 6);

    \node (S2Label) at (10, 21) {\bf External Yield};

    \node (T3Mutex) at (8, 20) {$T_1$};
    \node[begin, label=right:Claim] (T3claim) at (8, 19) {};
    \node[end] (T3endclaim) at (8, 15) {};
    \node[yield, label=right:Yield] (T3yield) at (8, 14) {};
    \node[lin] (T3linclaim) at (8, 10) {};
    \node[begin, label=right:Release] (T3release) at (8, 9) {};
    \node[lin] (T3linrelease) at (8, 8) {};
    \node[end] (T3endrelease) at (8, 7) {};
    \draw[dotted] (T3Mutex) -- (T3claim);
    \draw[dashed] (T3claim) -- (T3endclaim);
    \draw[dashed] (T3endclaim) -- (T3yield);
    \draw[dashed] (T3yield) -- (T3linclaim);
    \draw[very thick, double] (T3linclaim) -- (T3release);
    \draw[decorate, decoration=snake] (T3release) -- (T3linrelease);
    \draw[dashed] (T3linrelease) -- (T3endrelease);
    \draw[dotted] (T3endrelease) -- (8, 6);

    \node (T4Mutex) at (12, 20) {$T_2$};
    \node[begin, label=right:Claim] (T4claim) at (12, 18) {};
    \node[lin] (T4linclaim) at (12, 17) {};
    \node[end] (T4endclaim) at (12, 16) {};
    \node[begin, label=right:Release] (T4release) at (12, 14) {};
    \node[lin] (T4linrelease) at (12, 13) {};
    \node[end] (T4endrelease) at (12, 12) {};
    \draw[dotted] (T4Mutex) -- (T4claim);
    \draw[dashed] (T4claim) -- (T4linclaim);
    \draw[very thick, double] (T4linclaim) -- (T4endclaim);
    \draw[very thick, double] (T4endclaim) -- (T4release);
    \draw[decorate, decoration=snake] (T4release) -- (T4linrelease);
    \draw[dashed] (T4linrelease) -- (T4endrelease);
    \draw[dotted] (T4endrelease) -- (12, 6);

    \node (S2Label) at (19, 21) {\bf Single Linearization Point};

    \node (T5Mutex) at (16, 20) {$T_1$};
    \node[begin, label=right:Begin Claim] (T5begin) at (16, 19) {};
    \node[lin] (T5lin) at (16, 10) {};
    \node[end, label=right:End Release] (T5end) at (16, 7) {};
    \draw[dotted] (T5Mutex) -- (T5begin);
    \draw[decorate, decoration=snake] (T5begin) -- (T5lin);
    \draw[dashed] (T5lin) -- (T5end);
    \draw[dotted] (T5end) -- (16, 6);

    \node (T6Mutex) at (20, 20) {$T_2$};
    \node[begin, label=right:Begin Claim] (T6begin) at (20, 18) {};
    \node[lin] (T6lin) at (20, 17) {};
    \node[end, label=right:End Release] (T6end) at (20, 12) {};
    \draw[dotted] (T6Mutex) -- (T6begin);
    \draw[decorate, decoration=snake] (T6begin) -- (T6lin);
    \draw[dashed] (T6lin) -- (T6end);
    \draw[dotted] (T6end) -- (20, 6);

\end{tikzpicture}
}
\end{center}
\caption{Claim and Release as a Single Linearizable Operation.}
\label{fig:fail-claim}
\end{figure}
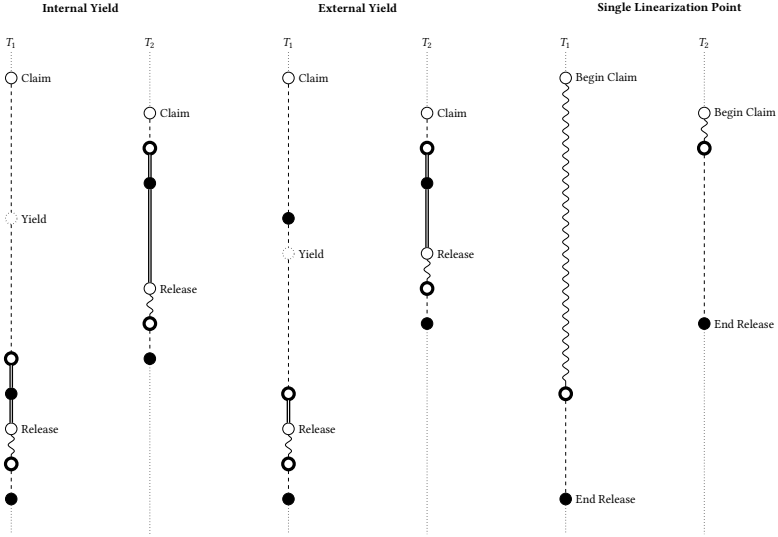

Consider Figure~\ref{fig:fail-claim}. On the left, we have added the {\it yield} point as modelled. Note that within our modelling, {\it yield} occurs before $end\_claim$. As such, we are stating that a claiming process will not $end\_claim$ until after a releasing process wakes it from the queue --- we extend the $end\_claim$ event to incorporate the \textit{yield point}. However, our algorithm as presented in Algorithm~\ref{alg:claim} does not behave in this manner. Rather, we end the claim and return false on an unsuccessful claim attempt, yielding after $end\_claim$. We present this as the middle scenario of Figure~\ref{fig:fail-claim}.

In the middle scenario, the question arises of where to place the linearization point. An operation cannot have a linearization point outside its lifetime ({\it begin} and {\it end} events), but the algorithm might not transfer ownership within the operation lifetime --- the {\it claim} has not claimed the resource but rather gone into a waiting state.

If we were to change our model to return true or false on $end\_claim$ we would face a challenge. Our mutex specification would have to deal with processes waiting in some form of queue on claim failure. This would add complexity to the specification and move towards something closer to the implementation model. The $lin\_claim$ event would have to add the claiming process to the internal queue, but this event is hidden, creating an unknown ordering of $end\_claim$ events. In the current specification, only one claiming process may proceed due to only one $lin\_claim$ occurring at a time. In a queue based specification, $lin\_claim$ can happen at any time after $begin\_claim$, creating a hidden order of queued processes. As each process must subsequently perform the externally observable $end\_claim$ event and potentially in a different order to the internal $lin\_claim$ events, the hidden internal ordering would have to be met by the implementation, which would be impossible. The observed order of $end\_claim$ events would not match the order of the internal queue.

Thus, although we have argued for linearizability and stability, we base the argument on extending where $end\_claim$ occurs to after a claiming process owns the resource rather than at the actual end point of the {\it claim} operation. We believe this limitation to our verification does not invalidate our claims as we have stated that the algorithm works if a yield will occur immediately after claim failure. Thus, yielding is an integral part of the overall {\it claim} operation.

\subsubsection{Mutex as a Guarantor of Linearizability}

We could take the view from the right of Figure~\ref{fig:fail-claim}. Rather than consider {\it claim} and {\it release} as separate operations with their own linearization points, we consider the full {\it claim-release} cycle as a single operation and a linearization point must occur between $begin\_claim$ and $end\_release$. Taking this view would return us to a general concurrent object. We can argue that this single linearization point does occur from our analysis of stability and linearizability --- that linearization point is the critical section. However, taking this view would mean that between $begin\_claim$ and $end\_release$, the owning process can undertake other operations not linked directly to the mutex concurrent object.

So, although from a {\it mutex as pure concurrent object} point of view a {\it single linearization point} may look sensible, it does raise concerns about what else is occurring during the combined {\it claim/release} operation. Linearizability is about the concurrent object (the {\it mutex}) and not what threads do outside this interaction. However, linearizability considers a thread only engaging with the linearizable operation of a concurrent object and not undertaking interactions with other objects. Given that a mutex is used to enforce sequential interaction, considering the mutex as part of another object interaction to enforce linearization for said interaction (such as in {\tt synchronized} methods) is a useful analogy. Indeed, we have demonstrated that our mutex will enforce this linearizable interaction. As such, the mutex can be considered as a guarantor of safe concurrent behaviour for any unsafe operation included within the linearization point.

\subsection{CAS Requirements for Resolution}
\label{sec:resolution-cas}

Although we have verified that our protocol provides the behaviour we expect, we have noted that it does so by relying on the queue as a form of mutual exclusion object. Thus, our protocol can require more {\it CAS} operations to successfully claim the resource than a single atomic value in user-space. However, we believe that resolutions overall are responsive in most scenarios (as outlined in Section~\ref{sec:diagrammatic}). There are some edge cases to consider. First, let us analyse the likely costs of each scenario.

\begin{description}
    \item[Unclaimed] Figure~\ref{fig:claimed-resolution} (a) is a fast resolution of a successful claim requiring three {\it CAS} operations. Even with multiple claimers in operation, the winner will only require three {\it CAS} operations. If there is a concurrent {\it release} with {\it claim}, then there may be four {\it CAS} operations if {\it release} modifies the tail of the queue as {\it claim} also tries.
    \item[Claimed] Figure~\ref{fig:claimed-resolution} (b) is a fast resolution of an unsuccessful claim with three {\it CAS} operations. Multiple claimers acting at once may result in multiple failed enqueue attempts, leading to further {\it CAS} operations.
    \item[Released-in-time] Figure~\ref{fig:claimed-resolution} (c) is a variation of \textit{claimed} where the releasing process notifies the claiming process before it moves into a \textit{waiting} state. It also has three {\it CAS} operations, but will not have clashed so has a maximum of three {\it CAS} operations.
    \item[Owner \texttt{null}] Figure~\ref{fig:concurrent-release} (a) occurs when the copied \texttt{owner} field is \texttt{null}. The claiming process performs no additional \textit{CAS} operations after the copy, and so has the same \textit{CAS} operation count as \textit{unclaimed}.
    \item[Unscheduled owner transfer] Figure~\ref{fig:concurrent-release} (b) happens when the releasing process has set a new owner, but has yet to schedule the claiming process. We have a slow resolution of a successful claim. The claiming process must wait for {\it release} to signal it by returning to the end of the run queue, or spinning waiting for the signal. There is one additional \textit{CAS} attempt on the process state.
    \item[Scheduled owner transfer] Figure~\ref{fig:concurrent-release} (c) is when the schedule signal from release arrives such that the claiming process does not have to wait for it. This situation has the same number of potential \textit{CAS} operations as the \textit{unscheduled} scenario.
    \item[Ownership stolen] Figure~\ref{fig:concurrent-release} (d) incurs just one additional \textit{CAS} than the fast \textit{unclaimed} resolution. This is the \textit{CAS} to steal ownership. Thus, this scenario has four or five \textit{CAS} operations.
    \item[Released, owner \texttt{null}] Figure~\ref{fig:release-completed} (a) occurs when the \texttt{owner} field has changed after the claiming process copied it, and thus contains all \textit{CAS} operations up to \textit{ownership stolen}. An additional \textit{CAS} occurs to change the \texttt{owner} to the claiming process.
    \item[Released, unscheduled] Figure~\ref{fig:release-completed} (b) is the next check where the claiming process has failed the \textit{CAS} to set the claiming process from \texttt{null} and tries to change its state to \textit{WAITING}, incurring an additional \textit{CAS}. This \textit{CAS} will succeed and lead to either a return to the run queue for slow resolution, or can be resolved via a busy wait for the scheduled signal.
    \item[Released, scheduled] Figure~\ref{fig:release-completed} (c) occurs when the release signal has been received by the claiming process. In this scenario, the \textit{CAS} count is the same, but we have a fully successful claim.
\end{description}

In Table~\ref{tab:resolution} we summarise this analysis. The one area of slow resolution is when a {\it claim} is active and {\it release} attempts to schedule it. This scenario occurs because {\it release} does not know whether the claiming process at the head of the queue is engaging or waiting.

\begin{table}
    \caption{Outcomes and Resolution Performance of Different {\it claim} Scenarios}
    \begin{tabular}{|l|l|r|r|l|}
        \hline
        {\bf Scenario} & {\bf Outcome} & {\bf Min {\it CAS}} & {\bf Max {\it CAS}} & {\bf Resolution} \\
        \hline
        (1) {\bf Unclaimed} & Claimed & 3 & 4 & Fast \\
        \hline
        (2) {\bf Claimed resource} & Unclaimed & 3 & Unknown & Fast \\
        \hline
        (3) {\bf Released-in-time} & Claimed & 3 & 3 & Fast \\
        \hline
        (4) {\bf Owner \texttt{null}} & Claimed & 3 & 4 & Fast \\
        \hline
        (5) {\bf Unscheduled owner transfer} & Claimed & 4 & 5 & Slow or busy \\
        \hline
        (6) {\bf Scheduled owner transfer} & Claimed & 4 & 5 & Fast \\
        \hline
        (7) {\bf Ownership stolen} & Claimed & 4 & 5 & Fast \\
        \hline
        (8) {\bf Released, owner \texttt{null}} & Claimed & 5 & 6 & Fast \\
        \hline
        (9) {\bf Released, unscheduled} & Claimed & 6 & 7 & Slow or busy \\
        \hline
        (10) {\bf Released, scheduled} & Claimed & 6 & 7 & Fast \\
        \hline
    \end{tabular}
    \label{tab:resolution}
\end{table}

Outside the slow resolution scenarios, the {\it claim} will at most require seven {\it CAS} operations. There are other atomic get and set operations on both \texttt{owner} and \texttt{state}, but these should have less impact on algorithm performance.

It should be noted that this analysis presents idealised scenarios. For example, it is possible for various other claim-release pairs to occur on the queue as a {\it claim} enqueues, causing further {\it CAS} operations. This scenario would require a claiming process, $P$, to constantly lose the enqueuing race to other process, and for the winner of the races to {\it release} before $P$ can complete enqueuing on subsequent attempts. As such, it is unlikely that such a scenario can occur, and occur continuously. For example, in ProcessJ, $P$ will not give up its runner resource, and will always be able to continue progress. Other processes claiming would likewise have to be active, and release the lock immediately. Therefore, we believe Table~\ref{tab:resolution} provides the most likely outcomes under normal operating conditions.

\subsection{Use of FDR for Iterative Lock-Free Algorithm Development}

We would also make note of using FDR for this style of lock-free algorithm development. Developing and refining the algorithm has been greatly simplified through having a specification to verify against with FDR. The below development stages highlight our process supported by FDR:
\begin{enumerate}
    \item With a specification in place, we attempted to implement the protocol with our scheduler. This allowed us to spot various bugs with our naive first attempts, and specifically the spurious scheduling issue.
    \item Our first working implementation would initially fail fully if the claiming process could not set the \texttt{owner} field, such as in the partially successful claim attempts. The \textit{claim} made no attempt to determine if it had already been notified.
    \item We added the fairness check and the lockable check.
    \item The second working implementation updated the fail points to return the value of the \textit{ready} flag to indicate if the claiming process had been scheduled by the releasing process.
    \item The third implementation improved the scheduling by creating a proper \textit{state} value for each process using atomic interactions.
    \item The fourth implementation refined further by using the result of a final \textit{CAS} on process \textit{state} to determine if a process should wait.
    \item We added the $SYSTEM\_SAFE$ test.
    \item We reordered the tests in \textit{claim} to improve readability and presentation.
\end{enumerate}

With each iteration, we could check the correctness via FDR, ensuring our modifications did not break the algorithm. As such, we believe using FDR as a tool for model checking algorithms with \textit{CAS}-like operations is very beneficial. Extending our approach to consider different memory semantics (e.g., relaxed, acquire-release) would allow further exploration of algorithms to improve performance.

\subsection{Java Reference Implementation}

To support algorithm implementers we have provided a reference Java implementation alongside the CSP models. This can be found in the GitHub repository for this project~\footnote{\url{https://github.com/kevin-chalmers/processj-csp/tree/main/claim-release}}.

The Java reference implementation relies on preemptive threads and busy waiting on process state to operate. The claim/release protocol will work under these conditions, but with a preemptive scheduler we must busy wait as we cannot control when a thread is ready-to-run without making it \texttt{wait}. However, we have no method of confirming when we should notify a waiting Java thread as values can change between testing and the thread entering the wait state.   
\section{Related Work}
\label{sec:related}

Our claim/release protocol is defined within a cooperatively scheduled context, relying on user mode threads rather than system level threads. There has been some work on defining locks in this context, and several implementations of note. We first explore published work and then provide an overview of implementations.

\subsection{Published Approaches}

There are three core approaches that can be applied to user mode mutexes. Mellor-Crummey and Scott~\cite{MCS91} provide a lock based on CAS operations and a lock-free queue, much like ours. However, the approach uses spin locks within a queue of waiters. So although lacking system-level locking, MCS locks are not spin free.

Mellor-Crummey and Scott~\cite{MCS91} also describe a ticket lock, where a thread entering a critical section aquires a numbered ticket. To claim the critical section, the thread must have the correct ticket number, and if not it spins until it does.

Craig~\cite{craig1993building} describes the Craig-Landin-Hagersen (CLH) lock, which is similar to an MCS lock. Processes spin on a flag within a queue of waiting processes.

In all three cases, spin locks are used to manage ownership of shared resources. Our claim/release protocol undertakes no such spinning, and therefore provides an approach that can potentially use fewer resources than other options.

\subsection{Implementations}

Table~\ref{tab:comparison} provides a comparison of other approaches to create user mode threading mutexes. We explore four properties that our claim/release protocol has:
\begin{enumerate}
    \item {\bf Lock-freedom}, insofar that the execution system is not locked or constrained through kernel level locking. Does the approach ensure that user mode threads do not invoke a system mode thread lock?
    \item {\bf Spin-freedom}, insofar that the locking does not attempt to busy wait (even briefly) on some atomic variable.
    \item {\bf Fairness}. Does the implementation ensure some form of FIFO ordering with no opportunity for a thread to steal a lock?
    \item {\bf Verification}. Has the implementation been formally verified?
\end{enumerate}

We include the Mellor-Crummey and Scott, and Craig-Landin-Hagersten lock in the table for completeness.

\begin{table}
    \small
    \caption{Comparison to Existing Cooperative Locking Approaches}
    \label{tab:comparison}
    \begin{center}
    \begin{tabular}{|l|l|l|l|l|}
        \hline
        \textbf{Implementation} & \textbf{Lock-free} & \textbf{Spin-free} & \textbf{Fair} & \textbf{Verified} \\
        \hline
        \textbf{Go} {\tt sync.Mutex} & No & No & No & No \\
        \hline
        \textbf{Go} {\tt sync.RWMutex} & No & No & No & No \\
        \hline
        \textbf{Kotlin} Kotlinx Mutex & Yes & Partial & Yes & No \\
        \hline
        \textbf{Rust} {\tt tokio::Mutex} & Yes & Partial & Yes & No \\
        \hline
        \textbf{Node.js} {\tt async-mutex} & Yes & Yes & Yes & No \\
        \hline
        \textbf{Python} {\tt asyncio.Lock} & Yes & Yes & Yes & No \\
        \hline
        \textbf{C++} {\tt Boost.Fiber} & Yes & No & No & No \\
        \hline
        \textbf{Win32} Fibers & No & No & No & No \\
        \hline
        \textbf{Mellor-Crummey and Scott} & Yes & No & Yes & Yes \\
        \hline
        \textbf{Craig, Landin, and Hagersten} & Yes & No & Yes & Partial \\
        \hline
    \end{tabular}
    \end{center}
\end{table}

Go's {\tt sync.Mutex}\footnote{See line 72 \url{https://cs.opensource.google/go/go/+/refs/tags/go1.9.7:src/sync/mutex.go}} is a wrapper around a state and a system level {\tt futex}. If a {\it CAS} on the state word fails (after a few spinning attempts), the OS level {\it futex} is used to park the {\it goroutine}. Therefore, {\tt sync.Mutex} is not lock free of spin free. The Go scheduler is not fair insofar that scheduling heuristics determine which {\it goroutines} to wake. There is no published verification of Go's {\tt sync.Mutex}. Likewise, Go's {\tt sync.RWMutex}\footnote{\url{https://github.com/golang/go/blob/master/src/sync/rwmutex.go}} uses the same principles. Go is no longer cooperatively scheduled since 1.14, although user-space scheduling is still used.

Kotlin's {\tt kotlinx.coroutines.Mutex}\footnote{\url{https://github.com/Kotlin/kotlinx.coroutines/blob/master/kotlinx-coroutines-core/common/src/sync/Mutex.kt}} provides a close comparison to our claim/release protocol for properties. The algorithm is almost spin-free, except in a small case when a {\tt release} runs ahead of a slow {\tt acquire}. However, the Kotlinx mutex has no published verification of correctness. The approach is somewhat based on Mellor-Crummey and Scott's~\cite{MCS91} lock approach.

Rust's {\tt tokio::Mutex}\footnote{\url{https://github.com/tokio-rs/tokio/blob/master/tokio/src/sync/mutex.rs}} uses a similar approach to Kotlinx, and thus does have a potential spin under certain conditions. No formal verification exists.

The Node.js {\tt async-mutex} package\footnote{\url{https://github.com/DirtyHairy/async-mutex}} allows control of asynchronous tasks in JavaScript. JavaScript is strictly single-threaded, so there is no system lock, and therefore no blocking of the thread runner resource. The system is built via {\tt Promise} resolution from the JavaScript event loop. Thus, all core criteria are met, but within a stictly single-threaded runtime. There is no formal verification of this system, although the implementation is a simple promise-queue pattern.

Python's {\tt asyncio.Lock} is similar to the Node.js {\tt async-mutex} package. {\tt asyncio} does allow multiple event queues to operate. However, the {\tt acquire} and {\tt release} operations must be executed from the same event queue. Internally, {\tt acquire} creates a future for the given loop, and if {\tt release} were called from a different loop then the operation would be unsafe (one run loop would be manipulating another). Our claim/release does not suffer from this sensitivity as our model has no event queues and all processes can always run. The {\tt asyncio.Lock} is likewise unverified.

The C++ {\tt Boost.Fiber}\footnote{\url{https://github.com/boostorg/fiber}} package provides lightweight threads (fibers) to C++ programmers using yielding much like ProcessJ, although the programmer must explicitly call {\tt yield}(). Boost fibers are also executed via runner threads. {\tt fiber\_mutex} in {\tt Boost.Fiber} is lock-free, although when syncrhonising between multiple runner threads, {\tt fiber\_mutex} is not spin-free as the processes will spin trying to communicate properly between runner threads. Our claim/release protocol has no such limitation. This also means outcomes depend on multiple threads spinning until those attempts are exhausted and the process gives up to the OS level scheduler, which is unfair. No published verification of {\tt fiber\_mutex} is provided.

Win32 Fibers are similar to {\tt Boost.Fiber}. A {\tt SRWLOCK} can operate purely in user mode (and thus is lock-free and spin-lock free). However, user mode contention can lead to spinning, and when contested between two kernel-level runners, there is a system level lock. There are also opportunities where threads may unfairly claim the lock. No published verification is available.

Mellor-Crummey and Scott~\cite{MCS91} queue-based lock relies on spinning on private per-thread flags. As such, although an MCS lock has no system locks, spin-lock behaviour is present. However, the internal queue ensures fairness. An MCS lock has been verified as part of the CertiKOS kernel~\cite{CertiKOS}.

Craig's~\cite{craig1993building} CLH lock also does not use system level locks, but it does rely on a spin lock on a flag similarly to an MCS lock. The lock is fair within the queue of waiting processes. CLH has been verified, including under weak memory effects~\cite{Colvin2024}. However, full liveness verification was not achieved.

\subsection{Summary of Related Work}

Although other lightweight, user-mode threading locks exist, no published article or implementation meets the same properties of our claim/release protocol. Either, the mechanism can fall back to system level blocking (Go, Win32 Fibers), requires some form of spin to acquire ({\tt Boost.Fiber}), or only operates within a single-threaded environment (Node.js, Python). Kotlinx and the Tokio package in Rust almost meet the requirements, but do have slight spinning on claiming. No implementation appears to be formally verified.

The MCS and CLH locks that many implementations are inspired from likewise have spin locks. Although there is some verification work, the CLH lock lacks the level of verification we have achieved in this paper.

Some claim algorithms spin test a value a few times to see if a claim can be made. In contrast, our algorithm is more deliberate in its \textit{CAS} operations. We transition through a series of tests to determine the \textit{release state} of the mutex when claiming, with each \textit{CAS} aimed at determining how the claiming process can safely claim ownership if it is the next owner. The claim algorithm performs no unnecessary \textit{CAS} operations and can determine the definite path the claiming process should take, including a fast resolution to waiting.

\subsection{Linearization Analysis with CSP}

We are not the first to use CSP to analyse linearization. Lowe~\cite{lock-free-queue-analysis} examined a lock-free queue using FDR. However, we model with {\it linearization points}, whereas Lowe used $Linearizer$ processes as external synchronizers of behaviour. Our approach eliminates the need for such processes, and simplifies proof obligations. Crucially, it allows us to model a linearizable mutex specification, as well as separate mutex FIFO fairness into an independent observation oracle that observes the queue's enqueue linearization event. Lowe's $Linearizer$ technique is powerful for pure data structures, but for a mutex scenario it would entangle scheduling events with object interaction events. Our hidden linearization point approach avoids that coupling.

Liu et al.~\cite{liu2009model, liu2012verifying} verify linearizability in PAT by viewing it as a trace refinement between an implementation a deterministic sequential specification. As Lowe, their method only relies on invocation-response events. By emitting an internal {\it linearization} event, we move linearizability into a single stable-failures refinement check and allow interaction with our scheduler and further proofs, making our proof easier to expand into other model checking scenarios and reusable specification development.      
\section{Conclusions, Limitations, and Future Work}
\label{sec:conclusion}

In this paper we presented a claim/release protocol for a cooperatively scheduled runtime of concurrent processes each trying to access a shared channel. The algorithm provides exclusive access to object data a based on a queue of access-requesting processes. This protocol is lock- and wait-free, and uses no mutexes or spin-locks; it is implemented using atomic variables and the {\it compare-and-set} operations on these atomics.

We have shown, using CSP and FDR, that our implementation is deadlock and live-lock free; furthermore, we have given a simple linearizable mutex specification and verified that our implementation behaves exactly like this specification. This means that our lock- and wait-free implementation can be used in place of mutexes and locks. 

Specifically, we have:
\begin{itemize}
    \item Provided a spin-free, CAS-only mutex for cooperatively scheduled runtimes (Section~\ref{sec:protocol}), including a Java reference implementation available for runtime implementers.
    \item Proving our mutex interacting with a cooperative scheduler preserves mutual exclusion and FIFO fairness (Sections~\ref{sec:spec}, \ref{sec:discussion}, and Appendix~\ref{app:proof}).
    \item Defined a reusable CSP mutex specification that supports concurrent claim and release operations (Section~\ref{sec:spec}).
    \item Defined a verification strategy from linearization analysis, and proving commit point equivalence (Section~\ref{sec:spec} and \ref{sec:discussion}). Our {\it begin-lin-end} triple, with {\it lin} hidden, collapses linearizability checking to a single stable-failures refinement, whereas prior CSP work (e.g., Lowe~\cite{lock-free-queue-analysis}) required an extra {\it Linearizer} process.
    \item Development of a fairness checking oracle external to the specification (Section~\ref{sec:spec}).
    \item CAS-cost analysis of different claim scenarios (Section~\ref{sec:discussion}).
    \item Survey of different approaches to locking in cooperative runtimes (Section~\ref{sec:related}).
\end{itemize}

\subsection{Limitations}

Our claim/release protocol meets its correctness properties (kernal-lock-freedom, spinlock-freedom, fair) under specific runtime environment conditions as described in Section~\ref{sec:runtime}. Although our protocol will ensure mutual exclusion outside these conditions --- as demonstrated in our reference Java threads implementation --- it will require softening of the correctness properties. We have specifically designed the algorithm for cooperatively scheduled environments --- a scheduling environment requiring processes to explicitly \textit{yield} to return execution resources.

Our direct FDR verification is limited to four processes contesting for the claimed resource. This is due to the growing state space as the number of processes --- and subsequent wait queue length --- increases. To mitigate, we have provided an inductive proof arguing that the protocol will manage larger number of processes. We believe there should be no implications in how the protocol operates outside this proof.

Our verification --- and specifically the linearizability claim --- is limited with regard to the inclusion of \textit{yielding} within the models checked. We have argued that rather than consider the individual operations of the mutex as linearizable, the entire claim-release cycle is the operation of note. However, although we have used a hidden linearization point modelling approach, the \textit{claim} operation only meets this behaviour with the inclusion of a \textit{yield} point.

There is also a potential starvation scenario within our protocol. It is possible (although unlikely) that a process can continuously fail to enqueue itself on the wait queue as it loses the race to modify the tail appropriately. Our core argument against this happening is that there will be a limited number of active processes based on the runner resources within the runtime. However, the scenario still remains and should be considered when increasing the number of active processes within the runtime, or if other scheduling algorithms are used.

We have not undertaken performance analysis of our protocol and this is left for future work. Benchmarking against Go or Rust approaches is inappropriate as ProcessJ targets the JVM. Likewise, comparison to event-queue approaches running in other virtual machines or interpreted language environments (Python, JavaScript) would not be a fair comparison. Our protocol is potentially slower in best-case scenarios due to our enqueue-first strategy.

\subsection{Future Work}

Our core aim was to develop a mechanism for protecting shared channel access in ProcessJ as we migrate to a lock-free runtime. Completing this development and the verification of behaviour is a key next step in our work. Furthermore, expanding our verification approach to other concurrency primitives is a worthwhile exploration of behaviour.

To understand protocol performance, we require microbenchmark analysis against other published lock types (e.g., MCS, CLH). Using a microbenchmark framework such as JMH would allow us to understand the cost of lock acquisition in comparison to other approaches. Also, examining different locking scenarios to test our assumption regarding starvation on enqueue is important.

Exploring the protocol in other memory models would also be beneficial. Indeed, expanding our simple atomic variable model to consider weaker memory models would be beneficial for analysis of other protocols.   

\bibliographystyle{elsarticle-num-names}
\bibliography{biblio.bib}

\appendix
\section{Concurrent Queue Module}
\label{app:conqueue}

\begin{tabbing}
==\===\===\===\===\===\kill
{\sf module} {\it ConcQueue(max)}\\
\>{\sf channel} {\it lin\_enqueue : NonNullPID.NonNullPID} \\
\>{\sf channel} {\it lin\_dequeue : NonNullPID.PID} \\
\>{\sf channel} {\it lin\_peek : NonNullPID.PID}\\
\\
\>$USER(id) =$\\
\>\>\>$enqueue.id?val \then lin\_enqueue.id.va \then end\_enqueue.id \then USER(id)$\\
\>\>$\extchoice$\>$dequeue.id \then lin\_dequeue.id?val \then return.id.val \then USER(id)$\\
\>\>$\extchoice$\>$peek.id \then lin\_peek.id?val \then return.id.val \then USER(id)$\\
\\
\>$QUEUE\_SPEC(q) = ({\sf length}(q) \leq max \& \ ($\\
\>\>\>$(enqueue?\_?\_ \then QUEUE\_SPEC(q))$\\
\>\>$\extchoice$\>$lin\_enqueue?proc?v \then QUEUE\_SPEC(q ^\frown \langle v \rangle)$\\
\>\>$\extchoice$\>$end\_enqueue?\_ \then QUEUE\_SPEC(q)$\\
\>\>$\extchoice$\>$dequeue?\_ \then QUEUE\_SPEC(q)$\\
\>\>$\extchoice$\>${\sf if} \ ({\sf length}(q) = 0) \ {\sf then} \ lin\_dequeue?proc!NULL \then QUEUE\_SPEC(q)$\\
\>\>\>${\sf else} \ lin\_dequeue?proc!head(q) \then QUEUE\_SPEC(tail(q))$\\
\>\>$\extchoice$\>$peek?\_ \then QUEUE\_SPEC(q)$\\
\>\>$\extchoice$\>${\sf if} \ {\sf length}(q) = 0) \ {\sf then} \ lin\_peek?proc!NULL \then QUEUE\_SPEC(q)$ \\
\>\>\>${\sf else} \ lin\_peek?proc!head(q) \then QUEUE\_SPEC(q)$\\
\>\>$\extchoice$\>$return?\_?\_ \then QUEUE\_SPEC(q))$\\
\\
\>$SYNC = \{ | enqueue, dequeue, peek, end\_enqueue, return, lin\_enqueue, lin\_dequeue, lin\_peek | \}$\\
\>$HIDE = \{ | lin\_enqueue, lin\_dequeue, lin\_peek | \}$\\
\\
$exports$\\
\>$QUEUE(users) = \Bigl( \ \big|\big|\big| \ id \in users \bullet USER(id) \underset{SYNC}{\big|\big|} QUEUE\_SPEC(\langle \rangle) \ \Bigr) \ \backslash \ HIDE$\\
\>$alpha = \{ | enqueue, dequeue, peek, end\_enqueue, return | \}$\\
$endmodule$
\end{tabbing}

\section{Claim and Release CSP Models}
\label{app:models}

\subsection{Claim Algorithm}

\begin{tabbing}
==\===\===\===\===\===\===\=\kill
$CLAIM(pid) =$\\
\>$begin\_claim.pid \then$\\
\>$setState.pid.ENGAGING \then$\\
\>$enqueue.pid.pid \then end\_enqueue.pid \then$\\
\>$peek.pid \then return.pid?headQ \then$\\
\>${\sf if} \ headQ = pid \ {\sf then}$\\
\>\>$casOwner!NULL!pid?succ \then \ {\sf if} \ succ \ {\sf then} \ setState.pid!ACTIVE \then end\_claim.pid \then \ {\sf SKIP}$\\
\>\>${\sf else}$\\
\>\>\>$getOwner?own \then$\\
\>\>\>${\sf if} \ own = NULL \ {\sf then} \ setOwner!pid \then setState.pid!ACTIVE \then end\_claim.pid \then \ {\sf SKIP}$\\
\>\>\>${\sf else \ if} \ own = pid \ {\sf then}$\\
\>\>\>\>$casState.pid!ENGAGING!WAITING?succ \then$\\
\>\>\>\>${\sf if} \ succ \ {\sf then} \ YIELD(pid); end\_claim.pid \then \ {\sf SKIP}$\\
\>\>\>\>${\sf else} \ setState.pid!ACTIVE \then end\_claim.pid \then \ {\sf SKIP}$\\
\>\>\>${\sf else}$\\
\>\>\>\>$casOwner!own!pid?succ \then$\\
\>\>\>\>${\sf if} \ succ \ {\sf then} \ setState.pid!ACTIVE \then end\_claim.pid \then \ {\sf SKIP}$\\
\>\>\>\>${\sf else}$\\
\>\>\>\>\>$casOwner!NULL!pid?succ \then$\\
\>\>\>\>\>${\sf if} \ succ \ {\sf then} \ setState.pid!ACTIVE \then end\_claim.pid \then \ {\sf SKIP}$\\
\>\>\>\>\>${\sf else}$\\
\>\>\>\>\>\>$casState.pid!ENGAGING!WAITING?succ \then$\\
\>\>\>\>\>\>${\sf if} \ succ \ {\sf then} \ YIELD(pid); end\_claim.pid \then \ {\sf SKIP}$\\
\>\>\>\>\>\>${\sf else} \ setState.pid!ACTIVE \then  end\_claim.pid \then \ {\sf SKIP}$\\
\>${\sf else}$\\
\>\>$casState.pid!ENGAGING!WAITING?succ \then$\\
\>\>${\sf if} \ succ \ {\sf then} \ YIELD(pid); end\_claim.pid \then \ {\sf SKIP}$\\
\>\>${\sf else} \ setState.pid!ACTIVE \then end\_claim.pid \then \ {\sf SKIP}$
\end{tabbing}

\subsection{Release Algorithm}

\begin{tabbing}
==\===\===\===\===\===\===\=\kill
$RELEASE(pid) =$\\
\>$begin\_release.pid \then$\\
\>$dequeue.pid \then return.pid.pid$\\
\>$peek.pid \then return.pid?headQ$\\
\>${\sf if} \ headQ = NULL \ {\sf then} \ casOwner!pid!NULL?succ \then end\_release.pid \then \ {\sf SKIP}$\\
\>${\sf else}$\\
\>\>$casOwner!pid!headQ?succ \then$\\\
\>\>\>${\sf if} \ succ \ {\sf then} \ SCHEDULE(headQ); end\_release.pid \then \ {\sf SKIP}$\\
\>\>\>${\sf else} \ end\_release.pid \then \ {\sf SKIP}$
\end{tabbing}

\section{Inductive Proof Sketch of Scalability}
\label{app:proof}

\theoremstyle{plain}
\newtheorem*{thm}{Theorem}
\newtheorem*{cor}{Corollary}
\newtheorem*{prop}{Proposition}
\newtheorem*{lem}{Lemma}

\theoremstyle{definition}
\newtheorem*{defn}{Definition}
\newtheorem*{hypothesis}{Hypothesis}
\newtheorem{assumption}{Assumption}
\newtheorem{assertion}{Assertion}
\newtheorem*{property}{Property}
\newtheorem*{goal}{Goal}

\theoremstyle{remark}
\newtheorem*{remark}{Remark}
\newtheorem{case}{Case}
\newtheorem{subcase}{Subcase}[case]
\newtheorem*{preliminaries}{Preliminaries}

In this section we present an inductive proof sketch arguing that the locking mechanism will hold for any number of processes.

\begin{preliminaries}
We have the following model components:
\begin{itemize}
    \item $P$ --- a set of processes, $\{ p_i, p_j, p_k, p_l, p_m, \dots \}$, that are executing within the runtime.
    \item $\bot$ --- the {\tt null} value.
    \item $Owner$ --- current process that owns the resource or $\bot$. $Owner \in P \cup \{ \bot \}$.
    \item $Queue$ --- set of processes currently claiming the resource.
    \item $Queue[] = \langle p_i, p_j, \dots \rangle$ --- logical order of entries in $Queue$ from $head$ to $tail$.
    \item $|Queue|$ --- the number of members in $Queue$.
    \item $pos(p_i)$ --- the position of process $p_i$ in $Queue[]$ from $1 \dots n$ where $n = |Queue|$. $pos(p_i) = \bot \iff p_i \notin Queue$.
    \item $next(p_i)$ --- the next process in the queue after $p_i$, such that $next(p_i) \in P \cup \{ \bot \}$. $next(p_i) = p_j \iff pos(p_j) = pos(p_i) + 1$.
    \item $head$ --- the first process in $Queue[]$. $head = p_i \implies pos(p_i) = 1$.
    \item $tail$ --- the last process in $Queue[]$. $tail = p_i \implies (pos(p_i) = |Queue| \land next(tail) = \bot)$.
    \item Events:
    \begin{itemize}
        \item $claim.p_i$, the combination of $begin\_claim.p_i, lin\_claim.p_i, end\_claim.p_i$.
        \item $release.p_i$, the combination of $begin\_release.p_i, lin\_release.p_i, end\_release.p_i$.
    \end{itemize}
    \item $claiming(p_i)$ is true while $claim.p_i$ is active.
    \item $releasing(p_i)$ is true when $p_i = Owner \land p_i \notin Queue$ while a new value of $Owner$ is established.
\end{itemize}

\begin{assumption}[Limited number of active processes]
\label{ass:scheduler}
There are at most $k$ active runner threads executing processes.
\end{assumption}

\begin{assumption}[Well behaved processes]
\label{ass:double}
We assume claiming processes are well-behaved such that if $Owner = p_i$ then $p_i$ will not invoke $claim.p_i$. Likewise, we assume no process $p_i$ will invoke $release.p_i$ unless $Owner = p_i$.
\end{assumption}

\begin{assertion}[Waiting processes]
\label{ass:waiting}
No process $p_i$ can invoke $claim.p_i$ while $p_i \in Queue$. We define $waiting(p_i) = p_i \in Queue \land \lnot claiming(p_i) \land Owner \neq p_i$.

\end{assertion}

\begin{goal}
We want to show that for any $n \geq 1$ {\it waiters} currently enqueued, the following two properties hold:
\begin{enumerate}
    \item {\bf Exclusive ownership} --- at most one process is in the critical section. $Owner \neq \bot \implies (Owner = head \lor releasing(Owner))$.
    \item {\bf FIFO order} --- the process that joins the queue first will exit the critical section first.
\end{enumerate}

{\it Progression} is covered by Assumption~\ref{ass:scheduler}.
\end{goal}

\end{preliminaries}

\begin{defn}[Mutex invariant $INV(n)$]
Let $Qstate = Qstate(n, releasing(Owner))$ be a reachable global state, with $n = |Queue|$ and $(n \geq 0)$ with $n$ waiters queued. $Qstate(n, releasing(Owner))$ is a pair such that $Qstate(n, releasing(Owner)) = (Queue[], Owner)$. Given $releasing(p_i)$ is true when $Owner = p_i \land p_i \notin Queue$, we can simply enumerate $Qstate(n, releasing(Owner))$:
\begin{align*}
    Qstate(0, false) &= (Queue[] = \langle \rangle, Owner = \bot)\\
    Qstate(0, true) &= (Queue[] = \langle \rangle, Owner = p_i)\\
    Qstate(1, false) &= (Queue[] = \langle p_i \rangle, Owner = p_i)\\
    Qstate(1, true) &= (Queue[] = \langle p_j \rangle, Owner = p_i)\\
    Qstate(2, false) &= (Queue[] = \langle p_i, p_j \rangle, Owner = p_i)\\
    Qstate(2, true) &= (Queue[] = \langle p_j, p_k \rangle, Owner = p_i)\\
    \dots &= \dots
\end{align*}

\noindent
for arbitrary processes $p_i, p_j, p_k, \dots$.

$Qstate' = (Queue'[], Owner')$ after an operation step. For $INV(n)$ to hold the following properties must hold:

\begin{property}[(S) Structure] $Queue$ and $Queue[]$ provide a well-formed singly linked list. Considered under one of two conditions:
\begin{enumerate}
    \item \label{S1} (\textit{Claim condition}) $|Queue'| = |Queue| + 1 \implies tail' \neq \bot \land (tail \neq \bot \implies next(tail) = tail')$.
    \item \label{S2} (\textit{Release condition}) $|Queue'| = |Queue| - 1 \implies head' = next(head)$ when $|Queue| \geq 1$.
\end{enumerate}

$|Queue| = 0 \implies Queue[] = \langle \rangle$ is considered trivially well-formed. $Queue'[] = Queue[]$ means $Queue'$ is also considered trivially well-formed as $Queue$ was well-formed.

Each process, $p_i$, appears at most once in $Queue[]$ --- $\forall p_i, p_j \in P \ \bullet \ pos(p_i) \neq pos(p_j) \implies p_i \neq p_j$. That is, $Queue$ is a set of claiming processes. No duplicates can appear in the queue due to Assumption~\ref{ass:double} and Assertion~\ref{ass:waiting}.
\end{property}

\begin{property}[(O) Ownership]
Exactly one of the following holds:
\begin{enumerate}
    \item {\bf Unlocked} --- $Owner = \bot \land |Queue| = 0$.
    \item {\bf Claiming} --- $Owner = \bot \land |Queue| \neq 0 \land claiming(head)$.
    \item {\bf Locked} --- $Owner = head$.
    \item {\bf Releasing} --- $Owner \neq \bot \land Owner \notin Queue$.
\end{enumerate}
\end{property}

\begin{property}[(F) FIFO promise]
    Considered under two conditions:
    \begin{enumerate}
        \item \label{F1} (\textit{Claim enqueued condition}) On successful {\it CAS} of $claim.p_i$, $tail' = p_i$ where $tail'$ is the last element of $Queue'[]$ on completion of {\it CAS} operation.
        \item \label{F2} (\textit{Released condition}) Whenever $releasing(Owner)$ transfers from true to false, $(Owner' \neq \bot \land Owner' = head') \implies |Queue| \geq 1$, otherwise $Owner' = \bot$ where $Owner'$ is the next resource owner on completion after transition step from $releasing(Owner)$ is true to false.
    \end{enumerate}

    In all other situations, $Queue$ is not changed in a manner that impacts FIFO ordering and so (F) trivially holds.  $|Queue| = 0 \implies Queue[] = \langle \rangle$ is empty and is trivially FIFO ordered.
\end{property}

$INV(n)$ holds when $(S) + (O) + (F)$ all hold. $Qstate$ must meet $(S) + (O) + (F)$ during all operation steps and thus all values of $Qstate$ and transitions between values of $Qstate$.
\end{defn}

\begin{thm}[$INV(n)$ holds $\forall n \geq 1 \implies$ mutual exclusion and FIFO fairness]
Demonstrate that $(S) + (O) + (F)$ for all cases.
\begin{proof}
We prove by induction.

\begin{hypothesis}
Assume $INV(n)$ holds after an arbitrary sequence of operations for some $n \geq 1$.
\end{hypothesis}

\begin{case}[Base case $n = 0$]
We have the initial state:
\begin{itemize}
    \item $Qstate = Qstate(0, false) = (Queue[] = \langle \rangle, Owner = \bot)$. 
    \item Thus, $|Queue| = 0, head = \bot, tail = \bot$. Currently {\it Unlocked}.
\end{itemize}

$INV(0)$ trivially satisfies $(S) + (O) + (F)$. Case verified in FDR as initial state of model.
\end{case}

\begin{case}[Claim step, $INV(n) \implies INV(n + 1)$]
\label{case:claim}
Three cases to consider: $n = 0, n = 1$, and $n \geq 2$.

Event $claim.p_l$ occurs. $claiming(p_l)$ is true until stated otherwise or until end of subcase outcome.

\begin{subcase}[$n = 0, INV(0) \implies INV(1)$]
$Qstate = Qstate(0, false) = (Queue[] = \langle \rangle, Owner = \bot)$. $|Queue| = 0, head = \bot, tail = \bot$. $claim.p_l$ will add $p_l$ to $Queue$. Two potential outcomes.
\begin{enumerate}
    \item {\it CAS} succeeds. $Qstate' = Qstate(1, false) = (Queue'[] = \langle p_l \rangle, Owner' = p_l)$. $|Queue'| = 1, head' = p_l, tail' = p_l$.
    \begin{description}
        \item[(S)] holds under claim condition --- $|Queue'| = |Queue| + 1 \implies tail' \neq \bot \land tail = \bot$.
        \item[(O)] {\it Locked}; $Owner' = p_l = head'$.
        \item[(F)] holds under claim enqueued condition. $tail' = p_l$.
    \end{description}
    As |$Queue'| = 1$, $INV(0) \implies INV(1)$ in this outcome.
    \item $claim.p_m$ occurs and {\it CAS} for $claim.p_n$ fails. Three steps:
    \begin{enumerate}[1. ]
        \item {\it CAS} for $p_m$ completes. Assume full completion before {\it CAS} for $p_l$ for simplicity ($p_l$ could retry {\it CAS} prior to $Owner' = p_m$, but outcome is the same). $Qstate' = Qstate(1, false)$. $(S) + (O) + (F)$ hold as this is the same as the first outcome of this subcase.
        \item {\it CAS} for $p_l$ fails. $n'' = n' = 1$ and $Qstate'' = Qstate'$. Therefore, $(S) + (O) + (F)$ still holds.
        \item As $n'' = |Queue''| = 1$, continued proof for {\it CAS} reattempt covered in Subcase~\ref{sub:cinv1} $n = 1, INV(1) \implies INV(2)$. However, $INV(0) \implies INV(1)$ is met.
    \end{enumerate}
    As $|Queue''| = 1$, $INV(0) \implies INV(1)$ in this outcome.
\end{enumerate}

Therefore, $INV(0) \implies INV(1)$ satisfies $(S) + (O) + (F)$ for all outcomes. Case also verified mechanically in FDR for $n = 0$.
\end{subcase}

\begin{subcase}[$n = 1, INV(1) \implies INV(2)$]
\label{sub:cinv1}
$Qstate = Qstate(1, false) = (Queue[] = \langle p_i \rangle, Owner = p_i)$. $|Queue| = 1, head = p_i, tail = p_i$. Five possible outcomes for {\it CAS} attempt:
\begin{enumerate}
    \item {\it CAS} succeeds; $Qstate' = Qstate(2, false) = (Queue'[] = \langle p_i, p_l \rangle, Owner' = p_i)$. $|Queue'| = 2, head' = p_i, tail' = p_l$. $waiting(p_l)$ is true.
    \begin{description}
        \item[(S)] holds under claim condition --- $|Queue'| = |Queue| + 1, tail' = p_l, tail = p_i, next(p_i) = p_l$.
        \item[(O)] trivially holds as $Owner' = Owner$; {\it Locked}.
        \item[(F)] holds under claim enqueued condition. $tail' = p_l$.
    \end{description}
    As $|Queue'| = |Queue| + 1$, $INV(1) \implies INV(2)$ in this outcome.
    \item $claim.p_m$ occurs and {\it CAS} for $claim.p_l$ fails. Three steps:
    \begin{enumerate}[1. ]
        \item {\it CAS} for $p_m$ completes. $Qstate' = Qstate(2, false) = (Queue'[] = \langle p_i, p_m \rangle, Owner' = p_i)$. $|Queue'| = 2, head' = p_i, tail' = p_m$. $waiting(p_m)$ is true.
        \begin{description}
            \item[(S)] holds under claim condition --- $|Queue'| = |Queue| + 1, tail' = p_m, tail = p_i, next(p_i) = p_m$.
            \item[(O)] trivially holds as $Owner' = Owner$; {\it Locked}.
            \item[(F)] holds under claim enqueued condition. $tail' = p_m$.
        \end{description}
        \item {\it CAS} for $p_l$ fails. $n'' = n' = 2$ and $Qstate'' = Qstate'$. Therefore, $(S) + (O) + (F)$ still holds.
        \item As $n'' = |Queue''| = 2$, proof for reattempt of {\it CAS} continued under Subcase~\ref{sub:cinv2}. However, $INV(1) \implies INV(2)$ met.
    \end{enumerate}
    As $|Queue''| = |Queue| + 1$, $INV(1) \implies INV(2)$ in this outcome.
    \item $release.p_i$ occurs and changes $Owner$ to $\bot$ before $p_l$ sets it to $p_l$. Three steps:
    \begin{enumerate}[1. ]
        \item $release.p_i$ dequeues $p_i$. $Qstate' = Qstate(0, true) = (Queue'[] = \langle \rangle, Owner' = p_i)$. $|Queue'| = 0, head' = \bot, tail' = \bot$. $releasing(p_i)$ is true.
        \begin{description}
            \item[(S)] holds under release condition --- $|Queue'| = |Queue| - 1, head' = \bot = next(head)$.
            \item[(O)] {\it Releasing}; $Owner' = p_i \neq \bot \land p_i \notin Queue'[]$.
            \item[(F)] trivially holds as $|Queue'| = 0$.
        \end{description}
        \item $Owner$ set to $\bot$. $Qstate'' = Qstate(0, false) = (Queue''[] = \langle \rangle, Owner'' = \bot)$. \\$|Queue''| = 0, head'' = \bot, tail'' = \bot$. $releasing(p_i)$ becomes false.
        \begin{description}
            \item[(S)] trivially well-formed.
            \item[(O)] {\it Unlocked}; $Owner'' = \bot \land |Queue''| = 0$.
            \item[(F)] holds under released condition. $Owner'' = \bot, |Queue'| = 0$.
        \end{description}
        \item $claim.p_l$ succeeds. As $|Qstate''| = 0$, this is the same outcome as the first where $claim$ succeeds. Therefore, $(S) + (O) + (F)$ hold.
    \end{enumerate}
    As $|Queue'''| = 1$, $INV(1) \implies INV(1)$ in this outcome, which is obvious.
    \item $release.p_i$ occurs and sets $Owner = p_l$. Three steps:
    \begin{enumerate}[1. ]
        \item {\it CAS} succeeds for $p_l$. $Qstate' = Qstate(2, false) = (Queue'[] = \langle p_i, p_l \rangle, Owner' = p_i)$. $|Queue'| = 2, head' = p_i, tail' = p_l$. $claiming(p_l)$ becomes false and $waiting(p_l)$ is true. $claiming(p_l)$ becoming false could occur after $Owner$ is set to $p_l$ but does not affect outcome.
        \begin{description}
            \item[(S)] holds under claim condition --- $|Queue'| = |Queue| + 1, tail' = p_l, tail = p_i \neq \bot, next(p_i) = p_l$
            \item[(O)] trivially holds as $Owner' = Owner$; {\it Locked}.
            \item[(F)] holds under claim enqueued condition. $tail' = p_l$.
        \end{description}
        \item $release.p_i$ continues. $Qstate'' = Qstate(1, true) = (Queue''[] = \langle p_l \rangle, Owner'' = p_i)$. $|Queue''| = 1, head'' = p_l, tail'' = p_l$. $releasing(p_i)$ becomes true.
        \begin{description}
            \item[(S)] holds under release condition --- $|Queue''| = |Queue'| - 1, head = p_i, head' = p_l = next(p_i)$.
            \item[(O)] {\it Releasing}; $Owner'' = p_i \neq \bigotimes \land p_i \notin Queue''[]$.
            \item[(F)] trivially holds.
        \end{description}
        \item Release completes setting $p_l$ as owner. $Qstate''' = Qstate(1, false) = (Queue'''[] = \langle p_l \rangle, Owner''' = p_l)$. $|Queue'''| = 1, head''' = p_l, tail''' = p_l$. $waiting(p_l)$ becomes false and $releasing(p_i)$ becomes false.
        \begin{description}
            \item[(S)] trivially well-formed as $Queue'''[] = Queue''[]$.
            \item[(O)] {\it Locked}; $Owner''' = p_l = head'''$.
            \item[(F)] holds under released condition. $Owner''' = p_l \neq \bot \land |Queue''| \geq 1$.
        \end{description}
    \end{enumerate}
    As $|Queue'''| = 1$, $INV(1) \implies INV(1)$ which is obvious.
    \item $release.p_i$ occurs and $p_l$ steals ownership. Three steps:
    \begin{enumerate}[1. ]
        \item $p_i$ removes itself from $Queue$. $Qstate' = Qstate(0, true) = (Queue'[] = \langle \rangle, Owner' = p_i)$. $|Queue'| = 0, head' = \bot, tail' = \bot$. $releasing(p_i)$ is true.
        \begin{description}
            \item[(S)] holds under release condition --- $|Queue'| = |Queue| - 1, head = p_i, head' = \bot = next(head)$.
            \item[(O)] {\it Releasing}; $Owner' = p_i \neq \bot \land Owner' \notin Queue'[]$.
            \item[(F)] trivially holds as $|Queue'| = 0$.
        \end{description}
        \item {\it CAS} succeeds for $p_l$. $Qstate'' = Qstate(1, true) = (Queue''[] = \langle p_l \rangle, Owner'' = p_i)$. $|Queue''| = 1, head'' = p_l, tail'' = p_l$. $releasing(p_i)$ is true.
        \begin{description}
            \item[(S)] holds under claim condition --- $|Queue'| = |Queue| + 1, tail' = p_l, tail = p_i \neq \bot, next(p_i) = p_l$.
            \item[(O)] {\it Releasing}; $Owner' = p_i \neq \bot \land Owner' \notin Queue'[]$.
            \item[(F)] holds under claim enqueued condition. $tail'' = p_l$.
        \end{description}
        \item $p_l$ sets $Owner$ to itself. $Qstate''' = Qstate(1, false) = (Queue'''[] = \langle p_l \rangle, Owner''' = p_l)$. $|Queue'''| = 1, head''' = p_l, tail''' = p_l$. $releasing(p_i)$ becomes false.
        \begin{description}
            \item[(S)] trivially holds as $Queue'''[] = Queue''[]$.
            \item[(O)] {\it Locked}; $Owner''' = p_l = head'''$.
            \item[(F)] holds under released condition. $Owner''' = p_l \land |Queue''| \geq 1$.
        \end{description}
    \end{enumerate}
    As $|Queue'''| = 1$, $INV(1) \implies INV(1)$ in this outcome,  which is obvious.
\end{enumerate}

Therefore, $INV(1) \implies INV(2)$ satisfies $(S) + (O) + (F)$ under all outcomes. Case also verified mechanically in FDR for $n = 2$.
\end{subcase}

\begin{subcase}[$n \geq 2, \forall n \geq 2 \bullet INV(n) \implies INV(n + 1)$]
\label{sub:cinv2}
$Qstate = Qstate(n, false) = (Queue[] = \langle p_i, p_j, \dots, p_k \rangle, Owner = p_i)$. $|Queue| = n, head = p_i, tail = p_k$. Three possible outcomes for {\it CAS} attempt:
\begin{enumerate}
    \item {\it CAS} succeeds. $Qstate' = Qstate(n + 1, false) = (Queue'[] = \langle p_i, p_j, \dots, p_k, p_l \rangle, Owner' = p_i)$. $|Queue'| = |Queue| + 1 = n + 1, head' = p_i, tail' = p_l$. $claiming(p_l)$ becomes false and $waiting(p_l)$ becomes true.
    \begin{description}
        \item[(S)] holds under claim condition --- $|Queue'| = |Queue| + 1, tail' = p_l, tail = p_k \neq \bot, next(p_k) = p_l$.
        \item[(O)] {\it Locked}; $Owner' = p_i = head'$. $Owner$ is unchanged.
        \item[(F)] holds under claim enqueued condition. $tail'' = p_l$.
    \end{description}
    As $|Queue'| = |Queue| + 1$, $INV(n) \implies INV(n + 1)$ in this outcome.
    \item $claim.p_m$ occurs and {\it CAS} for $claim.p_l$ fails. Three steps:
    \begin{enumerate}[1. ]
        \item {\it CAS} $p_m$ completes. $Qstate' = Qstate(n + 1, false) = (Queue'[] = \langle p_i, p_j, \dots, p_k, p_m \rangle,\\ Owner' = p_i)$. $|Queue'| = |Queue| + 1 = n + 1, head' = p_i, tail' = p_m$. $waiting(p_m)$ is true.
        \begin{description}
            \item[(S)] holds under claim condition --- $|Queue'| = |Queue| + 1, tail' = p_m, tail = p_k, next(p_k) = p_m$.
            \item[(O)] trivially holds as $Owner' = Owner$; {\it Locked}.
            \item[(F)] holds under claim enqueued condition. $tail' = p_m$.
        \end{description}
        \item {\it CAS} for $p_l$ fails. $n'' = n' = n + 1$ and $Qstate'' = Qstate'$. Therefore, $(S) + (O) + (F)$ still holds.
        \item As $n'' = |Queue''| = n + 1$, proof for reattempt of {\it CAS} continued under this subcase. However, $INV(n) \implies INV(n + 1)$ met.
    \end{enumerate}
    As $|Queue''| = |Queue| + 1$, $INV(n) \implies INV(n + 1)$.
    \item $release.p_i$ occurs. $p_l$ cannot become the owner (next owner is $p_j$) so only outcome of note is $release$ completing before $claim.p_l$. Two steps.
    \begin{enumerate}[1. ]
        \item $release.p_i$ completes. $Qstate' = Qstate(n - 1, false) = (Queue'[] = \langle p_j, \dots, p_k \rangle, \\Owner' = p_j)$. $|Queue'| = |Queue| - 1 = n - 1, head' = p_j, tail' = p_k$.
        \begin{description}
            \item[(S)] holds under release condition --- $|Queue'| = |Queue| - 1, head = p_i, head' = p_j = next(head)$.
            \item[(O)] {\it Locked}; $Owner' = p_j = head'$.
            \item[(F)] holds under released condition. When $releasing(p_i)$ changed from true to false, $Owner' = p_j \neq \bot$, $Owner' = p_j = head'$, $|Queue| \geq 1$.
        \end{description}
        \item $claim.p_l$ completes. $Qstate'' = Qstate(n, false) = (Queue''[] = \langle p_j, \dots, p_k, p_l \rangle, \\Owner'' = p_j)$. $|Queue''| = |Queue'| + 1 = |Queue| = n, head'' = p_j, tail'' = p_l$.
        \begin{description}
            \item[(S)] holds under claim condition --- $|Queue''| = |Queue'| + 1, tail'' = p_l, tail = p_k, next(p_k) = p_l$.
            \item[(O)] trivially holds as $Owner'' = Owner'$; {\it Locked}.
            \item[(F)] holds under claim enqueued condition. $tail'' = p_l$.
        \end{description}
    \end{enumerate}
    As $|Queue''| = |Queue| = n$, $INV(n) \implies INV(n)$ in this outcome, which is obvious.
\end{enumerate}

Therefore, $\forall n \geq 2 \ \bullet \ INV(n) \implies INV(n + 1)$ satisfies $(S) + (O) + (F)$. Case also verified mechanically in FDR for $n = 2$ and $n = 3$.
\end{subcase}

As $INV(0) \implies INV(1) \implies INV(2)$ and $\forall n \geq 2 \ \bullet \ INV(n) \implies INV(n + 1)$, then $INV$ holds for all Claim step cases.
\end{case}

\begin{case}[Release step, $INV(n) \implies INV(n - 1)$] 
\label{case:release}
Release cannot occur when $n = 0$ due to Assumption~\ref{ass:double}. Two subcases to consider: $n = 1$ and $n \geq 2$.

Event $release.p_l$ occurs.

\begin{subcase}[$n = 1, INV(1) \implies INV(0)$]
$Qstate = Qstate(1, false) = (Queue[] = \langle p_i \rangle, Owner = p_i)$. $|Queue| = 1, head = p_i, tail = p_i$. Four possible outcomes:
\begin{enumerate}
    \item $release.p_i$ completes without $claim.p_l$ occurring. $Qstate' = Qstate(0, false) = (Queue[] = \langle \rangle, Owner = \bot)$$head' = \bot$; $tail' = \bot$; $|Queue'| = 0$. $releasing(p_i)$ being true not considered as steps trivial.
    \begin{description}
        \item[(S)] trivially holds as $|Queue'| = 0$.
        \item[(O)] {\it Unlocked}; $Owner' = \bot \land |Queue'| = 0$.
        \item[(F)] holds under released condition. When $releasing(p_i)$ became false, $|Queue'| = 0$, and $Owner' = \bot$.  
    \end{description}
    As $|Queue'| = 0$, $INV(1) \implies INV(0)$ met.
    \item $release.p_i$ completes without observing an ongoing $claim.p_l$. Assume $claiming(p_l)$ at start, although can happen at any point before $p_l$ attempts {\it CAS} at step 3. Five steps.
    \begin{enumerate}[1. ]
        \item $p_i$ is dequeued. $Qstate' = Qstate(0, true) = (Queue'[] = \langle \rangle, Owner' = p_i)$. $|Queue'| = 0, head' = \bot, tail' = \bot$. $releasing(p_i)$ is true.
        \begin{description}
            \item[(S)] holds under release condition --- $|Queue'| = |Queue| - 1, head = p_i, head' = \bot = next(head)$.
            \item[(O)] {\it Releasing}; $releasing(p_i) \land p_i \notin Queue'$.
            \item[(F)] trivially holds.
        \end{description}
        \item $p_i$ peeks at $head'$. No state change, but $p_i$ committed to set $Owner' = \bot$. $Qstate'' = Qstate'$ so $(S) + (O) + (F)$ hold from previous step.
        \item $p_l$ successful {\it CAS} on queue. $Qstate''' = Qstate(1, true) = (Queue'''[] = \langle p_l \rangle, Owner''' = p_i)$. $|Queue'''| = 1, head''' = p_l, tail''' = p_l$.
        \begin{description}
            \item[(S)] holds under claim condition --- $|Queue'''| = |Queue''| + 1, tail''' = p_l \neq \bot, tail'' = \bot$.
            \item[(O)] {\it Releasing}; $Owner' = p_i \neq \bot \land Owner' \notin Queue'[]$.
            \item[(F)] holds under claim enqueued condition, $tail''' = p_l$.
        \end{description}
        \item $p_i$ sets $Owner$ to $\bot$. $Qstate'''' = Qstate(1, false) = (Queue''''[] = \langle p_l \rangle, Owner'''' = \bot)$ $|Owner''''| = 1, head'''' = p_l, tail'''' = p_l$. $releasing(p_i)$ becomes false.
        \begin{description}
            \item[(S)] trivially holds as $Queue[]'''' = Queue'''$.
            \item[(O)] {\it Claiming}; $Owner = \bot \land claiming(p_l)$.
            \item[(F)] trivially holds.
        \end{description}
        \item $p_l$ will set $Owner$ to $p_l$. $Qstate''''' = Qstate(1, false) = (Queue'''''[] = \langle p_l \rangle, Owner''''' = p_l)$. $|Owner'''''| = p_l, head''''' = p_l, tail''''' = p_l$. $claiming(p_l)$ becomes false.
        \begin{description}
            \item[(S)] trivially holds as $Queue[]''''' = Queue''''$.
            \item[(O)] {\it Locked}; $Owner''''' = p_l = head'''''$.
            \item[(F)] trivially holds.
        \end{description}
    \end{enumerate}
    As $|Queue'''''| = 1$, $INV(1) \implies INV(1)$ which is obvious.
    \item $release.p_i$ completes and observes ongoing $claim.p_l$. Assume $claiming(p_l)$ at start but can occur any point before $p_l$ attempts {\it CAS} on step~ii. Three steps.
    \begin{enumerate}[1. ]
        \item $p_i$ is dequeued. $Qstate' = Qstate(0, true) = (Queue'[] = \langle \rangle, Owner' = p_i)$. $|Queue'| = 0, head' = \bot, tail' = \bot$. $releasing(p_i)$ is true.
        \begin{description}
            \item[(S)] holds under release condition --- $|Queue'| = |Queue| - 1, head = p_i, head' = \bot = next(head)$.
            \item[(O)] {\it Releasing}; $Owner' = p_i \neq \bot \land Owner' \notin Queue'[]$.
            \item[(F)] trivially holds.
        \end{description}
        \item {\it CAS} for $p_l$ succeeds. $Qstate'' = Qstate(1, true) = (Queue''[] = \langle p_l \rangle, Owner'' = p_i)$. $|Queue''| = 1, head'' = p_l, tail'' = p_l$. $releasing(p_i)$ is true.
        \begin{description}
            \item[(S)] holds under claim condition --- $|Queue''| = |Queue'| + 1, tail'' = p_l \neq \bot, tail' = \bot$.
            \item[(O)] {\it Releasing}; $Owner' = p_i \neq \bot \land Owner' \notin Queue'[]$.
            \item[(F)] holds under claim enqueued condition, $tail'' = p_l$.
        \end{description}
        \item $p_i$ observes new $head$ and sets $Owner$. $Qstate'' = Qstate(1, true) = (Queue'''[] = \langle p_l \rangle, Owner''' = p_l)$. $|Queue'''| = 1, head''' = p_l, tail''' = p_l$. $releasing(p_i)$ becomes false.
        \begin{description}
            \item[(S)] trivially holds as $Queue'''[] = Queue''[]$.
            \item[(O)] {\it Locked}; $Owner''' = p_l = head'''$.
            \item[(F)] trivially holds as $Queue'''[] = Queue[]$.
        \end{description}
    \end{enumerate}
    As $|Queue'''| = 1$, $INV(1) \implies INV(1)$ which is obvious.
    \item $claim.p_l$ steals ownership during $release.p_i$. Assume $claiming(p_l)$ at start but can occur any point before $p_l$ attempts {\it CAS} on step~ii. Three steps.
    \begin{enumerate}[1. ]
        \item $p_i$ is dequeued. $Qstate' = Qstate(0, true) = (Queue'[] = \langle \rangle, Owner' = p_i)$. $|Queue'| = 0, head' = \bot, tail' = \bot$. $releasing(p_i)$ is true.
        \begin{description}
            \item[(S)] holds under release condition --- $|Queue'| = |Queue| - 1, head = p_i, head' = \bot = next(head)$.
            \item[(O)] {\it Releasing}; $Owner' = p_i \neq \bot \land Owner' \notin Queue'[]$.
            \item[(F)] trivially holds.
        \end{description}
        \item {\it CAS} for $p_l$ succeeds. $Qstate'' = Qstate(1, true) = (Queue''[] = \langle p_l \rangle, Owner'' = p_i)$. $|Queue''| = 1, head'' = p_l, tail'' = p_l$. $releasing(p_i)$ is true.
        \begin{description}
            \item[(S)] holds under claim condition --- $|Queue''| = |Queue'| + 1, tail'' = p_l \neq \bot, tail' = \bot$.
            \item[(O)] {\it Releasing}; $Owner' = p_i \neq \bot \land Owner' \notin Queue'[]$.
            \item[(F)] holds under claim enqueued condition, $tail'' = p_l$.
        \end{description}
        \item $p_l$ sets $Owner$ to itself. $Qstate'' = Qstate(1, true) = (Queue'''[] = \langle p_l \rangle, Owner''' = p_l)$. $|Queue'''| = 1, head''' = p_l, tail''' = p_l$. $releasing(p_i)$ becomes false.
        \begin{description}
            \item[(S)] trivially holds as $Queue'''[] = Queue''[]$.
            \item[(O)] {\it Locked}; $Owner''' = p_l = head'''$.
            \item[(F)] trivially holds.
        \end{description}
    \end{enumerate}
    As $|Queue'''| = 1$, $INV(1) \implies INV(1)$ which is obvious.
\end{enumerate}

If more than one $claim$ events occurs during $release.p_i$, then $|Queue| \geq 2$. As only the first $claim$ {\it CAS} can modify $head$, subsequent $claim$ events do not influence the ongoing interaction between a concurrent $claim$ and $release$. Thus, we do not have to explicitly consider multiple $claim$ events occurring in our proof. $INV(n + 1)$ has been demonstrated in Case~\ref{case:claim}.

Therefore, $INV(1) \implies INV(0)$ satisfies $(S) + (O) + (F)$. Case also verified mechanically in FDR for $n = 1$.
\end{subcase}

\begin{subcase}[$n \geq 2, \forall n \geq 2 \bullet INV(n) \implies INV(n - 1)$]
\label{sub:rinv2}
$Qstate = Qstate(n, false) = (Queue[] = \langle p_i, p_j, \dots, p_k \rangle, Owner = p_i)$. $|Queue| = n, head = p_i, tail = p_k$. Two possible outcomes:
\begin{enumerate}
    \item $release.p_i$ completes. $Qstate' = Qstate(n - 1, false) = (Queue[] = \langle p_j, \dots, p_k \rangle, Owner = \bot)$. $head' = p_j$; $tail' = p_k$; $|Queue'| = n - 1$; $Queue'[] = \langle \rangle$; $Owner' = \bot$. $releasing(p_i)$ being true not considered as steps trivial.
    \begin{description}
        \item[(S)] holds under release condition --- $|Queue'| = |Queue| - 1, head = p_i, head' = p_j = next(head)$.
        \item[(O)] {\it Locked}; $Owner' = p_j = head'$.
        \item[(F)] holds under released condition. When $releasing(p_i)$ changed from true to false, $Owner' = p_j \neq \bot$, $Owner' = p_j = head'$, $|Queue| \geq 1$.
    \end{description}
    As $|Queue'| = |Queue| - 1 = n - 1$, $INV(n) \implies INV(n - 1)$ met.
    \item $claim.p_l$ occurs during $release.p_i$. $p_l$ cannot become the owner (next owner is $p_j$) so only outcome of note is $release$ completing after $claim.p_l$. Two steps.
    \begin{enumerate}[1. ]
        \item $claim.p_l$ completes. $Qstate' = Qstate(n + 1, false) = (Queue'[] = \langle p_i, p_j, \dots, p_k, p_l \rangle, \\Owner' = p_i)$. $|Queue'| = |Queue| + 1, head' = p_i, tail' = p_l$.
        \begin{description}
            \item[(S)] holds under claim condition --- $|Queue'| = |Queue| + 1, tail' = p_l, tail = p_k, \\next(p_k) = p_l$.
            \item[(O)] trivially holds as $Owner' = Owner$; {\it Locked}.
            \item[(F)] holds under claim enqueued condition. $tail' = p_l$.
        \end{description}
        \item $release.p_i$ completes. $Qstate'' = Qstate(n, false) = (Queue''[] = \langle p_j, \dots, p_k, p_l \rangle, \\Owner'' = p_j)$. $|Queue''| = |Queue'| - 1 = n, head'' = p_j, tail'' = p_l$.
        \begin{description}
            \item[(S)] holds under release condition --- $|Queue''| = |Queue'| - 1, head' = p_i, head'' = p_j = next(head')$.
            \item[(O)] {\it Locked}; $Owner'' = p_j = head''$.
            \item[(F)] holds under released condition. When $releasing(p_i)$ changed from true to false, $Owner'' = p_j \neq \bot$, $Owner'' = p_j = head''$, $|Queue'| \geq 1$.
        \end{description}
    \end{enumerate}
    As $|Queue''| = |Queue| = n$, $INV(n) \implies INV(n)$ in this outcome, which is obvious.
\end{enumerate}

Therefore, $\forall n \geq 2 \ \bullet \ INV(n) \implies INV(n - 1)$ satisfies $(S) + (O) + (F)$. Case also verified mechanically in FDR for $2 \leq n \leq 4$.
\end{subcase}

As $INV(1) \implies INV(0)$ and $\forall n \geq 2 \bullet INV(n) \implies INV(n - 1)$, then $INV$ holds for all Release step cases. 
\end{case}

As $\forall n \geq 0 \ \bullet \ INV(n) \implies INV(n + 1) \land \forall m \geq 1 \ \bullet \ INV(m) \implies INV(m - 1)$ holds for $(S) + (O) + (F)$, the mutex maintains mutual exclusion and FIFO fairness for any length of queue of waiting processes.
\end{proof}
\end{thm}

\end{document}